\documentclass[a4paper,11pt]{article}
\usepackage[maxbibnames=99]{biblatex}
\usepackage[text={190mm,240mm},centering]{geometry}
 
\usepackage{graphics,graphicx}
\usepackage{subfig}
\usepackage{float}
\usepackage{booktabs}
\usepackage{longtable}
\usepackage{supertabular}
\usepackage{tabularx}
\usepackage{multirow}
\usepackage{array}
\usepackage{tabu}
\usepackage{mathtools}
\usepackage{hyperref}
\usepackage{accents} 

\usepackage[ruled,vlined]{algorithm2e}

\usepackage{amsmath}
\usepackage{amsfonts}
\usepackage{amssymb}
\usepackage{siunitx} 

\usepackage{enumerate}
\usepackage{sectsty}
\usepackage[affil-it,auth-sc]{authblk}
\usepackage{color}
\usepackage[dvipsnames]{xcolor}

\sectionfont{\normalsize}
\subsectionfont{\normalsize}

\usepackage{tikz}
\usetikzlibrary{arrows,decorations,backgrounds,shapes, snakes}
\usetikzlibrary{positioning}
\usetikzlibrary{matrix} 
\usepackage{varwidth}
\usepackage[most]{tcolorbox}
\usetikzlibrary{shapes.geometric,arrows.meta,decorations.markings}

\usepackage{fancyhdr}            
\fancypagestyle{plain}{%
  \fancyfoot[CE,CO]{\thepage}       
  \fancyhead[CE,CO]{\textcolor[rgb]{0.64,0.15,0.15}{Published in \emph{Journal of the Mechanics and Physics of Solids}  \textbf{188}: 105673
  \\doi: https://doi.org/10.1016/j.jmps.2024.105673
  }}
\setlength{\headheight}{14.5pt}}

\begin{document}

\newcommand{\singlespace}{\baselineskip=12pt\lineskiplimit=0pt\lineskip=0pt}
\def\ds{\displaystyle}

\def\salto#1#2{
[\mbox{\hspace{-#1em}}[#2]\mbox{\hspace{-#1em}}]}

\tikzstyle{every picture}+=[remember picture]

\newcommand{\beq}{\begin{equation}}
\newcommand{\eeq}{\end{equation}}
\newcommand{\lb}{\label}
\newcommand{\ph}{\phantom}
\newcommand{\beqar}{\begin{eqnarray}}
\newcommand{\eeqar}{\end{eqnarray}}
\newcommand{\barr}{\begin{array}}
\newcommand{\earr}{\end{array}}
\newcommand{\jump}{\parallel}
\newcommand{\Ehat}{\hat{E}}
\newcommand{\That}{\hat{\bf T}}
\newcommand{\Ahat}{\hat{A}}
\newcommand{\chat}{\hat{c}}
\newcommand{\shat}{\hat{s}}
\newcommand{\khat}{\hat{k}}
\newcommand{\muhat}{\hat{\mu}}
\newcommand{\mc}{M^{\scriptscriptstyle C}}
\newcommand{\mei}{M^{\scriptscriptstyle M,EI}}
\newcommand{\mec}{M^{\scriptscriptstyle M,EC}}
\newcommand{\hbeta}{{\hat{\beta}}}
\newcommand{\rec}[2]{\left( #1 #2 \ds{\frac{1}{#1}}\right)}
\newcommand{\rep}[2]{\left( {#1}^2 #2 \ds{\frac{1}{{#1}^2}}\right)}
\newcommand{\derp}[2]{\ds{\frac {\partial #1}{\partial #2}}}
\newcommand{\derpn}[3]{\ds{\frac {\partial^{#3}#1}{\partial #2^{#3}}}}
\newcommand{\dert}[2]{\ds{\frac {d #1}{d #2}}}
\newcommand{\dertn}[3]{\ds{\frac {d^{#3} #1}{d #2^{#3}}}}
\newcommand{\ct}{\captionof{table}}
\newcommand{\cf}{\captionof{figure}}

\def\c{{\circ}}
\def\bob{{\, \underline{\overline{\otimes}} \,}}
\def\ob{{\, \underline{\otimes} \,}}
\def\scalp{\mbox{\boldmath$\, \cdot \, $}}
\def\gdp{\makebox{\raisebox{-.215ex}{$\Box$}\hspace{-.778em}$\times$}}
\def\daa{\makebox{\raisebox{-.050ex}{$-$}\hspace{-.550em}$: ~$}}
\def\mK{\mbox{${\mathcal{K}}$}}
\def\cK{\mbox{${\mathbb {K}}$}}

\def\Xint#1{\mathchoice
   {\XXint\displaystyle\textstyle{#1}}%
   {\XXint\textstyle\scriptstyle{#1}}%
   {\XXint\scriptstyle\scriptscriptstyle{#1}}%
   {\XXint\scriptscriptstyle\scriptscriptstyle{#1}}%
   \!\int}
\def\XXint#1#2#3{{\setbox0=\hbox{$#1{#2#3}{\int}$}
     \vcenter{\hbox{$#2#3$}}\kern-.5\wd0}}
\def\ddashint{\Xint=}
\def\fpint{\Xint=}
\def\dashint{\Xint-}
\def\cpvint{\Xint-}
\def\intl{\int\limits}
\def\cpvintl{\cpvint\limits}
\def\fpintl{\fpint\limits}
\def\ointl{\oint\limits}
\def\bA{{\bf A}}
\def\ba{{\bf a}}
\def\bB{{\bf B}}
\def\bb{{\bf b}}
\def\bc{{\bf c}}
\def\bC{{\bf C}}
\def\bD{{\bf D}}
\def\bE{{\bf E}}
\def\be{{\bf e}}
\def\bbf{{\bf f}}
\def\bF{{\bf F}}
\def\bG{{\bf G}}
\def\bg{{\bf g}}
\def\bi{{\bf i}}
\def\bI{{\bf I}}
\def\bH{{\bf H}}
\def\bK{{\bf K}}
\def\bL{{\bf L}}
\def\bM{{\bf M}}
\def\bN{{\bf N}}
\def\bn{{\bf n}}
\def\bm{{\bf m}}
\def\b0{{\bf 0}}
\def\bo{{\bf o}}
\def\bX{{\bf X}}
\def\bx{{\bf x}}
\def\bP{{\bf P}}
\def\bp{{\bf p}}
\def\bQ{{\bf Q}}
\def\bq{{\bf q}}
\def\bR{{\bf R}}
\def\bS{{\bf S}}
\def\bs{{\bf s}}
\def\bT{{\bf T}}
\def\bt{{\bf t}}
\def\bU{{\bf U}}
\def\bu{{\bf u}}
\def\bv{{\bf v}}
\def\bV{{\bf V}}
\def\bw{{\bf w}}
\def\bW{{\bf W}}
\def\by{{\bf y}}
\def\bz{{\bf z}}
\def\T{{\bf T}}
\def\Te{\textrm{T}}
\def\Id{{\bf I}}
\def\bxi{\mbox{\boldmath${\xi}$}}
\def\balpha{\mbox{\boldmath${\alpha}$}}
\def\bbeta{\mbox{\boldmath${\beta}$}}
\def\bepsilon{\mbox{\boldmath${\epsilon}$}}
\def\bvarepsilon{\mbox{\boldmath${\varepsilon}$}}
\def\bomega{\mbox{\boldmath${\omega}$}}
\def\bphi{\mbox{\boldmath${\phi}$}}
\def\bsigma{\mbox{\boldmath${\sigma}$}}
\def\bfeta{\mbox{\boldmath${\eta}$}}
\def\bDelta{\mbox{\boldmath${\Delta}$}}
\def\btau{\mbox{\boldmath $\tau$}}
\def\tr{{\rm tr}}
\def\dev{{\rm dev}}
\def\div{{\rm div}}
\def\Div{{\rm Div}}
\def\Grad{{\rm Grad}}
\def\grad{{\rm grad}}
\def\Lin{{\rm Lin}}
\def\Sym{{\rm Sym}}
\def\Skw{{\rm Skew}}
\def\abs{{\rm abs}}
\def\Re{{\rm Re}}
\def\Im{{\rm Im}}
\def\capB{\mbox{\boldmath${\mathsf B}$}}
\def\capC{\mbox{\boldmath${\mathsf C}$}}
\def\capD{\mbox{\boldmath${\mathsf D}$}}
\def\capE{\mbox{\boldmath${\mathsf E}$}}
\def\capG{\mbox{\boldmath${\mathsf G}$}}
\def\tcapG{\tilde{\capG}}
\def\capH{\mbox{\boldmath${\mathsf H}$}}
\def\capK{\mbox{\boldmath${\mathsf K}$}}
\def\capL{\mbox{\boldmath${\mathsf L}$}}
\def\capM{\mbox{\boldmath${\mathsf M}$}}
\def\capR{\mbox{\boldmath${\mathsf R}$}}
\def\capW{\mbox{\boldmath${\mathsf W}$}}

\def\i{\mbox{${\mathrm i}$}}
\def\mC{\mbox{\boldmath${\mathcal C}$}}
\def\mB{\mbox{${\mathcal B}$}}
\def\mE{\mbox{${\mathcal{E}}$}}
\def\mL{\mbox{${\mathcal{L}}$}}
\def\mK{\mbox{${\mathcal{K}}$}}
\def\mV{\mbox{${\mathcal{V}}$}}
\def\C{\mbox{\boldmath${\mathcal C}$}}
\def\E{\mbox{\boldmath${\mathcal E}$}}

\def\AAM{{\it Advances in Applied Mechanics }}
\def\ACME{{\it Arch. Comput. Meth. Engng.}}
\def\ARMA{{\it Arch. Rat. Mech. Analysis}}
\def\AMR{{\it Appl. Mech. Rev.}}
\def\ASCEEM{{\it ASCE J. Eng. Mech.}}
\def\ACTA{{\it Acta Mater.}}
\def\CMAME {{\it Comput. Meth. Appl. Mech. Engrg.}}
\def\CRAS{{\it C. R. Acad. Sci. Paris}}
\def\CRM{{\it Comptes Rendus M\'ecanique}}
\def\EFM{{\it Eng. Fracture Mechanics}}
\def\EJMA{{\it Eur.~J.~Mechanics-A/Solids}}
\def\IJES{{\it Int. J. Eng. Sci.}}
\def\IJF{{\it Int. J. Fracture}}
\def\IJMS{{\it Int. J. Mech. Sci.}}
\def\IJNAMG{{\it Int. J. Numer. Anal. Meth. Geomech.}}
\def\IJP{{\it Int. J. Plasticity}}
\def\IJSS{{\it Int. J. Solids Structures}}
\def\IngA{{\it Ing. Archiv}}
\def\JAM{{\it J. Appl. Mech.}}
\def\JAP{{\it J. Appl. Phys.}}
\def\JAE{{\it J. Aerospace Eng.}}
\def\JE{{\it J. Elasticity}}
\def\JM{{\it J. de M\'ecanique}}
\def\JMPS{{\it J. Mech. Phys. Solids}}
\def\JSV{{\it J. Sound and Vibration}}
\def\MACRO{{\it Macromolecules}}
\def\MMT{{\it Mech. Mach. Th.}}
\def\MOM{{\it Mech. Materials}}
\def\MMS{{\it Math. Mech. Solids}}
\def\MMT{{\it Metall. Mater. Trans. A}}
\def\MPCPS{{\it Math. Proc. Camb. Phil. Soc.}}
\def\MSE{{\it Mater. Sci. Eng.}}
\def\NATURE{{\it Nature}}
\def\NATUREM{{\it Nature Mater.}}
\def\PHIL{{\it Phil. Trans. R. Soc.}}
\def\PMPS{{\it Proc. Math. Phys. Soc.}}
\def\PNAS{{\it Proc. Nat. Acad. Sci.}}
\def\PRE{{\it Phys. Rev. E}}
\def\PRL{{\it Phys. Rev. Letters}}
\def\PRSL{{\it Proc. R. Soc.}}
\def\RIIT{{\it Rozprawy Inzynierskie - Engineering Transactions}}
\def\ROCK{{\it Rock Mech. and Rock Eng.}}
\def\QAM{{\it Quart. Appl. Math.}}
\def\QJMAM{{\it Quart. J. Mech. Appl. Math.}}
\def\SCIENCE{{\it Science}}
\def\SCRMAT{{\it Scripta Mater.}}
\def\SM{{\it Scripta Metall.}}
\def\ZAMM{{\it Z. Angew. Math. Mech.}}
\def\ZAMP{{\it Z. Angew. Math. Phys.}}
\def\ZVDI{{\it Z. Verein. Deut. Ing.}}

\def\salto#1#2{
[\mbox{\hspace{-#1em}}[#2]\mbox{\hspace{-#1em}}]}

\renewcommand\Affilfont{\itshape}
\setlength{\affilsep}{1em}
\renewcommand\Authsep{, }
\renewcommand\Authand{ and }
\renewcommand\Authands{ and }
\setcounter{Maxaffil}{2}

\title{Elastic solids under
frictionless rigid contact and configurational force}
\author[]{Francesco Dal Corso}
\author[]{Marco Amato}
\author[]{Davide Bigoni\footnote{Corresponding author: bigoni@ing.unitn.it}}
\affil[]{DICAM, University of Trento, via~Mesiano~77, I-38123 Trento, Italy}

\maketitle

\begin{abstract}
A  homogeneous elastic solid, bounded by a flat surface in its unstressed configuration, 
undergoes a finite strain when in 
frictionless contact against a rigid and rectilinear constraint, ending with a rounded or sharp corner, in a two-dimensional formulation. With a strong analogy to fracture mechanics, it is shown that  (i.)  a path-independent $J$--integral can be defined for frictionless contact problems,  (ii.) which is    
equal to the energy release rate $G$ associated with an infinitesimal growth in the size of the frictionless constraint, and thus gives  the value of the configurational force component along the sliding direction. Furthermore, it is found that (iii.) such a configurational sliding force is the  Newtonian force component exerted by the elastic solid on the constraint at the frictionless contact. 
Assuming the kinematics of an Euler-Bernoulli rod for an elastic body of rectangular shape, the 
results (i.)--(iii.) lead to a new interpretation from a nonlinear  solid mechanics perspective  of 
the configurational forces recently  disclosed for one-dimensional structures of variable length. 
Finally, approximate but closed-form solutions (validated with finite element simulations) are exploited to provide further insight into the effect of configurational forces. 
In particular, two applications are presented which show  that 
a transverse compression can lead to Eulerian  buckling or to  longitudinal dynamic motion, 
both realizing novel examples of soft actuation mechanisms. 
As an application to biology, our results may provide a mechanical explanation for the observed phenomenon of negative durotaxis, where  cells migrate from stiffer to softer environments. 
\end{abstract}

\section{Introduction}

Initiated by Eshelby \cite{eshelby1956lattice, eshelby1975, eshelby1951force}, configurational mechanics provides a groundbreaking insight into problems where  a defect can change its position or increase in size and release  energy, which is associated to a force, called \lq configurational',  acting on the defect and causing its movement. In the specific case of a 
rectilinear crack in a linear elastic material, the energy release rate $G$ associated with a crack advancement was  found by Cherepanov \cite{Cherepanov}
and Rice \cite{rice1968path, rice1968} to be given by  a  path-independent integral, the so-called $J$--integral. The latter author involved the energy-momentum tensor $\mathbf{P}$ introduced by Eshelby \cite{eshelby1951force}, so that a crack driving force can be related  to fracture growth. 

Historically, configurational forces were assumed to be different in nature from Newtonian forces, which enter the equations of motion of a solid  \cite{gurtin1999nature,  kienzler}. 
However, a number of elastic structures with variable length has been recently investigated to show  
that a special class of configurational forces are Newtonian forces
and, as such, can even be determined experimentally. 
These structures include a rod with one end sliding inside a frictionless sleeve (in both quasi-static \cite{bigoni2015eshelby, Liakou2018} and dynamic \cite{ARMANINI201982,koutso2023, Wang2022} settings), a rod subjected to torsion \cite{bigonitorsional} and a rod moving inside a frictionless, rigid and curved channel \cite{dalcorsosnake}.
Remarkably, a common feature of these structures is the possibility of a free movement in a certain direction, to which the configurational force becomes energetically conjugate.

Inspired by these results in structural mechanics, Ballarini and Royer-Carfagni  
proposed an interpretation of  configurational forces as resultants of Newtonian contact forces acting on defects, through the solution of simplified models, representative of a solid containing an edge dislocation or a crack \cite{ballarini}. 

\begin{figure}[!h]
\centering
\includegraphics[width=1\textwidth]{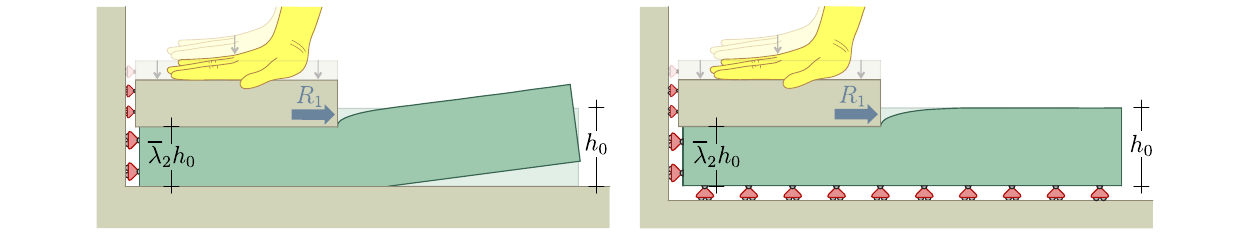}
\caption{\label{panino_intro} An elastic solid of rectangular shape (green) of initial height $h_0$ is deformed through a transverse compression 
(of nominal stretch $\overline{\lambda}_2<1$) 
against a flat, rigid, and frictionless punch (brown) ending with a sharp corner.
Rollers denote bilateral, while brown elements unilateral, frictionless contact. The transverse compression generates a horizontal reaction force $R_1$, shown in this article to be coincident with the negative of the $J$--integral, which in turn defines a configurational force, eqn (\ref{primaeq}), which can be evaluated with an excellent approximation through eqn (\ref{solettina}).}
\end{figure}

Along the same research line, the frictionless  contact is addressed in the present article of a homogeneous elastic solid, bounded with a planar surface and undergoing large deformations 
against a flat and rigid indenter, ending with a rounded or sharp corner. In this situation,  it is shown that a path-independent $J$--integral can be defined (so far restricted to small strain states \cite{Ma2008, Ma2006, Xie2009}) and corresponds to the energy release rate $G$, also known as configurational force, associated with  the constraint growth. In turn, the configurational force is shown to coincide with the negative of the reaction force 
$R_1$ (parallel to the undeformed flat boundary of the solid)  
which the corner of the constraint transmits to the elastic solid, in summary, 
\begin{equation}\label{primaeq}
    G=J=-R_1. 
\end{equation}
Two paradigmatic examples of generation of a horizontal configurational force are systematically referred throughout the article, Fig. \ref{panino_intro} (the rollers visualize bilateral smooth contact, while the brown element symbolizes unilateral contact), where an elastic solid of undeformed rectangular shape of height $h_0$ is subjected to a nominal transverse  stretch  $\overline{\lambda}_2<1$ on one of its end portions. 

It is shown that in both cases the horizontal reaction  force $R_1$ at the corner is provided  with an excellent approximation by
\begin{equation}
\label{solettina}
R_1\approx\dfrac{\Phi^l\, h_0}{\lambda_1^l},
\end{equation}
where $\Phi^l$ is the strain energy density and $\lambda_1^l$ the stretch, both evaluated at the left edge $\partial\mathcal{B}_0^l$ of the elastic rectangular domain,  where these are assumed constant.
The simple approximate expression (\ref{solettina}) is obtained within a large deformation framework for hyperelastic materials and is shown to remain valid for both rounded and sharp corners, as well as for both types of boundary conditions (unilateral or bilateral frictionless contact) applied at the lower side of the rectangular domain.

With the purpose of connecting the present solid mechanics framework   with the recent results  obtained in configurational structural mechanics \cite{bigoni2015eshelby,bigonitorsional,dalcorsosnake, Liakou2018, o2015some}, 
an elastic solid is analyzed, on which the kinematics of an Euler-Bernoulli rod is enforced. 

In this way, 
a novel derivation from a nonlinear solid mechanics perspective is obtained for the outward tangential reaction, generated at the end of a  sliding sleeve constraining an elastic rod, previously disclosed only through one-dimensional models. 

The relevance of our results to the design of new soft actuation mechanisms is demonstrated by two applications, whose approximate solution is obtained analytically and validated by finite element simulations. In particular, it is shown that the configurational forces induced by a transverse compression may lead in one case to Eulerian buckling and in the other to the longitudinal motion of an elastic layer. 
The latter result may introduce a mechanical explanation to the so-called \emph{negative durotaxis}, 
a biological process in which cells migrate from a stiffer to a softer environment \cite{cellulona, benvenuti}, an unexpected response opposite to the more common durotaxis phenomena \cite{locascio}.

\section{Prologue: non-accidental coincidences in contact mechanics at small strain}\label{sezJ+ciava}

Linear elastic solutions available in the literature for contact problems {\cite{barbercontactbook,ciavarella1998,ciavarella2002,Johnsonbook,Ma2006} are used to show that a  horizontal reaction force $R_1$ is generated at each  
(smooth, but even sharp)
corner of a frictionless, rigid and flat punch, 
pressed with a vertical load $P$  against the horizontal surface of an elastic solid.  
The horizontal reaction $R_1$ is 
nonlinear in $P$ and its presence is
 particularly  surprising because it can be evaluated  within the context of infinitesimal elasticity and even when the corner is sharp. 

More specifically, with reference to the indentation of an elastic half space, it is shown that the horizontal reaction force $R_1$ acting at each corner of the punch is quadratic in  the external vertical load $P$ and coincides with the negative of the path-independent $J$--integral evaluation at the corresponding corner, which in turn is equal to the energy release rate $G$ associated with an infinitesimal growth of an edge of the punch, namely, 
\begin{equation}
\boxed{
R_1=\dfrac{(1-\nu^2) P^2}{2\pi a E}=-J=-G,
}
\end{equation}
where $a$ is the punch half-width, while  $E$ and $\nu$ are  the Young's modulus and the Poisson's ratio of the indented half-space, respectively.
The coincidence of the horizontal reaction $R_1$ with the negative of the $J$--integral finds an explanation in the use of the energy-momentum tensor for  frictionless contact problem, as shown in Sect. \ref{sezintermezzo}. Moreover, the interpretation of $J$ as the energy release rate $G$ in a configurational mechanics framework is shown in Sect. \ref{eshelby_generalizzato}.

\subsection{$J$--integral and energy release rate $G$ for the indentation of a linear elastic material with a frictionless, rigid, flat punch with sharp corners}

The two-dimensional (plane strain) problem is considered in the $x_1$--$x_2$ plane for a frictionless, rigid, and flat punch indenting a linear  elastic isotropic solid on its surface, straight in the undeformed configuration and defined by $x_2=0$.

Restricting the attention to the right corner of the indenter (located at coordinate $x_1=a, x_2=0$), the leading-order term in the asymptotic expansion at this point for the components of the Cauchy stress tensor $\mathbf{T}$ in polar coordinates [$\rho>0, \theta\in(0,\pi)$, so that $x_1=a-\rho\cos\theta,x_2=\rho \sin\theta$, Fig. \ref{sist}, left] is given by \cite{barbercontactbook,Giannakopoulos1998,Xie2003} 
\begin{equation}\label{Tasymp}
\left\{
\begin{array}{cc}
     T_{\rho\rho}(\rho,\theta)\\
    T_{\theta\theta}(\rho,\theta) \\
      T_{\rho\theta}(\rho,\theta) 
\end{array}    
\right\}
=
\dfrac{K_I}{\sqrt{2\pi \rho}}\left\{
\begin{array}{cc}
     \cos\dfrac{\theta}{2}\left(1+\sin^2\dfrac{\theta}{2}\right) \\ \cos^3\dfrac{\theta}{2}\\
    \sin\dfrac{\theta}{2}\cos^2\dfrac{\theta}{2} 
\end{array}    
\right\},
\end{equation}
where $K_I$ is the Stress Intensity Factor (SIF)  representing the magnitude of the singular fields \lq condensing' the boundary conditions as
\begin{equation}\label{KIdef}
K_I=\lim_{\rho\rightarrow0}\sqrt{2\pi \rho}\,T_{\theta\theta}(\rho,\theta=0).
\end{equation}
It is noted that the square root singular stress asymptotics (\ref{Tasymp}) present at the sharp corner of a frictionless rigid   punch  coincides with the analogue holding for a crack tip under Mode I loading conditions, when a proper linear transformation of the angular coordinate is applied. 
\begin{figure}[!h]
\renewcommand{\figurename}{\footnotesize{Figure}}
    \begin{center}
   \includegraphics[width=1\textwidth]{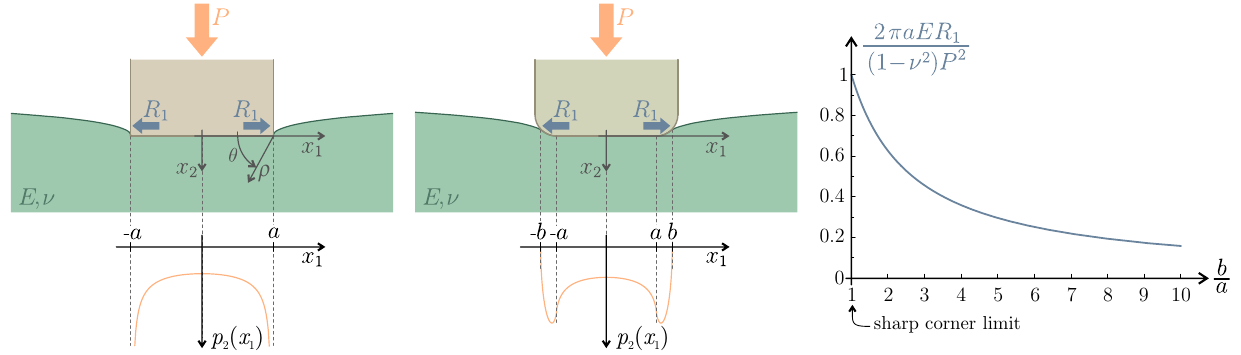}
    \caption{\footnotesize  The planar contact of a flat punch with (left) sharp and (center) rounded corners. In both cases the contact 
    with a linear elastic isotropic half space 
    is frictionless and provided by a vertical force $P$. The vertical component $p_2$ of the pressure along the contact surface (marked by the horizontal coordinate $x_1$) is reported below the sketches.  Right: Horizontal component $R_1$ of the contact force reaction present at each rounded corner  (made dimensionless through division by $(1-\nu^2)P^2/(2\pi a E)$ and 
    obtained from Ciavarella et al. \cite{ciavarella1998}), 
    as a function of the parameter $b/a\geq 1$, describing the ratio between the width of the contact region, $2b$, 
     and the width of its flat portion, $2a$.}
    \label{sist}
    \end{center}
\end{figure}

Moreover, the free surface ahead of the punch tip ($\theta=\pi$) displays the following first-order term for the normal displacement $u_\theta$ \cite{barbercontactbook}
\begin{equation}\label{uasymp}
    u_\theta(\rho,\theta=\pi)=-\dfrac{4 (1-\nu^2)\, K_I}{E}\sqrt{\dfrac{\rho}{2\pi}},
\end{equation}
where $E>0$ and $\nu\in(-1,1/2]$ are the  Young's modulus  and Poisson's ratio, respectively.

Introducing  the elastic strain energy density $\Phi$ and the displacement field  $u_i$, the path-independent $J$--integral used in fracture mechanics at small strains is defined as \cite{rice1968path}
\begin{equation}  \label{Jintinf}  
J=\int_{\Gamma_0} \left(\Phi n_1-T_{ij}n_j\dfrac{\partial u_i}{\partial x_1}\right)\mbox{d}\gamma_0,\qquad i,j=1,2,
\end{equation}
where $\Gamma_0$ is a continuous counter-clockwise path with outward unit normal  $n_i$. In the context of linear elasticity, the path-independence of the $J$--integral, eqn (\ref{Jintinf}), has been extended to flat punch problems by 
assuming the path $\Gamma_0$ starting from beneath the indenter and ending at the free surface \cite{Ma2006, Xie2009}.

Reducing $\Gamma_0$ to a semi-circular path of infinitesimal radius $r$ and centered at the right corner, the $J$--integral can be rewritten in terms of polar components as
\begin{equation}  \label{Jintinf2}  
J=\lim_{r \to 0}\int_0^\pi \left[\Phi(r,\theta) \cos\theta-T_{i\rho}(r,\theta)\left(
\frac{\sin\theta}{r}\dfrac{\partial u_i(r,\theta)}{\partial \theta}-
\cos\theta\dfrac{\partial u_i(r,\theta)}{\partial r}\right)\right]r\mbox{d}\theta,\qquad i=\rho,\theta.
\end{equation}
Considering the asymptotic expressions \eqref{Tasymp} and the linear constitutive relations, the $J$--integral \eqref{Jintinf2} for the flat rigid indentation  problem results 
\begin{equation}
\label{Jpunch}
J=-\dfrac{1-\nu^2}{2 E}K_I^2< 0,
\end{equation}
which differs by a factor $-2$ from the $J$--integral found for Mode I fracture. Similarly to rigid line inclusion problems \cite{bigoniRLI, goudarzi},  the $J$--integral associated to the flat punch with 
sharp
corners is always  non-positive.

Following Rice \cite{rice1968}, the energy release rate $G$,  associated with  a  growth $\Delta \xi$ of the punch corner and defined as the negative of the derivative of the potential energy $\mathcal{V}$ with respect to the configurational parameter $\xi$, can be evaluated as
\begin{equation}
    G=-\dfrac{\mbox{d}\mathcal{V}(\xi)}{\mbox{d}\xi}=\lim_{\Delta \xi\rightarrow 0} \dfrac{1}{2 \Delta \xi}\int_0^{\Delta \xi} T_{\theta\theta}(\Delta \xi-r,0) u_\theta(r,\pi) \mbox{d} r,
\end{equation}
which, considering the asymptotic expansions (\ref{Tasymp}) and (\ref{uasymp}), equals the $J$--integral
\begin{equation}\label{GequalJinf}
    G=J<0.
\end{equation}
Equation (\ref{GequalJinf}) 
shows that a growth in the punch size leads to an increase of the total potential energy of the system, implying that the process is not favorable, as in the stiffener problem, but opposite to crack growth where the energy release is always positive for an advance of the tip.

It is noted that, although the path-independent $J$--integral, eqn (\ref{Jintinf}),   
was already known in flat punch problems of linear elasticity
it has never been related to the energy release rate $G$ associated with a flat punch growth. Indeed, the $J$--integral has been so far used only in the investigation of failure mechanisms connected with crack initiation \cite{Ma2006} or dislocation nucleation \cite{Ma2008} at the sharp corners of flat punches. Moreover, in \cite{Xie2009} the $J$--integral was found to be null for a rigid-body sliding of  the whole punch, a result which is correct, but trivial because the two opposite  forces $R_1$ cancel each other (Fig. \ref{sist}).

\paragraph{Indenting an elastic half space.} The above results can be  used to analyze a linear elastic isotropic  half space ($x_1\in(-\infty,\infty)$, $x_2>0$) indented by a (frictionless, rigid, and flat) punch, with  horizontal base of width $2a$ (and centered at $x_1=x_2=0$). 
When the punch is subjected to a given compressive normal force $P$  (Fig. \ref{sist}, left), the  pressure  distribution $p(x_1)$ (positive when compressive) at the contact has only a vertical component ($p_1(x_1)=0$, $p_2(x_1)>0$) given  by \cite{Johnsonbook}
\begin{equation}\label{eqpres}
 p(x_1)= p_2(x_1)=\dfrac{P}{\pi\sqrt{a^2-x_1^2} },
\end{equation}
which approaches an infinite  value at the two sharp corners ($x_1=\pm a$) and leads to the following stress $T_{ij}$ ($i,j=1,2$)  distribution \cite{Saddbook}
\begin{equation}\label{stressflatind}
\left\{
\begin{array}{cc}
     T_{11}(x_1,x_2)  \\[5mm]
    T_{22}(x_1,x_2)\\[5mm]
      T_{12}(x_1,x_2)
\end{array}    
\right\}
=-\dfrac{2P}{\pi^2}\left\{
\begin{array}{cc}
     \displaystyle x_2\int_{-a}^a\dfrac{(x_1-s)^2}{\sqrt{a^2-s^2}\left[(x_1-s)^2+x_2^2\right]^2}\mbox{d}s \\[5mm]
    \displaystyle x_2^3\int_{-a}^a\dfrac{1}{\sqrt{a^2-s^2}\left[(x_1-s)^2+x_2^2\right]^2}\mbox{d}s\\[5mm]
    \displaystyle x_2^2\int_{-a}^a\dfrac{x_1-s}{\sqrt{a^2-s^2}\left[(x_1-s)^2+x_2^2\right]^2}\mbox{d}s
\end{array}    
\right\}.
\end{equation}

Considering the full-field representation (\ref{stressflatind}) for the stress $\mathbf{T}$, the Stress Intensity Factor (SIF) $K_I$ (\ref{KIdef}) for a flat punch of width $2a$ subject to  a vertical load $P$ indenting an elastic half space results to be \cite{Giannakopoulos1998,Xie2003}
\begin{equation}
\label{Kprimopunch}
    K_I=-\dfrac{P}{\sqrt{\pi a}},
\end{equation}
and the $J$--integral (\ref{Jpunch}) reduces to 
\begin{equation}
\label{Jpunch2}    
J=-\dfrac{1-\nu^2 }{2\pi a E}P^2< 0.
\end{equation}
Exploiting the path-independence of the $J$--integral and the null value of its integrand at the punch contact and at the free surface (namely, $x_2=0$ and excluding the corner point) and at infinite (namely, $\sqrt{x_1^2+x_2^2}\rightarrow\infty$),  the $J$--integral  (\ref{Jpunch2}) can be evaluated as
\begin{equation}  \label{Jintinf3}  
J=-\int_0^\infty \left.\left[\Phi(x_1,x_2)-T_{11}(x_1,x_2)\dfrac{\partial u_1(x_1,x_2)}{\partial x_1}\right]\right|_{x_1=0}\mbox{d} x_2.
\end{equation}

Within the context of configurational mechanics for an hyperelastic solid  undergoing large deformations, the $J$--integral, eqn \eqref{Jintinf}, is proven in Sect. \ref{eshelby_generalizzato} to equal the energy release $G$ associated with an increase in the size of the frictionless straight constraint with a corner and therefore to correspond to the horizontal force exerted by the elastic solid on the rigid constraint.
This result is anticipated below for the rigid flat punch, by showing that the negative of the $J$--integral \eqref{Jpunch} matches  the horizontal reaction force $R_1$ at its corner. To this purpose, 
an indenter with rounded corners is considered in the Sect. \ref{Hr1sectrounded}, including the  limit of vanishing curvature radius.

\subsection{Horizontal contact reaction force $R_1$ at the indenter with rounded corner in linear elasticity}\label{Hr1sectrounded}

A rigid punch with rounded corners 
is considered (Fig. \ref{sist}, center), with a central flat portion of width 2$a$, rounded at  both ends with a parabola of radius of curvature $R$, described by  $x_2=h(x_1)$. The latter function has the following derivative   
\beq
\label{positionfield0}
h'(x_1)=\left\{
\begin{array}{lll}
0, &\,\,\mbox{if}\,\, x_1\in[-a,a],\\[2mm]
-\dfrac{x_1\mp a}{R}, &\,\,\mbox{if}\,\, x_1\in[\pm a, \pm b],
\end{array}
\right.
\eeq
where $2b\geq 2a$ defines the unknown contact width, measured as the projection of the contact zone onto $x_1$. 
On introduction of a mapping for the horizontal coordinate $x_1\in[-b,b]$ in terms of the angle $\phi\in[-\pi/2,\pi/2]$ as
\begin{equation}
 x_1(\phi)=\dfrac{\sin \phi}{\sin \phi_0} a,\qquad
 \mbox{with}\qquad
 b=\dfrac{a}{\sin\phi_0},
\end{equation}
the component $p_2(x_1)$ of the pressure distribution $p$ at the contact  
is  evaluated for an applied vertical force $P$ as \cite{ciavarella1998,ciavarella2002,kim2023}
\begin{equation}\label{ciavaciava}
 p_2(\phi)=\dfrac{2P}{\pi (\pi-2\phi_0-\sin2\phi_0)b}\left\{
 (\pi-2\phi_0)\cos\phi+\ln\left[
 \left|\dfrac{\sin(\phi+\phi_0)}{\sin(\phi-\phi_0)}\right|^{\sin\phi}  \left|\tan\dfrac{\phi+\phi_0}{2}\tan\dfrac{\phi-\phi_0}{2}\right|^{\sin\phi_0}
 \right]
 \right\}. 
\end{equation}

The unknown angle $\phi_0\in(0,\pi/2]$ (and therefore the corresponding detachment semi-distance $b\geq a$) can be evaluated as the solution of the following nonlinear equation
\begin{equation}\label{curvature}
    \dfrac{(1-\nu^2) P R}{a^2 E}=\dfrac{\pi-2\phi_0}{4\sin^2\phi_0}-\dfrac{\cot\phi_0}{2}.
\end{equation}

Note that the pressure distribution $p_2(x_1)$,
 eqn (\ref{ciavaciava}), 
has never been exploited to evaluate the horizontal resultant force $R_1$ of the contact pressure at each  rounded  corner, 
where the two forces have opposite directions and thus satisfy equilibrium.
Such horizontal resultant $R_1$ can be calculated as the following positive quantity  
\begin{equation}\label{eqR10}  
R_1=-\int_{a}^{b}p_2(x_1) h'(x_1)\mbox{d}x_1=\dfrac{a^2}{R\sin\phi_0}\int_{\phi_0}^{\frac{\pi}{2}}p_2(\phi)\left(\dfrac{\sin\phi}{\sin\phi_0}-1\right)\cos\phi\,\mbox{d}\phi>0,
\end{equation}
confirming that the  horizontal reaction $R_1$  has an outward direction at each rounded corner.
Exploiting eqns (\ref{ciavaciava})  and (\ref{curvature}), the  horizontal force $R_1$ (\ref{eqR10}) can be rewritten as
\begin{equation}
\label{R1conintegrale}
\begin{split}
R_1=\dfrac{8 (1-\nu^2) P^2 \sin \phi_0}{\pi a E \left(\pi - 2\phi_0 - \sin 2 \phi_0  \right)^2}\int_{\phi_0}^{\frac{\pi}{2}}
\left\{
 (\pi-2\phi_0)\cos\phi+\ln\left[
 \left|\dfrac{\sin(\phi+\phi_0)}{\sin(\phi-\phi_0)}\right|^{\sin\phi}  \left|\tan\dfrac{\phi+\phi_0}{2}\tan\dfrac{\phi-\phi_0}{2}\right|^{\sin\phi_0}
 \right]
 \right\}\\
\left(\sin \phi -\sin \phi_0\right)
\mbox{d}\phi.
\end{split}
\end{equation}

The  horizontal reaction force $R_1$, present at each rounded corner, can be evaluated through a numerical integration of equation (\ref{R1conintegrale}). The result is reported in Fig. \ref{sist} (right), where the force is represented as a function of the ratio $b/a\geq 1$.

The expression for the horizontal force $R_1$, eq. (\ref{R1conintegrale}), can be expanded by assuming a vanishing small contact region at the rounded corner ($\phi_0\rightarrow \pi^-/2$, $b\rightarrow a^+$) as follows
\begin{equation}
\label{wlfwlf}
R_1\left(b/a\right)=\dfrac{(1-\nu^2) P^2}{2\pi a E}\left[1-\dfrac{3}{5}\left(\dfrac{b}{a}-1\right)
\right]+o(b/a-1),
\end{equation}
which shows that the horizontal reaction attains a  non-null finite value in the case of non-rounded, and therefore sharp, corner ($b\rightarrow a^+$),
\begin{equation}\label{R1limitcorner}
\lim_{b\rightarrow a^+} R_1\left(b/a\right)=\dfrac{(1-\nu^2) P^2}{2\pi a E}.
\end{equation}
The above equation confirms the presence of a non-null horizontal reaction force $R_1$ at each sharp corner of a frictionless rigid indenter, which is quadratic in $P$,  similarly to the configurational force  acting on an inextensible rod constrained with a sliding sleeve \cite{ARMANINI201982, bigoni2015eshelby,dalcorsosnake,o2015some}. 
Interestingly, the  limit value of the horizontal reaction $R_1$, eqn  (\ref{R1limitcorner}), 
equals 
the negative of the $J$--integral and the energy release rate $G$, 
evaluated for the flat punch problem, eqs.  \eqref{Jpunch} and \eqref{GequalJinf}, namely
\begin{equation}\label{JR_infinitesimal}
\lim_{b\rightarrow a^+} R_1\left(b/a\right)= -J=-G.
\end{equation}
The coincidence of the reaction force component $R_1$ with the negative of the $J$--integral is proven in the next Section within a finite elasticity framework, where both cases of contact with a sharp corner or a rounded surface are addressed. Moreover, through the evaluation of energy  variation for an increase of the frictionless rigid surface of a flat indenter, the $J$--integral is found 
in Sect. \ref{eshelby_generalizzato} 
to coincide with the energy release rate $G$ and therefore  representative of a   configurational force component (called $F_1^c$).

\section{Frictionless  contact reaction component $R_1$ through energy-momentum tensor(s) and $J$--integral}\label{sezintermezzo}

It is shown that the reaction component force $R_1$ acting at the 
contact between a frictionless constraint and an elastic solid 
coincide 
with the negative of  the $J$--integral, even when large deformations occur and the end of the constraint is both a smooth or sharp corner. To this purpose, 
 the definition (\ref{Jintinf}) for the $J$--integral is extended as follows
\begin{equation}  \label{Jintfinite}  
J=\int_{\Gamma_0} \left(\Phi n_1^0-S_{ij}n_j^0\dfrac{\partial u_i}{\partial x_1^0}\right)\mbox{d}\gamma_0,\qquad i,j=1,2,
\end{equation}
where $\mathbf{S}$ is the first Piola-Kirchhoff stress tensor, $\Gamma_0$ is a counter-clockwise path with initial and final points selected on 
the boundary in contact, and the superscript $0$
stands for quantities evaluated in the undeformed configuration.

After recalling concepts of finite elasticity, frictionless contact, and energy momentum tensors, the $J$--integral is shown to provide the reaction force component $R_1$ acting on a generic portion of a smooth contact region $\partial\mathcal{P}_0^{\scriptsize \mbox{tou}}$ of an elastic solid defined as the undeformed domain $\mathcal{B}_0$, namely,
\begin{equation}
\label{R1generica0}
\boxed{
    R_1\left(\partial\mathcal{P}_0^{\scriptsize \mbox{tou}}\right)= -J\left(\Gamma_0\equiv\partial\mathcal{P}_0\backslash\partial\mathcal{P}_0^{\scriptsize \mbox{tou}}\right),\qquad \forall \,\,\mathcal{P}_0 \in \mathcal{B}_0,
    }
\end{equation}
and therefore the path independence of $J$ holds only for every path $\Gamma_0$ 
emanating from a selected point ($A_0$ in Fig. \ref{pathindependenceROUND}) and ending to another fixed
point ($B_0$ in Fig. \ref{pathindependenceROUND}). Points $A_0$ and $B_0$ enclose a given portion 
of the boundary $\partial\mathcal{P}_0^{\scriptsize \mbox{tou}}$.

Assuming proper regularity conditions, equation (\ref{R1generica0}) holds true even for a flat indenter with a sharp corner, where point $A_0$ is located in the contact region, while point $B_0$ on the right of the corner, on a free boundary. Therefore, the reaction force $R_1$ at the sharp corner can be evaluated as the negative of the $J$--integral, 
\begin{equation}\label{R1genericacorner}
\boxed{
    R_1= -J\left(\Gamma_0\right),\qquad \forall \,\,\Gamma_0 ,
    }
\end{equation}
where $\Gamma_0$ is any contour enclosing the corner, so that $J$ is path-independent with regards to every pair of points $A_0$ and $B_0$.

\begin{figure}[!h]
\centering
\includegraphics[width=0.8\textwidth]{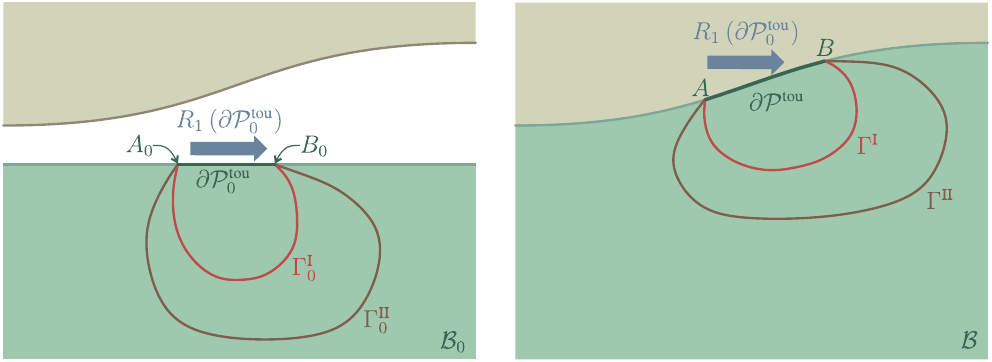}
\caption{\label{pathindependenceROUND} Undeformed (left) and deformed (right) configurations for an elastic solid (green) having its initially flat boundary in frictionless contact with a rigid constraint (brown) with smooth boundary. The contact reaction force  $R_1$ associated with the contact region $\partial\mathcal{P}_0^{\scriptsize \mbox{tou}}$ can be evaluated through  the $J$--integral, whose path-independence is restricted to all paths $\Gamma_0$ ($\Gamma_0^I$ and 
$\Gamma_0^{II}$ in the reference configuration,
$\Gamma^I$ and $\Gamma^{II}$ in the current one) 
emanating from the same initial point ($A_0$ in the reference configuration and $A$ in the current) and terminating at the same final ($B_0$ and $B$) point.  
}
\end{figure}

Anticipating results obtained at the end of this Section, 
it can be pointed out that 
the application of eqn (\ref{R1generica0}) to rectangular elastic solids with edges subject to uniform loading conditions, as sketched in both parts of Fig. \ref{panino_intro}, provides the estimation of the reaction force component $R_1$ at both sharp and rounded corner as given by eqn (\ref{solettina}).

\subsection{A premise on finite elasticity}
A solid undergoing large deformations is considered, in which the point $\mathbf{x}_0$ in the  reference configuration $\mathcal{B}_0$ is transformed into the point $\mathbf{x}=\mathbf{g}(\mathbf{x}_0)$ in the  current configuration $\mathcal{B}$, through the deformation function $\mathbf{g}$.
The displacement field $\mathbf{u}$ and the deformation gradient $\mathbf{F}$ follow as
\begin{equation}
\label{solitabazza}
\mathbf{u}=\mathbf{x}-\mathbf{x}_0,
\qquad
\mathbf{F}=\nabla \mathbf{g}= \mathbf{I}+\nabla \mathbf{u}, 
\end{equation}
where the gradient $\nabla$ is evaluated with respect to $\mathbf{x}_0$. The  unit  vectors $\mathbf{n}_0$ and $\mathbf{t}_0$, normal and tangential to a surface, are transformed into the corresponding unit vectors $\mathbf{n}$ and $\mathbf{t}$ as
\begin{equation}
\mathbf{n}=\dfrac{\mathbf{F}^{-T}\mathbf{n}_0}
{\left|\mathbf{F}^{-T}\mathbf{n}_0\right|},\qquad
\mathbf{t}=\dfrac{\mathbf{F}\,\mathbf{t}_0}
{\left|\mathbf{F}\,\mathbf{t}_0\right|},
\label{enneeti}
\end{equation}
so that $\mathbf{n}_0\scalp\mathbf{t}_0=\mathbf{n}\scalp\mathbf{t}=0$. 
By assuming a hyperelastic response and introducing the strain energy density $\Phi$ for a unit volume in the reference state $\mathcal{B}_0$, the first Piola-Kirchhoff stress tensor $\mathbf{S}$ can be derived as
\begin{equation}\label{const}
    \mathbf{S}= \dfrac{\partial\Phi(\mathbf{F})}{\partial \mathbf{F}}.
\end{equation}
The first Piola-Kirchhoff stress tensor $\mathbf{S}$, eqn \eqref{const}, is related to the Cauchy stress tensor
$\mathbf{T}$ through
\begin{equation}\label{PK1def}
    \mathbf{S}= \mathcal{J} \mathbf{T} \mathbf{F}^{-T},
\end{equation}
where $\mathcal{J}= \det \mathbf{F}$ (thus $\mathcal{J}=1$ for incompressible materials) and the superscript $T$ denotes the transpose operator, so that the resultant force acting on an infinitesimal area $\mbox{d}a_0$ in the reference configuration is equal to the force acting  on the area element $\mbox{d}a$ in the current state 
\begin{equation}\label{tractionequivalence}
    \mathbf{S}\mathbf{n}_0\mbox{d}a_0= \mathbf{T}\mathbf{n}\mbox{d}a,
\end{equation}
where $\mathbf{n}_0$  and $\mathbf{n}$ are the unit vectors orthogonal to the two area elements.

In the absence of body forces, equilibrium can be written in terms of first Piola-Kirchhoff stress tensor as
\begin{equation}\label{eommm}
    \mbox{Div}\,\mathbf{S} = \mathbf{0},
\end{equation}
where the divergence operator $\mbox{Div}$ is evaluated with respect to $\mathbf{x}_0$. Assuming continuity of $\mathbf{S}$ and therefore \emph{excluding the presence of concentrated forces within  the generic volume $\mathcal{P}_0\subseteq\mathcal{B}_0$ described by its boundary $\partial \mathcal{P}_0$}, the divergence theorem yields
\begin{equation}
\label{solenoidati1}
    \int_{\partial \mathcal{P}_0} \mathbf{Sn}_0 =\mathbf{0},
\end{equation}
showing that the first Piola-Kirchhoff stress tensor $\mathbf{S}$ is solenoidal.

\subsection{Frictionless contact problem: target and contactor}

The boundary of a rigid and frictionless constraint, called \lq target', is described by the implicit surface (assumed here smooth 
for simplicity, Fig. \ref{target})
\begin{equation}
\label{straz}
    \Sigma (\mathbf{x}) = 0, 
\end{equation}
so that the points $\mathbf{x}$ in the current configuration can be divided in three disjoint sets as: 
\begin{equation}
\label{classconstr}
\Sigma(\mathbf{x}):\left\{    \begin{array}{ll}
 <0,
 & \mbox{points $\mathbf{x}$ inside  the constraint}, \\ 
  =0 , & \mbox{points $\mathbf{x}$ on the constraint boundary},    
\\ 
 >0,&\mbox{points $\mathbf{x}$ outside the constraint}. 
    \end{array}
    \right.
\end{equation}


\begin{figure}[!h]
\centering
\includegraphics[width=1\textwidth]{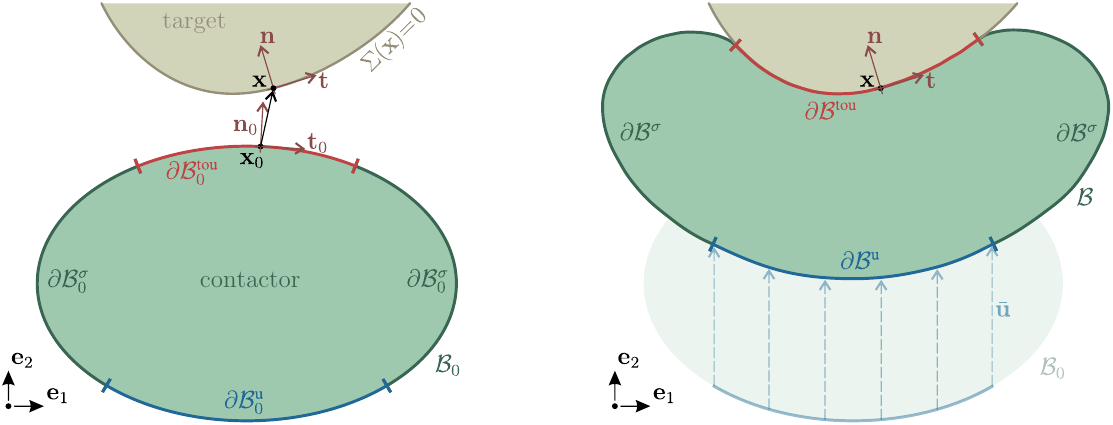}
\caption{\label{target} The contact problem between a \lq contactor' elastic body (green) and a rigid and frictionless \lq target' (brown). 
Left: The point $\mathbf{x}_0$ on the boundary of the body in the reference configuration has unit outward normal $\mathbf{n}_0$ and unit tangent 
$\mathbf{t}_0$. Right: The deformation transforms these quantities to 
$\mathbf{x}$ on the contact surface and to $\mathbf{n}$ and $\mathbf{t}$, which become the  unit normal (inward the target, outward the contactor) and the tangent to the target, respectively.
The contact is sketched as the result of an imposed displacement $\overline{\mathbf{u}}$
on $\partial \mathcal{B}_0$. 
}
\end{figure}

The \lq contactor' body, in its reference configuration $\mathcal{B}_0$, assumed undeformed, is transformed through a sufficiently regular deformation function $\mathbf{g}(\mathbf{x}_0)$, to become in frictionless contact with the target, thus reaching a deformed configuration $\mathcal{B}$
under the action of prescribed dead tractions on $\partial \mathcal{B}^\sigma$ and displacements 
on $\partial \mathcal{B}_0^u$. 
Therefore, points of boundary $\mathbf{x} = \mathbf{g}(\mathbf{x}_0) \in \partial \mathcal{B}$, transformed of the corresponding points in the reference configuration $\mathbf{x}_0 \in \partial \mathcal{B}_0$, 
can be classified as: 
\begin{itemize}
    \item  Points $\mathbf{x}$ (equivalently,  $\mathbf{x}_0$) belonging to $\partial \mathcal{B}^{\scriptsize\mbox{sep}}$ ($\partial \mathcal{B}_0^{\scriptsize\mbox{sep}}$) 
    separated from the constraint, when $\mathbf{x} = \mathbf{g}(\mathbf{x}_0)$ are outside the constraint, set (\ref{classconstr})$_3$;
    \item   
    Points $\mathbf{x}$ (equivalently,  $\mathbf{x}_0$) belonging to $\partial \mathcal{B}^{\scriptsize \mbox{tou}}$ ($\partial \mathcal{B}_0^{\scriptsize \mbox{tou}}$)
    touching the constraint, when $\mathbf{x} = \mathbf{g}(\mathbf{x}_0)$ is on the boundary of the constraint, set (\ref{classconstr})$_2$. 
\end{itemize}

The subset of separated points $\partial \mathcal{B}_0^{\scriptsize\mbox{sep}}$, and equivalently $\partial \mathcal{B}^{\scriptsize\mbox{sep}}$, can be partitioned as subject to prescribed loading (assumed dead for simplicity) or displacement
\begin{equation}
\partial \mathcal{B}_0^{\scriptsize \mbox{sep}} \equiv \partial \mathcal{B}_0^u \cup \partial \mathcal{B}_0^\sigma, ~~ \mbox{ and equivalently} ~~
\partial \mathcal{B}^{\scriptsize\mbox{sep}} \equiv \partial \mathcal{B}^u \cup \partial \mathcal{B}^\sigma.   
\end{equation}
It is assumed that a portion of the boundary 
outside $\partial \mathcal{B}^{\scriptsize \mbox{tou}}$ 
and bordering with it at its two edges exists, where tractions are null, so that the ends of the constraint can be moved on a free portion of the boundary of the elastic body.

The subset of touching points $\partial \mathcal{B}_0^{\scriptsize \mbox{tou}}$ can be subdivided into a subtle partition, with reference to the Cauchy stress $\mathbf{T}$ and its  spatial counterpart of the first Piola-Kirchhoff stress $\mathbf{S}$, eqn (\ref{const}), as
\begin{equation}
\label{grazing}
    \begin{array}{llcl}
\mbox{Grazing}
       &
       \partial \mathcal{B}^{G}  := \{ \left. \mathbf{x} \in \partial \mathcal{B}^{\scriptsize \mbox{tou}} \right| ~ \mathbf{T}\mathbf{n} = \mathbf{0} \}, & \mbox{and equivalently}
       &
       \partial \mathcal{B}_0^{G} : = \{ \left.\mathbf{x}_0 \in \partial \mathcal{B}_0^{\scriptsize \mbox{tou}} \right| \mathbf{S}\mathbf{n}_0 = \mathbf{0} \},  \\ [3 mm]
       \mbox{Full contact}
       &
\partial \mathcal{B}^{C} : = \{\left. \mathbf{x} \in \partial \mathcal{B}^{\scriptsize \mbox{tou}} \right| ~ \mathbf{n}\cdot\mathbf{T}\mathbf{n} < 0\}, & \mbox{and equivalently}
       &
  \partial \mathcal{B}_0^{C}  := \{ \left.\mathbf{x}_0 \in \partial \mathcal{B}_0^{\scriptsize \mbox{tou}} \right| ~ \mathbf{n}_0\cdot\mathbf{F}^{-1}\mathbf{S}\mathbf{n}_0 < 0  \},         
    \end{array}
\end{equation}
so that 
\begin{equation}
\partial \mathcal{B}_0^{\scriptsize \mbox{tou}} \equiv \partial \mathcal{B}_0^{G} \cup \partial \mathcal{B}_0^C, ~~ \mbox{ and equivalently} ~~
\partial \mathcal{B}^{\scriptsize \mbox{tou}} \equiv \partial \mathcal{B}^G \cup \partial \mathcal{B}^C.   
\end{equation}

In both the above cases along the touching boundary, the frictionless contact condition holds 
\begin{equation}
\label{pugnettone}
     (\mathbf{I}-\mathbf{n}\otimes\mathbf{n})\mathbf{T}\mathbf{n} = \mathbf{0} 
~~\mbox{ on }~~
    \partial \mathcal{B}^{\scriptsize \mbox{tou}},
     ~~
     \mbox{ and equivalently }
    ~~(\mathbf{I}-\mathbf{F}^{-T}\mathbf{n}_0\otimes\mathbf{F}^{-T}\mathbf{n}_0)\mathbf{S}\mathbf{n}_0 = \mathbf{0}
~~\mbox{ on }~~
    \partial \mathcal{B}_0^{\scriptsize \mbox{tou}},
\end{equation}
which can also be rewritten with reference to every tangent vectors $\mathbf{t}_0$ and $\mathbf{t}$ (see eqn (\ref{enneeti})) as
\begin{equation}
\label{pugnettone2}
     \mathbf{t}\scalp\mathbf{T}\mathbf{n} = 0
~~\mbox{ on }~~
    \partial \mathcal{B}^{\scriptsize \mbox{tou}},
     ~~
     \mbox{ and equivalently }
    ~~ \mathbf{t}_0\scalp\mathbf{F}^{T}\mathbf{S}\mathbf{n}_0 = 0
~~\mbox{ on }~~
    \partial \mathcal{B}_0^{\scriptsize \mbox{tou}}.
\end{equation}
Interestingly, eqn (\ref{solitabazza})$_2$ shows that the eqn (\ref{pugnettone2})$_2$ implies the validity of the following identity at every point of the frictionless contact surface in the undeformed configuration
\begin{equation}
\label{pazienza}
\mathbf{t}_0 \cdot \mathbf{Sn}_0 = -\mathbf{t}_0 \cdot (\nabla\mathbf{u})^T\mathbf{Sn}_0,
~~~\mbox{ on }~~~\partial \mathcal{B}_0^{\scriptsize \mbox{tou}}.
\end{equation}

An application of the virtual work principle to the mechanics of sliding contact is provided for completeness in Appendix \ref{AppendixA}.

\subsection{Two energy-momentum tensors}

Two different definitions of the energy-momentum tensor for solids subject to  large deformation can be found in the literature. 
In particular, Eshelby \cite{eshelby1951force} introduced the energy-momentum tensor $\mathbf{P}$ as 
\begin{equation}
\label{nonneshelby}
    \mathbf{P} = \Phi \, \mathbf{I} - (\nabla \mathbf{u})^T\mathbf{S},  
\end{equation}
while Gurtin \cite{gurtin1999nature}  defined a different  energy-momentum tensor $\mathbf{C}$ as\footnote{ The divergence operator here used is, in Cartesian rectangular coordinates, $
\left(\mbox{Div } \mathbf{C}\right)_i =\partial C_{ij}/\partial x^0_j. 
$
If the definition of divergence is changed, so that the first index is repeated, the transpose of $\mathbf{C}$ is accordingly used, as in \cite{chadwick}. 
A further definition of energy-momentum tensor has been introduced by Maugin \cite{mauginino}. 
}  
\begin{equation}
\label{em0}
    \mathbf{C} = \Phi \, \mathbf{I} - \mathbf{F}^T\mathbf{S},
\end{equation}
where the two tensors can easily be related using the definition \eqref{solitabazza}$_2$ of the deformation gradient $\mathbf{F}$ as
\begin{equation}
\label{duem}
 \mathbf{C} = \mathbf{P} - \mathbf{S}.  
\end{equation}
It is noted that the $J$--integral (\ref{Jintfinite}) involves the energy-momentum tensor $\mathbf{P}$, because it can be rewritten as
\begin{equation}  \label{Jintfinite2}  
\boxed{J=\mathbf{e}_1\scalp\int_{\Gamma_0} \mathbf{P}\mathbf{n}_0\,\mbox{d}\gamma_0.
    }
\end{equation}

The divergence of the energy-momentum tensor $\mathbf{C}$  \eqref{em0} can be evaluated as
\begin{equation}
    \frac{\partial C_{ij}}{\partial x^0_j} = \frac{\partial \Phi}{\partial x^0_i} - \frac{\partial F_{ki}}{\partial x^0_j} S_{kj}-F_{ki}\frac{\partial S_{kj}}{\partial x^0_j},
\end{equation}
which, recalling the constitutive relation \eqref{const},
simplifies to
\begin{equation}\label{intermezzo}
    \frac{\partial C_{ij}}{\partial x^0_j} = S_{hk}\frac{\partial F_{hk}}{\partial x^0_i} - \frac{\partial F_{ki}}{\partial x^0_j} S_{kj}-F_{ki}\frac{\partial S_{kj}}{\partial x^0_j}.
\end{equation}
Considering again the definition \eqref{solitabazza}$_2$ of the deformation gradient $\mathbf{F}$, the application of the Schwarz theorem implies
\begin{equation}
    S_{hk}\frac{\partial F_{hk}}{\partial x^0_i} - \frac{\partial F_{ki}}{\partial x^0_j} S_{kj} = 
    S_{hk}\frac{\partial^2 x_{h}}{\partial x^0_k \partial x^0_i} - 
\frac{\partial^2 x_{k}}{\partial x^0_i \partial x^0_j} S_{kj}   = 0,
\end{equation}
so that eqn \eqref{intermezzo} further simplifies as
\begin{equation}
\label{soccm0}
    \frac{\partial C_{ij}}{\partial x^0_j} = 
    -F_{ki}\frac{\partial S_{kj}}{\partial x_j}.    
\end{equation}
Due to equilibrium equation (\ref{eommm}), eqn (\ref{soccm0})  implies the null divergence of both  the energy momentum tensors  $\mathbf{C}$ and $\mathbf{P}$,
\begin{equation}\label{soccmp0}
    \mbox{Div}\,\mathbf{C} = \mathbf{0}, \qquad \mbox{Div}\,\mathbf{P} = \mathbf{0}.
\end{equation}

Assuming continuity of the  fields and therefore \emph{excluding material discontinuities and stress singularities within  the generic volume $\mathcal{P}_0\subseteq\mathcal{B}_0$ described by its boundary $\partial \mathcal{P}_0$}, the divergence theorem yields
\begin{equation}
\label{trallallero000}
        \int_{\partial \mathcal{P}_0} \mathbf{Cn}_0 = \mathbf{0},\qquad
    \int_{\partial \mathcal{P}_0} \mathbf{Pn}_0 = \mathbf{0},
\end{equation}
showing that both the energy momentum tensors  $\mathbf{C}$ and $\mathbf{P}$ are solenoidal, as the first Piola-Kirchhoff stress tensor $\mathbf{S}$, eqn  (\ref{solenoidati1}), is. 

The solenoidal property is now used to solve the equilibrium condition of a solid  loaded  through a generic pressure loading $p(\mathbf{x})$ on its boundary $\partial \mathcal{B}$, so that the static boundary condition is 
\begin{equation}\label{caricop}
    \mathbf{T}\mathbf{n}=- p \mathbf{n}, ~~~~~~
\mbox{ on } 
\partial \mathcal{B},
\end{equation}
which, by considering the traction equivalence (\ref {tractionequivalence}), implies
\begin{equation}
    \mathbf{Sn}_0 = - p \mathcal{J}\,\mathbf{F}^{-T} \mathbf{n}_0 = 
-p  \frac{\mbox{d}a}{\mbox{d}a_0} \mathbf{n} ,
~~~~~~
\mbox{ on } 
\partial \mathcal{B}_0
~~
\mbox{and} 
~~
\partial \mathcal{B},
\end{equation}
and therefore
\begin{equation}
\mathbf{F}^T\mathbf{Sn}_0 = - 
p
\mathcal{J}\, \mathbf{n}_0,
~~~~~~
\mbox{ on } 
\partial \mathcal{B}_0.    
\end{equation}
It follows that under the pressure loading of (\ref{caricop}), the solenoidal property (\ref{trallallero000})  for the energy-momentum tensor $\mathbf{C}$  can be expressed for $\partial \mathcal{P}_0\equiv\partial \mathcal{B}_0$ as
\begin{equation}
\label{trallalleropress}
    \int_{\partial \mathcal{B}_0} \left(\Phi  + p \mathcal{J} \right) \mathbf{n}_0= \mathbf{0}. 
\end{equation}

Equation (\ref{trallalleropress}) applies to any (non-singular) solid boundary $\partial \mathcal{B}_0$ and any non-uniform distribution of the pressure $p$. Equation (\ref{trallalleropress}) relates the elastic energy to the pressure (multiplied by $\mathcal{J}$) on the boundary and is trivially satisfied when $\Phi$ and $p\mathcal{J}$ are uniform. 
It is noted that the pressure loading   $p(\mathbf{x})$ on the boundary can be realized through the 
contact with both a unilateral or a bilateral 
 frictionless constraint. While $p\geq 0$ for unilateral contact, $p$ may have any sign when the contact becomes bilateral. The latter contact condition will be visualized in the following as obtained with rollers.

If the boundary $\partial \mathcal{B}_0$ is subjected to a pressure $p$ only on its portion $\partial \mathcal{B}_0^p\subset\partial \mathcal{B}_0$, equation (\ref{trallalleropress}) changes into
\begin{equation}
\label{trallalleropress23}
    \int_{\partial \mathcal{B}_0^p} \left(\Phi  + p \mathcal{J} \right) \mathbf{n}_0
    + \int_{\partial \mathcal{B}_0  \backslash  \partial \mathcal{B}_0^p} \mathbf{C}\mathbf{n}_0= \mathbf{0}, 
\end{equation}
where the appropriate boundary conditions have to be imposed on $\partial \mathcal{B}_0  \backslash  \partial \mathcal{B}_0^p$.
In terms of tensor $\mathbf{P}$, an equivalent of equation (\ref{trallalleropress23}) is obtained as 
\begin{equation}
\label{trallalleropress2333}
    \int_{\partial \mathcal{B}_0^p} \left[\Phi \mathbf{I} + p \mathcal{J}(\mathbf{I} - \mathbf{F}^{-T}) \right] \mathbf{n}_0
    + \int_{\partial \mathcal{B}_0  \backslash  \partial \mathcal{B}_0^p} \mathbf{P}\mathbf{n}_0= \mathbf{0}. 
\end{equation}

\subsection{Different energy-momentum tensors in the solution of rectangular elastic domains under pressure loading}\label{rettoangolo}

Attention is now restricted to plane problems of solids with an undeformed rectangular domain $\mathcal{B}_0$, having  sides parallel and orthogonal to the two unit vectors $\mathbf{e}_1$ and $\mathbf{e}_2$ defining the Cartesian reference system. Thus the domain is described as  
\begin{equation}\label{rectangulardomain}
    \mathcal{B}_0:=\left\{x_1^0\in[0,\ell_0],x_2^0\in[h_0/2,-h_0/2]\right\},
\end{equation}
where $\ell_0$ and $h_0$ are respectively the length of the sides parallel to $\mathbf{e}_1$ and $\mathbf{e}_2$.
The boundary  $\partial \mathcal{B}_0$ is given by the union of the four rectangle sides  
$
    \partial \mathcal{B}_0\equiv
    \partial \mathcal{B}_0^l
    \cup
    \partial \mathcal{B}_0^a
    \cup
    \partial \mathcal{B}_0^r
    \cup
    \partial \mathcal{B}_0^b,
$
with corresponding outward unit normal $\mathbf{n}_0$ respectively equal to 
$-\mathbf{e}_1$,
$\mathbf{e}_2 $,
$\mathbf{e}_1$, 
and 
$-\mathbf{e}_2$, 
 Fig. \ref{presssss}. 
\begin{figure}[!htb]
\centering
\includegraphics[width=0.9\textwidth]{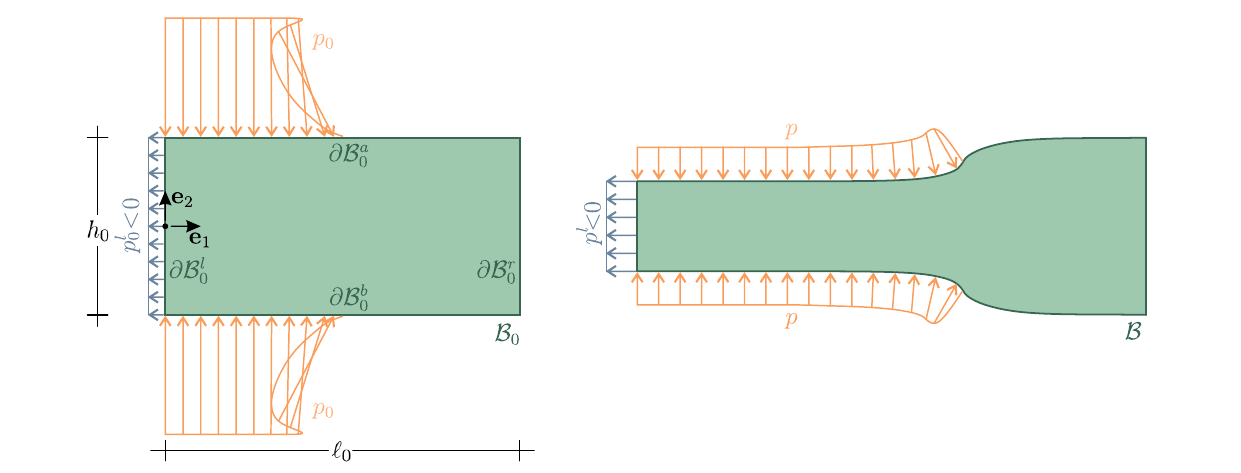} 
\caption{Left: An elastic solid of undeformed rectangular shape $\partial\mathcal{B}_0$ with the image of a pressure loading distribution $p$, symmetric with respect to $\mathbf{e}_1$,  on the boundary portions $\partial \mathcal{B}_0^l$, $\partial \mathcal{B}_0^a$, and $\partial \mathcal{B}_0^r$. Right: Deformed configuration. Exploiting the concept of  energy-momentum tensor $\mathbf{C}$, the resultant of the unknown loading pressure $p^l$, enforcing equilibrium, can be evaluated with an excellent approximation  through eqn (\ref{solettina}).
\label{presssss}
}
\end{figure}
A pressure loading condition $p$ is considered on the boundary portions $\partial \mathcal{B}_0^a$ and $\partial \mathcal{B}_0^b$, constraining the traction vector  to
\begin{equation}\label{ppartizionato}
    \mathbf{T}\mathbf{n}=- p \mathbf{n}, ~~~~~~
\mbox{ on } 
\partial \mathcal{B}_0^a\cup \partial \mathcal{B}_0^b.
\end{equation}

\subsubsection{Reaction forces $R_1^a+R_1^b$ and $R_1(\partial\mathcal{P}_0^{\scriptsize \mbox{tou}})$ from the energy momentum tensor $\mathbf{P}$}
The projection along $\mathbf{e}_1$ of the solenoidal property of $\mathbf{P}$, eqn (\ref{trallallero000})$_2$, implies that
\begin{equation}
\label{figodiddio0}
    \mathbf{e}_1\scalp\int_{\partial \mathcal{B}_{0}^a\cup\partial\mathcal{B}_{0}^b} 
    \mathbf{P}\mathbf{n}_0=
    -
     \mathbf{e}_1\scalp\int_{\partial \mathcal{B}_{0}^l\cup\partial\mathcal{B}_{0}^r} 
    \mathbf{P}\mathbf{n}_0,
\end{equation}
where the left hand side, 
because of the applied pressure loading (\ref{ppartizionato}) and the related property (\ref{pazienza}), can be rewritten as
\begin{equation}
\label{figodiddio}
    \mathbf{e}_1\scalp\int_{\partial \mathcal{B}_{0}^a\cup\partial\mathcal{B}_{0}^b} 
    \mathbf{P}\mathbf{n}_0= \mathbf{e}_1\scalp\int_{\partial \mathcal{B}_{0}^a\cup\partial\mathcal{B}_{0}^b} 
   \mathbf{S}\mathbf{n}_0.
\end{equation}
Introducing the  contact force components $R_1^a$ and $R_1^b$ along $\mathbf{e}_1$ on the two respective boundary portions $\partial \mathcal{B}_0^a$ and $\partial \mathcal{B}_0^b$ as \begin{equation}
    R_1^a=\mathbf{e}_1\scalp\int_{\partial \mathcal{B}_{0}^a} 
   \mathbf{S}\mathbf{n}_0,\qquad
   R_1^b=\mathbf{e}_1\scalp\int_{\partial \mathcal{B}_{0}^b} 
   \mathbf{S}\mathbf{n}_0 ,
\end{equation}
and considering the equilibrium equation (\ref{solenoidati1}) and eqn (\ref{figodiddio}), 
leads to 
\begin{equation}\label{R1daP}
    R_1^a+R_1^b=- \mathbf{e}_1\scalp\int_{\partial \mathcal{B}_{0}^l\cup\partial\mathcal{B}_{0}^r} 
    \mathbf{P}\mathbf{n}_0.
\end{equation}

A generalization of eqn (\ref{R1daP}) can be obtained for 
any arbitrary surface $\partial \mathcal{P}_0$ having a non-null portion in contact $\partial \mathcal{P}_0^{\scriptsize \mbox{tou}}\in \partial \mathcal{P}_0$ with outward normal $\mathbf{n}_0=\pm\mathbf{e}_2$.  The component $R_1(\partial\mathcal{P}_0^{\scriptsize \mbox{tou}})$ of the resultant force of the pressure distribution acting on $\partial\mathcal{P}_0^{\scriptsize \mbox{tou}}$ can be computed using any line integral with initial and ending point coincident with the limit points of the contact region for which the reaction force is evaluated
\begin{equation}\label{R1generica}
    R_1\left(\partial\mathcal{P}_0^{\scriptsize \mbox{tou}}\right)=\mathbf{e}_1\scalp\int_{\partial \mathcal{P}_0^{\scriptsize \mbox{tou}}} 
   \mathbf{S}\mathbf{n}_0=
   \mathbf{e}_1\scalp\int_{\partial \mathcal{P}_0^{\scriptsize \mbox{tou}}} 
   \mathbf{P}\mathbf{n}_0=-\mathbf{e}_1\scalp\int_{\partial \mathcal{P}_0\backslash\partial \mathcal{P}_0^{\scriptsize \mbox{tou}}} 
   \mathbf{P}\mathbf{n}_0.
\end{equation}
From eqn (\ref{R1generica}) it can be concluded that, even in the case of smooth constraints, the contact reaction force component $R_1$ transmitted to the body from the contact region $\partial \mathcal{P}_0^{\scriptsize \mbox{tou}}$  coincides with the negative of the $J$--integral (\ref{Jintfinite}), evaluated for a path $\Gamma_0 \equiv \partial \mathcal{P}_0\backslash\partial \mathcal{P}_0^{\scriptsize \mbox{tou}}$ as expressed by eqn (\ref{R1generica}). It follows that  
 the $J$--integral path-independence is preserved only for all paths with the same initial and final points,
 because the reaction force $R_1$ depends on the extension of the specific contact region,
  Fig. \ref{pathindependenceROUND}. 
In the case of a flat constraint ending with a sharp corner, the initial and final points  of $\Gamma_0$ can be chosen on the left and on the right of  the discontinuity in curvature, respectively, in the contact and in the traction-free surface. Thus, assuming a sufficiently regular behavior, the equation (\ref{R1generica}) becomes equation (\ref{R1generica0}) and a path-independence of the $J$--integral is found.

\subsubsection{Reaction force $R_1^a+R_1^b$ from the energy momentum tensor $\mathbf{C}$} 

It is now interesting to readdress the equilibrium of an elastic rectangular undeformed domain subject to either a pressure loading $p$ on $\partial \mathcal{B}_0^a$ and $\partial \mathcal{B}_0^b$,   by exploiting the solenoidal property of $\mathbf{C}$.
Considering the normal direction $\mathbf{n}_0=\pm \mathbf{e}_2$ and the property (\ref{pugnettone2})$_2$, it follows that
\begin{equation}
    \mathbf{e}_1\scalp \mathbf{C}\mathbf{n}_0=0,\qquad \mbox{on} \,\,\partial \mathcal{B}_0^a \cup \partial \mathcal{B}_0^b,
\end{equation}
and therefore taking the scalar product with $\mathbf{e}_1$, the solenoidal property of $\mathbf{C}$ reduces to
\begin{equation}
\label{stock84}
     \int_{\partial \mathcal{B}_0^l} \left(\Phi  - \mathbf{F}\mathbf{e}_1\scalp \mathbf{S}\mathbf{e}_1 \right) -
      \int_{\partial \mathcal{B}_0^r} \left(\Phi -  \mathbf{F}\mathbf{e}_1\scalp \mathbf{S}\mathbf{e}_1 \right)     = 0.
\end{equation}
If either a pressure $p$ or a dead loading $\mathbf{S}\mathbf{n}_0$ is applied on the boundary portions  $\partial \mathcal{B}_0^l$ and $\partial \mathcal{B}_0^r$, eqn (\ref{stock84}) simplifies as 
\begin{equation}
\label{stock69}
     \int_{\partial \mathcal{B}_0^l} \Big(\Phi   -
\Big\{
\begin{array}{lll}
     - p \mathcal{J}  \\
    S_{11}F_{11} +S_{21}F_{21} 
\end{array}
         \Big) -
 \int_{\partial \mathcal{B}_0^r} \Big(\Phi   -
\Big\{
\begin{array}{lll}
    - p \mathcal{J}  \\
      S_{11}F_{11} +S_{21}F_{21} 
\end{array}
         \Big)
= 0.
\end{equation}

Introducing the further assumption of homogeneous deformation gradient $\mathbf{F}$ in the neighborhood of the two boundaries $\partial \mathcal{B}_0^l$ and $\partial \mathcal{B}_0^r$, as sketched in Fig. \ref{presssss}, the integrals in equation (\ref{stock69}) can be trivially solved to yield
\begin{equation}
\label{stock79}
      \Phi^l - \Big\{
\begin{array}{lll}
     - p^l \mathcal{J}^l  \\
    S_{11}^lF_{11}^l +S_{21}^l F_{21}^l
\end{array}
         -
\Phi^r +\Big\{
\begin{array}{lll}
     - p^r \mathcal{J}^r  \\
    S_{11}^r F_{11}^r +S_{21}^r F_{21}^r
\end{array}
         = 0, 
\end{equation}
where the superscripts $l$ and $r$ respectively identify the relevant (constant) quantity evaluated on the boundaries $\partial \mathcal{B}_0^l$ and $\partial \mathcal{B}_0^r$. Interestingly,  the expression obtained by restricting  eqn (\ref{stock79}) to  only the terms in $p$,
\begin{equation}
\label{stock89}
     \Phi^l   +
     p^l \mathcal{J}^l =  \Phi^r   + p^r \mathcal{J}^r, 
\end{equation}
shares some similarities with Bernoulli’s equation for stationary flow in fluid mechanics.

It should be noted that $\Phi^j$, $\mathcal{J}^j$, $F_{11}^j$, and  $F_{21}^j$ ($j=l,r$) in eqns  (\ref{stock79}) are all functions of: (i.) the contact (pressure) distribution $p$ on the  boundaries $\partial \mathcal{B}_0^a$ and $\partial \mathcal{B}_0^b$, not explicitly appearing in eqns (\ref{stock84})--(\ref{stock79}) and (ii.) the pressure distribution $p$ or the dead loading $\mathbf{S}\mathbf{n}_0$  on the boundaries $\partial \mathcal{B}_0^l$ and $\partial \mathcal{B}_0^r$, as in Figs. \ref{panino_intro} and \ref{presssss}. 
Except for trivial cases, the pressure or dead load (ii.) cannot easily be related to the pressure distribution (i.), because equilibrium has to be satisfied, therefore eqns (\ref{stock79}) contain more than one unknown. 
However, assuming $S_{21}^l=S_{21}^r=0$ and that  the lateral load (ii.) is applied only on the boundary $\partial \mathcal{B}_0^j$  ($j=l$ or $r$) while the boundary  $\partial \mathcal{B}_0^i$  ($i=l$ or $r$, with $i\neq j$) remains  unloaded, equations (\ref{stock79}) can be used to define the unknown loading, either $p^j$ or $S_{11}^j$. In particular, eqns (\ref{stock79}) lead  to 
\begin{equation}
\label{strazalbal}
\begin{array}{rr}
     p^j \mathcal{J}^j \\
     - S_{11}^j F_{11}^j
\end{array}
\Big\} = \Phi^i -\Phi^j, 
\qquad 
i,j=l,r,\quad \mbox{with}\,\,i\neq j,
\end{equation}
so that, when the load  (i.) (the pressure distribution $p$ on the the  boundaries $\partial \mathcal{B}_0^a$ and $\partial \mathcal{B}_0^b$) is prescribed, the relevant equation 
becomes a nonlinear implicit equation in the variable representing the   
load (ii.), applied on the boundary   $\partial \mathcal{B}_0^j$, either $p^j$ or $S_{11}^j$ ($j=l,r$).  The sum of the two components  $R_1^a+R_1^b$ of the resultant force along $\mathbf{e}_1$ of the pressure $p$ applied on the  boundaries $\partial \mathcal{B}_0^a$ and $\partial \mathcal{B}_0^b$ can be obtained from equilibrium for the two loading cases as
\begin{equation}
\label{reactionR1smooth}
R_1^a+R_1^b=\Big\{
\begin{array}{rr}
     -p^l h^l,\\
     S_{11}^l h_0,
\end{array}, 
\qquad
\mbox{and}
\qquad
R_1^a+R_1^b=\Big\{
\begin{array}{ll}
     p^r h^r,\\
     -S_{11}^r h_0. 
\end{array}
\end{equation}
Assuming now $F_{21}^j=0$, so that $\lambda_1^j=F_{11}^j$, $\lambda_2^j=F_{22}^j$, and $h^j=\lambda_2^j h_0$ on the loaded boundary $\partial \mathcal{B}_0^j$ ($j=l$ or $r$), equation (\ref{strazalbal}) implies
\begin{equation}
\label{reactionR1smoothENERGY}
R_1^a+R_1^b=\dfrac{\Phi^l -\Phi^r}{\lambda_1^j}h_0,
\qquad
\mbox{with}\,\,\,j=l \,\mbox{ or }\,r,
\end{equation}
an expression that can alternatively be derived from eqn (\ref{R1daP}) by recalling from eqn (\ref{solitabazza}) that $\lambda_1=1+u_{1,1}$. 
Equation (\ref{reactionR1smoothENERGY}) shows that only a non-null difference in the strain energy $\Phi$ at the two boundaries $\partial \mathcal{B}_0^l$ and $\partial \mathcal{B}_0^r$ induces a force $R_1=R_1^a+R_1^b$ and reduces to eqn (\ref{solettina}) when the right edge is unloaded, $\Phi^r=0$, as is the case of the loading conditions sketched in Fig. \ref{panino_intro}.

\section{Energy release rate $G$ and the configurational nature of 
the frictionless contact force component $R_1$}\label{eshelby_generalizzato}

A rigid, frictionless, and flat constraint ending with a  rounded or sharp corner is in contact  against the boundary of a hyperelastic body with a flat surface in its reference configuration. 
The frictionless constraint is  assumed to be capable of altering its extension of contact 
by increasing the size of its flat surface 
through an horizontal growth of the position of its, say right, corner. 
Analogously to the concept of configurational force on defects or inhomogeneities introduced by Eshelby \cite{eshelby1975}, the idea of a configurational force acting on the  corner of frictionless and rigid  constraints can be introduced. 

For a growth $\delta \xi_0$ in  the size of the frictionless constraint along  $\mathbf{e}_1$, defined with respect to the undeformed configuration of a hyperelastic solid, the configurational force component $F_1^c$ parallel to the growth direction can be defined as an energy release rate $G$
\begin{equation}
\label{energymom0}
    F_1^c=G= -\frac{\partial \mathcal{V}}{\partial \xi_0},
\end{equation}
where $\mathcal{V}$ is the total potential energy of the mechanical system at equilibrium.
For both cases of sharp or rounded corner, it is shown  that the configurational force component $F_1^c$ equals the $J$--integral,
\begin{equation}
\label{JF1c}
    \boxed{
    F_1^c=  J. 
    }
\end{equation}

The treatment is restricted for simplicity to two-dimensions, where surfaces are curves and planes straight lines. Final applications are referred to a
rectangular undeformed shape $\mathcal{B}_0$ of the contactor, as described by eqn  (\ref{rectangulardomain}). 
As a generalization of the results presented in Sect. \ref{sezJ+ciava}, 
it is shown that a 
frictionless punch ending with a sharp corner can generate a horizontal configurational force even when in contact with a planar surface of an elastic solid.

\subsection{Variation in the length of a flat, frictionless, and rigid  constraint ending with a sharp corner}
\label{corneraz}

The contactor has an initially flat boundary $\partial\mathcal{B}_0^a$ having unit normal $\mathbf{n}_0=\mathbf{e}_2$, while the target has a rectilinear surface (with outward unit normal $-\mathbf{e}_2$) ending with a corner, located at point $\mathbf{y}$, Fig. \ref{variazione}. 
The frictionless constraint is in contact with the elastic body on the portion of the boundary $\partial\mathcal{B}_0^{a,C}$. 
The contact is 
assumed to be \lq full', so that grazing does not occur and all the points on the touching surface  $\mathbf{x} \in \partial \mathcal{B}_0^{a,\scriptsize \mbox{tou}}$ belong to $\partial \mathcal{B}^{a,C}_0 = \partial \mathcal{B}_0^{a,\scriptsize \mbox{tou}}$, including the corner point $\mathbf{y}$. On the solid, the latter point is back-transformed in the reference configuration into $\mathbf{y}_0$. 
The latter point is perturbed by postulating a small growth, parallel to the rectilinear contact surface of the constraint, to a neighboring point $\mathbf{y}_0 + \delta \xi_0 \be_1$.
It is therefore possible to strictly follow Eshelby \cite{eshelby1975}, thus defining a surface $\partial \mathcal{S}_0$ 
enclosing a region $\mathcal{S}_0$ which contains the corner $\mathbf{y}_0$ in the reference configuration $\mathcal{B}_0$, called \lq original surface' and 
introducing a \lq replica' region  equal to $\mathcal{S}_0$, but translated to 
the region $\mathcal{S}'_0$
with 
surface $\partial \mathcal{S}'_0$, obtained by applying a  rigid displacement vector $-\delta \xi_0 \mathbf{e}_1$ to $\mathcal{S}_0$. 
Note that the surfaces $\partial \mathcal{S}_0$ and $\partial \mathcal{S}'_0$ are punctured at the singular point $\mathbf{y}_0$. 
The steps below are followed.
\begin{figure}[!h]
\centering
\includegraphics[width=1\textwidth]{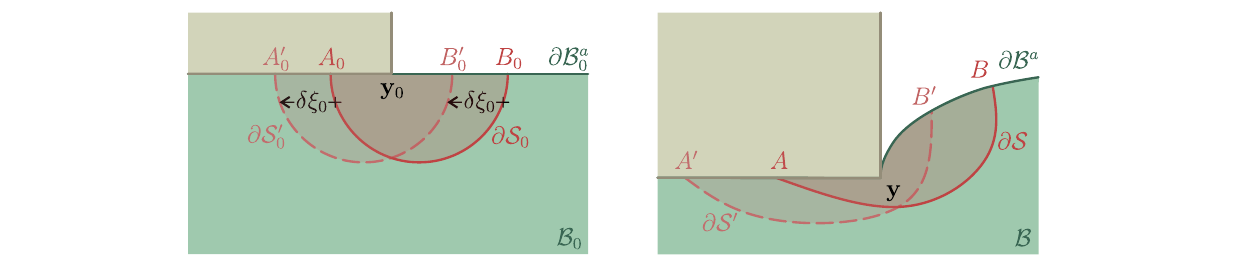}
\caption{\label{variazione} 
An elastic solid (green) with a planar surface is pressed against a flat, rigid, and frictionless constraint (brown). Left (Right): The constraint has a corner touching the elastic body at point $\mathbf{y}_0$ 
(at point $\mathbf{y}$) 
in the reference (the current) configuration $\mathcal{B}_0$ ($\mathcal{B}$). 
In a setting which follows Eshelby, the corner of the frictionless constraint is assumed to grow of an amount $\delta\xi_0\,\mathbf{e}_1$ in the reference configuration. 
Two identical regions $\mathcal{S}_0$ and $\mathcal{S}_0'$ are assumed in the reference configuration differing in a rigid horizontal shift $-\delta \xi_0\,\mathbf{e}_1$, both enclosing $\mathbf{y}_0$. The two regions are transformed by the deformation into the regions $\mathcal{S}$ and $\mathcal{S}'$, both enclosing the corner of the constraint at point $\mathbf{y}$. 
}
\end{figure}

\begin{enumerate}[(i.)]
    \item  In the reference  configuration $\mathcal{B}_0$, the material in the region $\mathcal{S}_0$ is cut out and kept aside. Both the latter and the rest of the body are considered to still be subject to the nominal tractions that were exchanged across the surface cut out of the body, in addition, the cut out piece is also assumed to be subjected to the surface forces transmitted by the constraint. 
    
    \item  Consider the material in the replica region, inside of $\mathcal{S}'_0$, and apply on its surface $\partial \mathcal{S}'_0$ the nominal tractions transmitted by the rest of the deformable body and by the constraint. Comparing the energies inside $\mathcal{S}'_0$ and $\mathcal{S}_0$ and taking the limit of vanishing
    $\delta \mathbf{\xi}_0$, 
the 
    Leibniz integral rule for a closed curve  $\partial \mathcal{S}_0$ in a two-dimensional domain, rigidly shifted inside $\mathcal{B}_0$, is obtained  \cite{flanders}
    \begin{equation}
    \label{pera}
        \frac{\mbox{d}}{\mbox{d} \xi_0} \int_{\mathcal{S}_0(\xi_0)} \Phi = - 
        \int_{\partial \mathcal{S}_0} \Phi \,\mathbf{n}_0
        \cdot \mathbf{e}_1,
    \end{equation}
where $\mathbf{n}_0$ is outward unit normal to $\partial \mathcal{S}_0$, so that the surface on the horizontal edge of $\partial \mathcal{S}_0$ does not contribute.
Equation (\ref{pera}) may be understood in a generalized sense,  depending on the kind of possible singularity present at the end of the target, and provides the differentiation of the elastic energy corresponding to an infinitesimal translation of $\mathcal{S}_0$, equivalent to an infinitesimal increase in the length of the target.

\item  Due to the deformation, the deformed replica   $\mathcal{S}'$ (transformed of $\mathcal{S}'_0$) does not fit into the hole left by the \lq excision' of $\mathcal{S}$ (transformed of $\mathcal{S}_0$). 
In particular, any point $\mathbf{r}_0$ inside the region of the replica equals a corresponding point $\mathbf{x}_0$ inside $\mathcal{S}_0$, plus the shift $-\delta \xi_0\mathbf{e}_1$. Therefore the displacement of $\mathbf{r}_0$ is 
$\mathbf{u}(\mathbf{r}_0) = \mathbf{u}(\mathbf{x}_0-\delta \xi_0 \mathbf{e}_1$), so that at first-order 
\begin{equation}
\label{soc-mel}
    \mathbf{u}(\mathbf{r}_0) = \mathbf{u}(\mathbf{x}_0) -\delta \xi_0 \nabla \mathbf{u}(\mathbf{x}_0)\mathbf{e}_1.
\end{equation}

It follows from eq. (\ref{soc-mel}) that, in addition to a rigid-body translation $\delta \xi_0 \mathbf{e}_1$ (which does not produce any work), to fit the deformed $\mathcal{S}'$ into the deformed hole left by $\mathcal{S}$, an additional displacement has to be added to the displacement $\mathbf{u}(\mathbf{x}_0)$ on the surface $\partial \mathcal{S}_0$ of the hole left in $\mathcal{B}_0$. In differential terms, the latter displacement satisfies  
\begin{equation}
    \frac{\partial \mathbf{u}}{\partial \xi_0} = 
    -(\nabla\mathbf{u})\mathbf{e}_1, 
\end{equation}
so that the amount of work done by the tractions on the surface of the hole $\partial \mathcal{S}_0$ is equal to 
\begin{equation}
\label{peretta}
    \frac{\partial W}{\partial \xi_0} =  \int_{\partial \mathcal{S}_0} 
   \mathbf{e}_1 
    \cdot  \nabla \mathbf{u}^T
    \mathbf{S} \mathbf{n}_0,
\end{equation}
where again $\mathbf{n}_0$ is the outward unit normal to $\partial \mathcal{S}_0$.

\item  The change in the total potential energy $\mathcal{V}$ is the sum of equations (\ref{pera}) and (\ref{peretta}),
\begin{equation}
    \frac{\partial \mathcal{V}}{\partial \xi_0} = 
    \frac{\partial W}{\partial \xi_0} 
    +\frac{\mbox{d}}{\mbox{d} \xi_0} \int_{\mathcal{S}_0(\xi_0)}\Phi,
\end{equation}
which, using equations (\ref{pera}) and (\ref{peretta}), yields to the energy release rate $G$ as 
\begin{equation}
\label{abelarda}
  G=  -\frac{\partial \mathcal{V}}{\partial \xi_0} =  \int_{\partial \mathcal{S}_0} 
        \mathbf{e}_1 \cdot  
        \left(
        \Phi \mathbf{I}
        - \nabla 
        \mathbf{u}^T
    \mathbf{S} \right) \mathbf{n}_0.
\end{equation}

Note that the surface $\partial \mathcal{S}_0$ comprises only the part inside the solid, while contributions on the flat boundary vanish. This statement is trivial for the term $\Phi$ because $\mathbf{n}_0$ is orthogonal to $\mathbf{e}_1$ on the flat contact edge of the elastic solid. 
Regarding the term 
$\nabla \mathbf{u}^T\mathbf{S} \mathbf{n}_0$, it may be observed that $\mathbf{Sn}_0=\mathbf{0}$ on the flat boundary outside the constraint, while inside the constraint equation (\ref{solitabazza}) together with the frictionless condition (\ref{pugnettone}) and the fact that $\mathbf{Fe}_1$ is parallel to $\mathbf{e}_1$ along the contact allow to conclude.

\item Now $\mathcal{S}'$ fits the hole left by $\mathcal{S}$ in the original body and can be welded in it. The nominal tractions on both sides of the surface $\partial \mathcal{S}_0$ differ on a force distribution which gives higher-order effects and can be neglected. 
{\it We are now left with the system as it was to begin with, except that the end of the constraint is shifted of an amount $\delta\xi_0\mathbf{e}_1$ with respect to its initial position defined in the reference configuration.}

\end{enumerate}

The operations (i.)--(iii.) involved in the Eshelby proof are summarized in Fig. \ref{eshelbone}.

\begin{figure}
\centering
\includegraphics[width=0.8\textwidth]{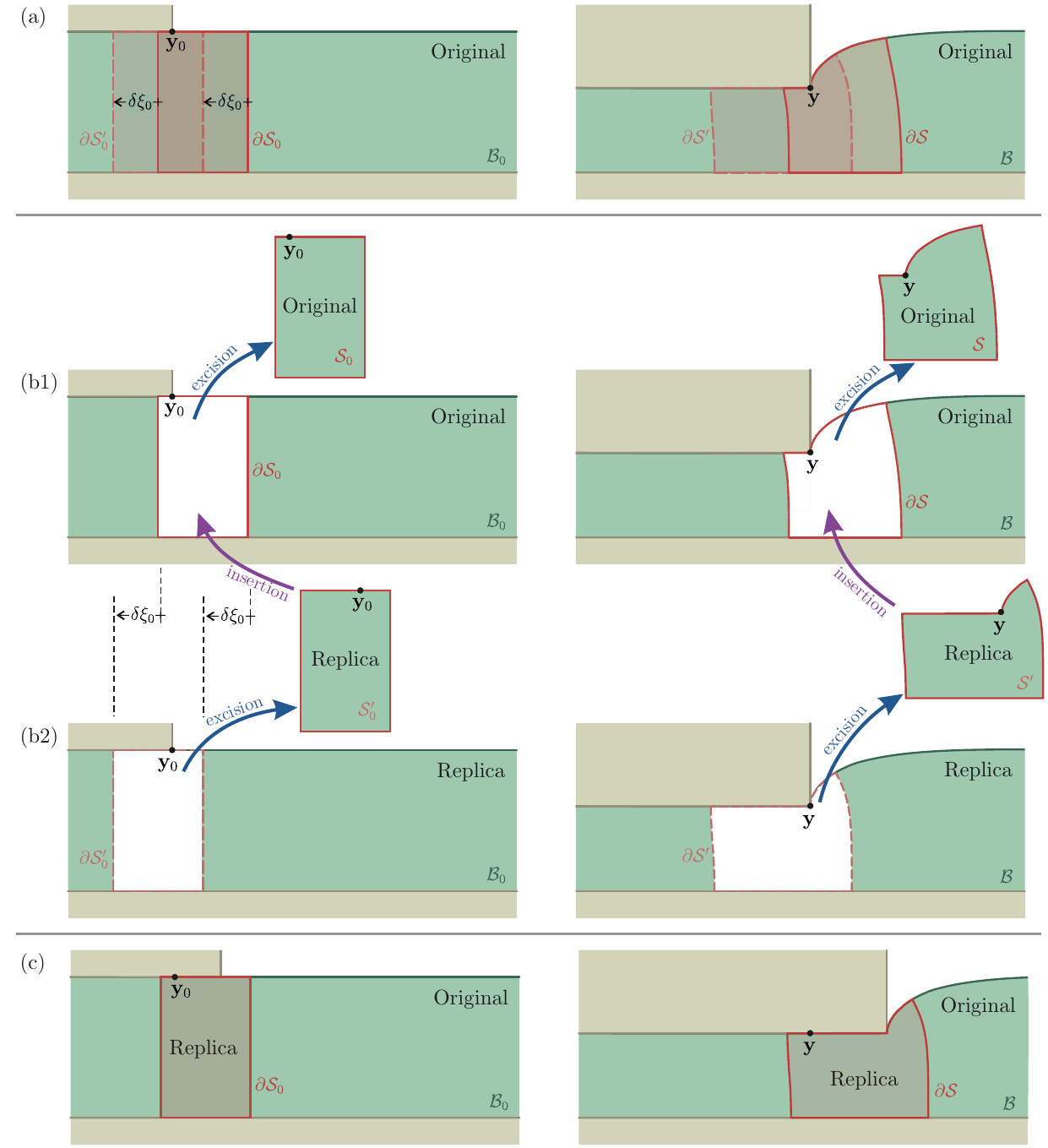}
\caption{\label{eshelbone} Sequence of operations in the Eshelby scheme for the deformation of an elastic body 
against a rigid and frictionless constraint, which increases its length of an amount $\delta \xi_0$ in the reference configuration.  Reference (current) configurations are shown on the left (on the right).
(a) The referential regions $\mathcal{S}_0$ and $\mathcal{S}_0'$ (the latter is the image of the former obtained through a shift of amount $-\delta \xi_0$) enclose the end of the constraint and are transformed by the deformation into $\mathcal{S}$ and $\mathcal{S}'$ in the current configuration.
(b1) The region $\mathcal{S}_0$ and its transformed counterpart $\mathcal{S}$ are ideally \lq excised' from the original body. 
(b2) The region $\mathcal{S}'_0$ and its transformed counterpart $\mathcal{S}'$ are ideally \lq excised' from a \lq replica' version of the original body.
The elastic energies contained within $\mathcal{S}_0$ and $\mathcal{S}'_0$ differ only in the crescent-shaped regions 
obtained by superposition of $\mathcal{S}_0$ and $\mathcal{S}_0'$, so that the derivative of the elastic energy with respect to the configurational parameter is given by eqn (\ref{pera}). 
The deformed \lq replica' region $\mathcal{S}'$ does not fit the hole left in the original region by the excision of $\mathcal{S}$, so that displacements have to be applied on the boundary of the hole, producing the increment of work expressed by eqn (\ref{peretta}).
(c) The replica finally fits the hole in the original body and the corner of the rigid constraint is advanced of an amount $\delta\xi_0$ with respect to the original reference configuration. The remaining mismatch in the traction vector at the boundary of the region is higher-order and can be neglected.
}
\end{figure}

Equation (\ref{abelarda})
 can be viewed as the work done by the {\it configurational force} $\mathbf{F}^c$ for a unit displacement in the direction $\mathbf{e}_1$, and therefore by the configurational force component $F_1^c= \mathbf{F}^c \cdot \mathbf{e}_1$
\begin{equation}
\label{energymom}
    \boxed{
    F_1^c=  G =        \mathbf{e}_1 \cdot \int_{\partial \mathcal{S}_0} 
                \left(
        \Phi \mathbf{I}
        -\nabla 
        \mathbf{u}^T
    \mathbf{S} \right) \mathbf{n}_0. 
    }
\end{equation}

Note that the configurational force  $\mathbf{F}^c$ remains determined only in its  component $F_1^c$ along $\mathbf{e}_1$, because the translation of $\mathcal{S}_0$ is not arbitrary, differently from  the original treatment by Eshelby, but prescribed parallel to the direction $\mathbf{e}_1$.

Finally, recalling the energy-momentum tensor $\mathbf{P}$, eqn (\ref{nonneshelby}), 
equation (\ref{energymom}) becomes
\begin{equation}
\label{sollievo}
\boxed{
F_1^c  =  
         \mathbf{e}_1 \cdot \int_{\partial \mathcal{S}_0} 
         \mathbf{P} \mathbf{n}_0, 
         }
\end{equation}
where the integrand is null on the upper flat portion of the boundary, except possibly at the point $\mathbf{y}_0$ where the corner of the constraint is present. 
The surface 
$\partial \mathcal{S}_0$ 
can be shrunk up to the limit of that point, without changing the value of the integral. This leads  to the path-independent $J$--integral, 
when the target is flat and ends with a corner, 
\begin{equation}
\label{eraora}
    J = F_1^c,
\end{equation}
so that the configurational force in the direction $\mathbf{e}_1$ is equal to the horizontal resultant of the force acting on the solid with reversed sign
\begin{equation}\label{fcmenor1}
    F_1^c = -R_1.
\end{equation}

\paragraph{Application to rectangular elastic domains.} It is interesting to note that, when a rectangular undeformed elastic solid is considered, the region  $\mathcal{S}_0$ can be assumed as illustrated in Fig. \ref{eshelbone}(a), namely, rectangular with boundary $\partial\mathcal{S}_0^l\cup\partial\mathcal{S}_0^b\cup\partial\mathcal{S}_0^r$, so that, assuming the frictionless condition on $\partial\mathcal{S}_0^b\subseteq\partial\mathcal{B}_0^b$, the configurational force component $F_1^c$ reduces to
\begin{equation}\label{errrettangolo}
\boxed{
    F_1^c = \int_{\partial \mathcal{S}_0^r} 
        \left[\Phi - u_{1,1}S_{11}- u_{2,1}S_{21}\right]-\int_{\partial \mathcal{S}_0^l} 
        \left[\Phi - u_{1,1}S_{11}- u_{2,1}S_{21}\right],
        }
\end{equation}
which is equivalent to eqns (\ref{R1daP}) and (\ref{stock84}) (respectively obtained through the solenoidal property of the energy-momentum tensors $\mathbf{P}$ and $\mathbf{C}$ in the absence of singularities) because equilibrium implies 
\begin{equation}
R_1=\int_{\partial\mathcal{S}_0^r} S_{11}-\int_{\partial\mathcal{S}_0^l} S_{11}.
\end{equation}

\subsection{Variation in the length of the constraint with a rounded corner} \label{tondo}

The presence of a smooth-end is now addressed.
Analogously to the treatment of the growth of a flat surface notch in a two-dimensional deformation field given by Rice \cite{rice1968}, the right end of the frictionless and straight constraint is considered to have  a smooth \lq cap', along which the contact with the elastic body is lost.

The end of the constraint is assumed to be able to rigidly translate in the direction $\mathbf{e}_1$, parallel to the constraint before the initiation of the smooth cap.  
The treatment developed in the previous Section still holds under  the caution that $\partial \mathcal{S}_0$ has to contain  all the zone contacting with the smooth \lq movable' cap. 
A repetition of the calculations developed in the previous Section leads now again to equation 
(\ref{sollievo}), where now $\partial \mathcal{S}_0$ is any surface enclosing all the zone in contact with the smooth \lq movable' cap. Consequently, equation (\ref{eraora}) is again obtained, in agreement with the  evaluation of $R_1$ provided  by eqn (\ref{R1generica}), following from the solenoidal property of the energy-momentum tensor $\mathbf{P}$.

\section{Connection with the  configurational sliding force acting on an unshearable rod constrained by a sliding sleeve}\label{sezstrutture}

The introduced theoretical framework, disclosing the development of configurational sliding forces at the (sharp or rounded) corner of a frictionless, rigid, and flat surface acting on an elastic body, is now used to  throw light on the akin problem of  elastic rods partially constrained with a sliding sleeve. The presence of  configurational forces at the end of a sliding sleeve  was disclosed by analyzing  one-dimensional flexible structures through a variational approach \cite{ARMANINI201982,bigoni2015eshelby,koutso2023} or by imposing jump conditions in the material momentum balance law \cite{hanna2018partial, o2017book, o2007material}. 
This result is  confirmed here through the application of the framework developed in the previous Sections to an elastic solid of rectangular shape in its undeformed configuration $\mathcal{B}_0$, as defined by eqn (\ref{rectangulardomain}), in contact with two frictionless rigid surfaces realizing a sliding sleeve constraint. 

\subsection{Rod's kinematics} 
According to  the kinematic assumptions usually made in rod mechanics \cite{MAGNUSSON20018441},  the deformation for an  elastic solid of rectangular shape is prescribed to provide null transverse strain and to be described by the following  expressions, linearized in the variable $x_2^0$   (Fig. \ref{travata}, left), 
\begin{equation}\label{rodpositionfield}
\left\{
\begin{array}{lll}
x_1\left(x_1^0,x_2^0\right)=x_1^0+u\left(x_1^0\right)-x_2^0 \sin\theta\left(x_1^0\right),\\[2mm]
x_2\left(x_1^0,x_2^0\right)=v\left(x_1^0\right)+x_2^0 \cos\theta\left(x_1^0\right),
\end{array}\right.
\qquad 
    \mbox{in}\,\,\,\mathcal{B}_0:=\left\{x_1^0\in[0,\ell_0],x_2^0\in\left[\dfrac{h_0}{2},-\dfrac{h_0}{2}\right]\right\} ,
\end{equation}
where $u(x_1^0)$, $v(x_1^0)$, and $\theta(x_1^0)$  are the three kinematic fields describing the deformed configuration of the solid, respectively, the displacement components along $\mathbf{e}_1$ and $\mathbf{e}_2$, and the inclination angle of the rod's axis 
with respect to the  direction $\mathbf{e}_1$, corresponding to the undeformed tangent.
Only two among the three kinematic descriptors $u$, $v$, and $\theta$ are independent, since the Euler–Bernoulli assumption implies 
\begin{equation}
\tan\theta\left(x_1^0\right)=\dfrac{v'\left(x_1^0\right)}{1+u'\left(x_1^0\right)},
\end{equation}
where   a prime denotes differentiation with respect to the axial coordinate $x_1^0$, while impenetrability imposes the constraint $u'\left(x_1^0\right)>-1$ on the displacement. 
\begin{figure}[!htb]
\centering
\includegraphics[width=1\textwidth]{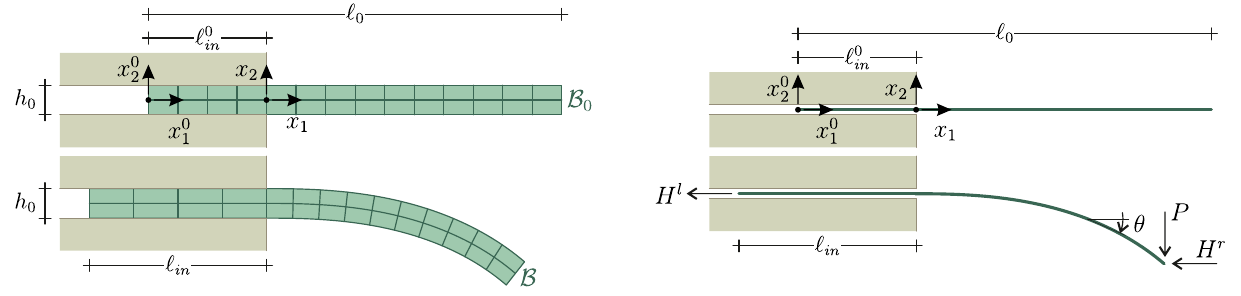}
\caption{\label{travata} 
Left: The kinematic assumptions, usually introduced for a rod, eqn (\ref{rodpositionfield}), are imposed to an elastic solid of rectangular shape in its undeformed configuration, constrained between two rigid and frictionless constraints. Right: the rod's model, representing an extensible  version of the variable-length elastica subject to end loads. 
}
\end{figure}

The two primary kinematic fields measuring the deformed state of the extensible elastica are the generalized curvature $\theta'\left(x_1^0\right)$  and the rod's axis axial deformation $\eta\left(x_1^0\right)$ (which satisfies $\eta\left(x_1^0\right)>-1$ because of the impenetrability constraint)
\begin{equation}
    \eta\left(x_1^0\right)=\sqrt{\left[1+u'\left(x_1^0\right)\right]^2+\left[v'\left(x_1^0\right)\right]^2}-1,
\end{equation}
through which the following geometrical relations can be derived
\begin{equation}\label{travegeometry}
\sin\theta\left(x_1^0\right)=\dfrac{v'\left(x_1^0\right)}{1+\eta\left(x_1^0\right)},
    \qquad     \cos\theta\left(x_1^0\right)=\dfrac{1+u'\left(x_1^0\right)}{1+\eta\left(x_1^0\right)}.
\end{equation}
The resultant  force components $N_1$ and $N_2$, respectively aligned parallel to $\mathbf{e}_1$ and $\mathbf{e}_2$, and the moment $M$ are given from the equilibrium equivalence, imposed for the cross section at the generic coordinate $x_1^0$, as
\begin{equation}
    N_1\left(x_1^0\right)=\int_{-\frac{h_0}{2}}^{\frac{h_0}{2}} S_{11}\mbox{d}x_2^0,\qquad
    N_2\left(x_1^0\right)=\int_{-\frac{h_0}{2}}^{\frac{h_0}{2}} S_{21}\mbox{d}x_2^0,\qquad
    M\left(x_1^0\right)=-\int_{-\frac{h_0}{2}}^{\frac{h_0}{2}}  \left[S_{11}\cos\theta+S_{21}\sin\theta\right]x_2^0\mbox{d}x_2^0,
\end{equation}
where the first two components can be composed to evaluate the axial and shear forces $N$ and $T$
\begin{equation}
    N\left(x_1^0\right)=N_1\left(x_1^0\right) \cos \theta\left(x_1^0\right)
    + N_2\left(x_1^0\right) \sin \theta\left(x_1^0\right),\qquad
    T\left(x_1^0\right)=-N_1\left(x_1^0\right) \sin \theta\left(x_1^0\right)
    + N_2\left(x_1^0\right) \cos \theta\left(x_1^0\right).
\end{equation}

\subsection{Sliding sleeve constraint and the evaluation of the configurational force component $F_1^c$}
The elastic rectangular solid under consideration is assumed in partial contact with two symmetric straight, frictionless, and rigid constraints, realizing  a sliding sleeve with sliding direction parallel to $\mathbf{e}_1$, in a setting similar to that reported in Fig. \ref{travata} (left). The sliding sleeve is assumed to have its  exit  point located at the  cross section marked by the coordinate $x_1^0=\ell_{in}^0$ (with $\ell_{in}^0\in[0,\ell_0]$), referred to the undeformed rod. Thus, the following constraints apply 
\begin{equation}\label{slidingsleeveconstraint}
    \theta\left(x_1^0\right)=
    v\left(x_1^0\right)=0,\qquad\mbox{for}\,\,x_1^0\in\left[0,\ell_{in}^0\right],
\end{equation}
which, considering  the origin of the reference system $x_1$--$x_2$ coincident with the sliding sleeve exit, 
\begin{equation}\label{exitpoint}
x_1\left(x_1^0=\ell_{in}^0,x_2^0\right)=0,
\end{equation}
imply the validity of the following relation
\begin{equation}
    u_1\left(x_1^0=\ell_{in}^0\right)=-\ell_{in}^0.
\end{equation}

From the deformation field, eqs. (\ref{rodpositionfield}), the  components of the displacement gradient $\nabla\mathbf{u}$, relevant for the application of the expression (\ref{errrettangolo}), provides the configurational force component $F_1^c$  as
\begin{equation}
\begin{array}{lll}
u_{1,1}\left(x_1^0,x_2^0\right)=u'\left(x_1^0\right)-x_2^0 \,\theta'\left(x_1^0\right)\,\cos\theta\left(x_1^0\right),\\[2mm]
u_{2,1}\left(x_1^0,x_2^0\right)=v'\left(x_1^0\right)-x_2^0 \,\theta'\left(x_1^0\right)\,\sin\theta\left(x_1^0\right).
\end{array}
\end{equation}
By considering the boundaries $\partial\mathcal{S}_0^r$ and $\partial\mathcal{S}_0^l$ coincident with the cross sections  respectively located at 
$x_1^{[r]0}$ and $x_1^{[l]0}$, the expression (\ref{errrettangolo})  for the configurational force  component $F_1^c$ reduces to
\begin{equation}\label{trave}
   F_1^c =  \Psi\left(x_1^{[r]0}\right) -
   \Psi\left(x_1^{[l]0}\right)-\left.\left[N_1\,u' +  N_2\, v' + M\,\theta' \right]\right|_{x_1^{[r]0}}+ \left.\left[N_1\,u' +  N_2\, v' + M\,\theta' \right]\right|_{x_1^{[l]0}},
\end{equation}
where $\Psi$ is the rod's elastic energy density, evaluated as
\begin{equation}
    \Psi\left(x_1^0\right)=\int_{-\frac{h_0}{2}}^{\frac{h_0}{2}} \Phi\,\mbox{d}x_2^0.
\end{equation}
Assuming that the two cross sections $\partial\mathcal{S}_0^r$ and $\partial\mathcal{S}_0^l$ are the cross sections respectively \lq just after' and 
\lq just before' the coordinate $\ell_{in}^0$, where the sliding sleeve exit is back transformed in the reference configuration, the configurational force component
$F_1^c$ (\ref{trave}) simplifies to
\begin{equation}\label{trave2}
   \boxed{
   F_1^c = \salto{.1}{
        \Psi\left(\ell_{in}^0\right)}  -\salto{.1}{N\left(\ell_{in}^0\right) \eta\left(\ell_{in}^0\right) } -M\left(\ell_{in}^{0+}\right) \theta'\left(\ell_{in}^{0+}\right),
        }
\end{equation}
where the brackets $\salto{.1}{\cdot}$   denote the jump of the relevant quantity at the sliding sleeve exit,
\begin{equation}
    \salto{.1}{
        f\left(\ell_{in}^0\right)}=
        f\left(\ell_{in}^{0+}\right)-
        f\left(\ell_{in}^{0-}\right),
\end{equation}
and, because of the constraint (\ref{slidingsleeveconstraint}), the following identities at the left and right limit points of the sliding sleeve exit hold
\begin{equation}
    \eta\left(\ell_{in}^{0\pm}\right)=u'\left(\ell_{in}^{0\pm}\right),\qquad
    N\left(\ell_{in}^{0\pm}\right)=N_1\left(\ell_{in}^{0\pm}\right), \qquad
    v'\left(\ell_{in}^{0\pm}\right)=0.
\end{equation}
It is highlighted that $N$, $M$,  $\eta$, $\theta'$, and  $\Psi$ may display a jump in their value at the coordinate $\ell_{in}^0$.

Equation (\ref{trave2}) 
provides the expression for the configurational force $F_1^c$ acting on the variable-length Euler  elastica by including  axial deformability and a generic (possibly non-quadratic) rod's  energy density $\Psi$. This equation reduces to that obtained in \cite{bigoni2015eshelby} when  axial inextensibility and  quadratic energy in the curvature, $\theta'$, are assumed.

It is finally observed that equation (\ref{trave2}) can be interpreted as a jump condition for the material momentum balance law, which was introduced in the  forerunning contribution  by O'Reilly \cite{o2007material}, who established a novel frontier in the configurational mechanics of structures, by enhancing a previous formulation by Kienzler and Herrmann \cite{kienzler}. More specifically, by introducing the concept of material force $\mathsf{C}(x_1^0)$,
\begin{equation}\label{Cforceoreilly}
    \mathsf{C}\left(x_1^0\right)=\Psi\left(x_1^0\right)  -N\left(x_1^0\right) \eta\left(x_1^0\right)  -M\left(x_1^0\right) \theta'\left(x_1^0\right),
\end{equation}
the jump condition at the singularity point $x_1^0=\ell_{in}^0$ is given by
\begin{equation}
     \salto{.1}{ \mathsf{C}\left(\ell_{in}^0\right)} =F_1^c, 
\end{equation}
which is coincident with eqn (\ref{trave2}), by recalling that $\theta'\left(\ell_{in}^{0-}\right)=0$,  due to the presence of the sliding sleeve.
Interestingly, the material force $\mathsf{C}(x_1^0)$, eqn (\ref{Cforceoreilly}), for the one-dimensional model can be obtained as the integral of the energy-momentum tensor component  $P_{11}=\mathbf{e}_1\scalp\mathbf{P}\mathbf{e}_1$  [expressed in a linearized  kinematics (\ref{rodpositionfield})], calculated on the cross section of the rod
\begin{equation}
    \mathsf{C}\left(x_1^0\right)=\int_{-\frac{h_0}{2}}^{\frac{h_0}{2}} P_{11}\left(x_1^0,x_2^0\right) \mbox{d}x_2^0=\int_{-\frac{h_0}{2}}^{\frac{h_0}{2}} \left[\Phi-u_{1,1}S_{11}-u_{2,1}S_{21}\right]\mbox{d}x_2^0.
\end{equation} 
In conclusion, the presence of the configurational force $F_1^c$ at the sliding sleeve exit so far  obtained for rod models \cite{ARMANINI201982,bigoni2015eshelby, hanna2018partial,koutso2023, o2017book, o2007material} is confirmed from a solid mechanics point of view.

\subsection{Configurational force via a variational approach}

Expression (\ref{trave2}) for the configurational force component $F_1^c$ is now derived through a variational approach. Attention is restricted to a specific loading condition, corresponding to dead loads at the two rod's ends, in particular a load $-H^l\mathbf{e}_1$ is assumed to be applied at  $x_1^0=0$, while a load  $-H^r\mathbf{e}_1-P\mathbf{e}_2$ at  $x_1^0=\ell_0$ (Fig. \ref{travata}, bottom right). The total potential energy $\mathcal{V}$ is given by the difference of strain energy stored within the rod and the work done by the dead loadings,
\begin{equation}\label{EPTtrave}
\mathcal{V}=\int_0^{\ell_{in}^{0-}} \Psi(\eta,\theta')\mbox{d}x_1^0+\int_{\ell_{in}^{0+}}^{\ell_0} \Psi(\eta,\theta')\mbox{d}x_1^0+H^l x_1(0,0)+H^r x_1(\ell_0,0)
+P x_2(\ell_0,0).
\end{equation}
Recalling eqns (\ref{travegeometry}), (\ref{slidingsleeveconstraint}),  and (\ref{exitpoint}),  the following kinematic relations hold
\begin{equation}
     \ds x_1(0,0)=
     -\int_0^{\ell_{in}^{0-}} (1+\eta)\mbox{d}x_1^0,  \qquad
    \ds x_1(\ell_0,0)=
    \int_{\ell_{in}^{0+}} ^{\ell_{0}} (1+\eta)\cos\theta\mbox{d}x_1^0 ,\qquad
    \ds x_2(\ell_0,0)=
    \int_{\ell_{in}^{0+}} ^{\ell_{0}} (1+\eta)\sin\theta\mbox{d}x_1^0,
\end{equation}
and the total potential energy $\mathcal{V}$, eqn (\ref{EPTtrave}), 
can be rewritten as
\begin{equation}
\begin{array}{ll}
\mathcal{V}\left(\eta,\theta,\ell_{in}^0\right) =  &
\displaystyle 
\int_0^{\ell_{in}^{0-}} \Psi(\eta,\theta')\mbox{d}x_1^0 
+\int_{\ell_{in}^{0+}}^{\ell_0} \Psi(\eta,\theta')\mbox{d}x_1^0 
\\ [5 mm]
&
\displaystyle
-H^l\int_0^{\ell_{in}^{0-}}(1+\eta)\mbox{d}x_1^0
+H^r\int_{\ell_{in}^{0+}}^{\ell_0}(1+\eta)\cos\theta\mbox{d}x_1^0
+P\int_{\ell_{in}^{0+}}^{\ell_0}(1+\eta)\sin\theta\mbox{d}x_1^0.
\end{array}
\end{equation}

A  variation in the configuration defined by the fields $\eta(x_1^0)$ and $\theta(x_1^0)$  and the configurational parameter $\ell_{in}^0$ is considered, through  the small positive parameter $\varepsilon$, as
\begin{equation}
\eta\left(x_1^0\right)\rightarrow \eta\left(x_1^0\right)+\varepsilon \,\delta \eta\left(x_1^0\right),
\qquad
\theta\left(x_1^0\right)\rightarrow  \theta\left(x_1^0\right)+\varepsilon\, \delta \theta\left(x_1^0\right),
\qquad
\ell_{in}^0   \rightarrow  \ell_{in}^0 +\varepsilon\, \delta \ell_{in}^0,
\end{equation}
where, from the sliding sleeve constraint (\ref{slidingsleeveconstraint}), the perturbations $\delta \ell_{in}^0$ and $\delta \theta\left(x_1^0\right)$ satisfy the following compatibility equation
\begin{equation}\label{compatibilityslidingsleeve}
    \delta \theta\left(\ell_{in}^0\right)=-\theta'\left(\ell_{in}^0\right)\delta \ell_{in}^0.
\end{equation}
Keeping into account that the axial  force $N$ and the bending moment $M$ are work-conjugate to the axial deformation $\eta$ and the generalized curvature $\theta'$, the following constitutive equations can be assumed 
\begin{equation}\label{constitutivetrave}
    N=\dfrac{\partial \Psi}{\partial \eta},\qquad
     M=\dfrac{\partial \Psi}{\partial \theta'},
\end{equation}
so that, integration by parts,  the sliding sleeve constraint conditions (\ref{slidingsleeveconstraint}), and the compatibility condition (\ref{compatibilityslidingsleeve}), allow to evaluate the first variation $\delta\mathcal{V}$ of the total potential energy as
\begin{equation}
\begin{array}{lll}
\delta\mathcal{V}\left(\eta,\theta,\ell_{in}^0,\delta \eta, \delta \theta, \delta \ell_{in}^0 \right)=
     & \displaystyle \int_0^{\ell_{in}^{0-}} N\delta\eta\mbox{d}x_1^0+\int_{\ell_{in}^{0+}}^{\ell_0} N\delta\eta\mbox{d}x_1^0-\int_{\ell_{in}^{0+}}^{\ell_0} M'\delta\theta\mbox{d}x_1^0-
        H^l\int_0^{\ell_{in}^{0-}}\delta\eta\mbox{d}x_1^0+H^r\int_{\ell_{in}^{0+}}^{\ell_0}\delta\eta\cos\theta\mbox{d}x_1^0
        \\[5mm]
     &\displaystyle -H^r\int_{\ell_{in}^{0+}}^{\ell_0}(1+\eta)\sin\theta\delta\theta\mbox{d}x_1^0
           -P\int_{\ell_{in}^{0+}}^{\ell_0}\delta\eta\sin\theta\mbox{d}x_1^0
        -P\int_{\ell_{in}^{0+}}^{\ell_0}(1+\eta)\cos\theta \delta\theta\mbox{d}x_1^0
       \\[5mm]
     &\displaystyle
       +\left\{-H^r\left[1+\eta\left(\ell_{in}^{0+}\right)\right]
        -  H^l\left[1+\eta\left(\ell_{in}^{0-}\right)\right]
                +M\left(\ell_{in}^{0+}\right)\theta'\left(\ell_{in}^{0+}\right)-\salto{.1}{
        \Psi(\ell_{in}^0)}\right\}
        \delta \ell_{in}^0.
\end{array}
\end{equation}
The annihilation of the first variation $\delta\mathcal{V}$ for every compatible perturbations $\delta \eta\left(x_1^0\right)$, $\delta \theta\left(x_1^0\right)$, and $\delta \ell_{in}^0$ provides the following equilibrium equations for the portion of the rod respectively  outside 
\begin{equation}\label{eqtravefuori}
   \left\{\begin{array}{lll}
        M'\left(x_1^0\right)+H^r\left[1+\eta\left(x_1^0\right)\right]\sin\theta\left(x_1^0\right)  +P\left[1+\eta\left(x_1^0\right)\right]\cos\theta\left(x_1^0\right)=0 , \\[3mm]
        N\left(x_1^0\right)=P \sin\theta\left(x_1^0\right)  - H^r \cos\theta\left(x_1^0\right),  
   \end{array} \right.\qquad x_1^0\in\left(\ell_{in}^0,\ell_0\right],
\end{equation}
and inside 
\begin{equation}\label{eqtravedentro}
N\left(x_1^0\right)=H^l,\qquad x_1^0\in\left[0,\ell_{in}^0\right),
\end{equation}
the sliding sleeve, together with the interface condition at the sliding sleeve end 
\begin{equation}\label{interfaceequilibriumslidingsleeve}
H^r\left[1+\eta\left(\ell_{in}^{0+}\right)\right]
        +  H^l\left[1+\eta\left(\ell_{in}^{0-}\right)\right]
                -M\left(\ell_{in}^{0+}\right)\theta'\left(\ell_{in}^{0+}\right)+\salto{.1}{
        \Psi(\ell_{in}^0)}=0.
\end{equation}
From eqns (\ref{eqtravefuori})$_2$ and (\ref{eqtravedentro}) it follows that the axial force $N(x_1^0)$ at the left and right limit points of the sliding sleeve exit are given by
\begin{equation}\label{Nsliding}
N\left(\ell_{in}^{0-}\right)=H^l,
\qquad
N\left(\ell_{in}^{0+}\right)=-H^r,
\end{equation}
so that the  jump in the normal force at the sliding sleeve end is provided by the reaction force component $R_1$, which  remains determined from equilibrium as
\begin{equation}\label{R1sliding}
R_1=H^l+ H^r.
\end{equation}
On account of equations (\ref{Nsliding}) and (\ref{R1sliding}), a comparison between the axial equilibrium at the sliding sleeve exit, eqn  (\ref{interfaceequilibriumslidingsleeve}), and  expression (\ref{trave2}), derived for the configurational force component $F_1^c$, implies the validity of equation (\ref{fcmenor1}),  obtained with reference to the frictionless contact conditions.

\subsection{Application to rods characterized by a quadratic elastic energy density} 

Assuming the usual quadratic expression for the  elastic energy density of the rod, 
\begin{equation}\label{ElEnQuadratic}
    \Psi\left(x_1^0\right)=\dfrac{B\left[\theta'\left(x_1^0\right)\right]^2}{2}+\dfrac{K \left[ \eta\left(x_1^0\right)\right] ^2}{2},
\end{equation}
with $B$ and $K$ representing the (constant and positive) bending and  axial stiffnesses, respectively, the application of equation (\ref{constitutivetrave}) provides the  constitutive relation 
\begin{equation}\label{constlinearextbeam}
    N\left(x_1^0\right)=K\, \eta\left(x_1^0\right),\qquad
     M\left(x_1^0\right)=B\, \theta'\left(x_1^0\right),
\end{equation}
so that the configurational force component $F_1^c$, eqn (\ref{trave2}), reduces to
\begin{equation}\label{trave20}
F_1^c = - \,\dfrac{B\left[\theta'\left(\ell_{in}^{0+}\right)\right]^2}{2}- \dfrac{K\salto{.1}{\eta\left(\ell_{in}^{0}\right)^2}}{2}. 
\end{equation}
Consequently the concentrated reaction force $R_1$ becomes
\begin{equation}\label{trave20r1}
R_1 =  \,\dfrac{B\left[\theta'\left(\ell_{in}^{0+}\right)\right]^2}{2}+ \dfrac{K\salto{.1}{\eta\left(\ell_{in}^{0}\right)^2}}{2}.
\end{equation}

As a consequence of the assumed  linear elastic axial behavior, eqn (\ref{constlinearextbeam}), the jump in the axial deformation $\eta$ at  the sliding sleeve exit satisfies the following identity
\begin{equation}
   K\left[\eta\left(\ell_{in}^{0+}\right)- \eta\left(\ell_{in}^{0-}\right)\right]=-R_1. 
\end{equation}
Introducing the average axial deformation $\langle \eta\left(\ell_{in}^{0}\right) \rangle$ at the sliding sleeve exit 
\begin{equation}
   \langle \eta\left(\ell_{in}^{0}\right) \rangle=\dfrac{\eta\left(\ell_{in}^{0+}\right)+\eta\left(\ell_{in}^{0-}\right)}{2},
\end{equation}
the concentrated reaction $R_1$, eqn (\ref{trave20r1}), can finally be obtained as
\begin{equation}\label{trave2024}
R_1 =  \,\dfrac{B\left[\theta'\left(\ell_{in}^{0+}\right)\right]^2}{2\left[
1+\langle \eta\left(\ell_{in}^{0}\right) \rangle
\right]},
\end{equation}
which approaches,   in the limit of vanishing axial deformation $\eta\left(\ell_{in}^0\right)$ at the end of the sliding sleeve, the value for  inextensible rods
\begin{equation}
    \lim_{\eta\left(\ell_{in}^0\right)\rightarrow0}R_1=\dfrac{B\left[\theta'\left(\ell_{in}^{0+}\right)\right]^2}{2},
\end{equation}
obtained in \cite{bigoni2015eshelby}.

\section{Reaction force $R_1$, Eulerian buckling, and dynamic ejection from a frictionless, rigid, flat punch partially compressing an elastic  rectangular block}

The plane strain problem of  symmetric (frictionless, flat, and rigid) punches 
indenting an elastic solid of rectangular shape 
is further addressed. The purpose is  to provide insight into the significance of the framework 
introduced in the previous Sections 
in terms of both reliability of the obtained expressions and applicability of the results to the design of novel soft mechanisms. 

In particular, the high reliability of the expression (\ref{reactionR1smoothENERGY}) for the contact reaction component $R_1$ (obtained under the approximate  assumption of uniform state at both the lateral edges of the rectangular domain) is assessed, through a comparison with results from Finite Element (FE) analyses at variable   geometric parameters. Moreover, actuation mechanisms from Eulerian buckling and longitudinal dynamic  ejection of the elastic solid due to  transverse compression  are presented, highlighting effects related to the presence of the reaction force $R_1$.

\paragraph{Constitutive hyperelastic material models.}
A specific hyperelastic material response is defined through the introduction of a specific strain energy density $\Phi$ as a function of the principal stretches $\lambda_i$ ($i$=I, II, III, so that $\mathcal{J}=\lambda_{\footnotesize{\mbox{I}}}\lambda_{\footnotesize{\mbox{II}}}\lambda_{\footnotesize{\mbox{III}}}$). 
It is assumed that plane strain prevails, so that the out-of-plane principal  stretch assumes a unit value,  $\lambda_{\footnotesize{\mbox{III}}}=\lambda_3=1$. 
The following two material models are analyzed.
\begin{itemize} 
\item  A compressible and initially isotropic material, 
obtained retaining only the first term ($N=1$) in the summation characterizing the strain energy density of 
the Stor{\aa}kers model \cite{storaakers1986material},
\begin{equation}
\mbox{\lq First-term' Stor{\aa}kers model:}\qquad    
\Phi (\lambda_{\footnotesize{\mbox{I}}},\lambda_{\footnotesize{\mbox{II}}})= \frac{2 \mu}{\alpha^2}\left[ \lambda_{\footnotesize{\mbox{I}}}^{\alpha}+\lambda_{\footnotesize{\mbox{II}}}^{\alpha} - 2 + \frac{1}{\beta} \left( \mathcal{J}^{-\alpha\beta} -1 \right) \right] ,
\label{compdens}   
\end{equation}
where  $\mu>0$ is the ground-state shear modulus, $\alpha\neq0$ is a parameter affecting the nonlinear response, and $\beta>-1/3$ is another parameter, related to the value of the ground-state Poisson's ratio $\nu\in(-1,1/2)$ as 
\begin{equation}
    \beta = \frac{\nu}{1-2\nu}.
\label{betapoisson}   
\end{equation}
The principal  components of the first Piola-Kirchhoff  stress tensor $\mathbf{S}$  can be obtained from eqn (\ref{const}) as
\begin{equation}\label{conststor}
    S_i=2 \mu\frac{  \lambda_i^{\alpha }-(\lambda_i \lambda_j)^{-\alpha  \beta }}{\alpha  \lambda_i},\qquad i\neq j,\qquad i,j=\mbox{I},\mbox{II}.
\end{equation}
By assuming $\alpha=2$, the strain energy density $\Phi$ (\ref{compdens}) reduces to that recently proposed by Pence and Gou \cite{pence2015compressible} as a compressible version of the neo-Hookean material model, \begin{equation}
\mbox{Pence and Gou model:}\qquad        \Phi(\lambda_{\footnotesize{\mbox{I}}},\lambda_{\footnotesize{\mbox{II}}}) = \frac{\mu}{2}\left[\lambda_{\footnotesize{\mbox{I}}}^{2}+\lambda_{\footnotesize{\mbox{II}}}^{2}-2+\frac{1}{\beta}\left(\mathcal{J}^{-2\beta}-1\right)\right].
\label{NHC}   
\end{equation}
\item The incompressible and isotropic neo-Hookean material model \cite{bigoni2012nonlinear}:
\begin{equation}
\mbox{neo-Hookean model:}\qquad
\Phi (\lambda_{\footnotesize{\mbox{I}}},\lambda_{\footnotesize{\mbox{II}}})= \frac{\mu}{2}\left(\lambda_{\footnotesize{\mbox{I}}}^2+\lambda_{\footnotesize{\mbox{II}}}^2-2\right), 
~~~~ \lambda_{\footnotesize{\mbox{I}}}\lambda_{\footnotesize{\mbox{II}}}=1,
\label{mrmat}   
\end{equation}

where $\mu>0$  is the  ground-state shear modulus.
In the case of incompressible materials, the constitutive relation, eqn (\ref{const}), becomes 
\begin{equation}
\label{constINCOMP}
   S_{i}=-\dfrac{\Pi}{\lambda_i}+\dfrac{\partial\Phi}{\partial \lambda_i},
\end{equation}
where $\Pi$  is the Lagrangian multiplier associated to the  incompressibility constraint. Equation (\ref{constINCOMP}) leads to the principal components
\begin{equation}
    S_{i}=-\dfrac{\Pi}{\lambda_i}+\mu\lambda_{i},\qquad
    \qquad  i=\mbox{I},\mbox{II}.
\end{equation}
\end{itemize}

It is noted that, assuming  $\alpha=2$, the compressible model (\ref{NHC}) approaches the incompressible one (\ref{mrmat}) in the limit $\beta\rightarrow\infty$ (corresponding to $\nu\rightarrow1/2$).

\paragraph{Common details of  FE simulations.} 
FE simulations are performed through the  commercial  code Abaqus 2023.
The rectangular elastic domain is modeled in both of the two compressible and  incompressible  versions,  eqs (\ref{NHC}) and (\ref{mrmat}). 
The elastic domain is meshed with bi-quadratic plane strain elements (CPE8) when the material is compressible, otherwise, a hybrid formulation is used (CPE8H). 
The boundary conditions prescribed  on the sides of the rectangle are:  (i.) free from tractions and  
constrained frictionless  
(ii.) bilateral (visualized with rollers) or (iii.) unilateral contact 
with a flat undeformable surface. 
In the numerical simulations the corner of the rigid constraint is smoothed 
with a quarter-of-circle arc of radius $r$.

 Where the rigid surface is present, it is meshed with a linear rigid link (R2D2). To prevent interpenetration, the mesh size of the elastic body (the slave, defined on nodes) is chosen to be finer than the mesh size of the rigid constraint (the master, defined on segments). 
The adopted solver implements nonlinear geometry, enhanced through the introduction of a moderate energy dissipation when dynamic conditions prevail, to overcome ill-posedness at contact.

\subsection{The horizontal reaction force $R_1$ at the corner of a frictionless flat punch}\label{R1sezione}

Two boundary value problems are considered for an elastic material occupying a rectangular domain in its  undeformed configuration, $\mathcal{B}_0$, eqn (\ref{rectangulardomain}). The   sides of the domain have lengths $h_0$ and $\ell_0$, Fig. \ref{presssss} (left). 
In both problems the lower boundary $\partial\mathcal{B}_0^b$ is entirely constrained by a horizontal bilateral frictionless constraint (a condition indicated with applied rollers) along a flat surface with unit normal $\mathbf{e}_2$. 
 The upper boundary of the elastic solid, $\partial\mathcal{B}_0^a$,  is partially constrained on its left part by a (frictionless, flat, and rigid) punch, defined by the outward unit normal $-\mathbf{e}_2$. The  punch pushes the elastic solid remaining aligned parallel to $\mathbf{e}_2$, until a thickness  $h=\overline{\lambda}_2 h_0$ is reached, corresponding to a nominal transverse stretch  $\overline{\lambda}_2<1$.

The two analyzed boundary value problems, called {\tt BVP1} and {\tt BVP2}, differ in the boundary conditions provided on the lateral sides $\partial\mathcal{B}^l$ (with normal $\mathbf{n}_0^l=-\mathbf{e}_1$) and $\partial\mathcal{B}^r$ (with normal $\mathbf{n}_0^r=\mathbf{e}_1$) as follows. 
\begin{enumerate}[]
    \item{- For \tt BVP1}: the boundary $\partial\mathcal{B}_0^l$ is loaded through the application of a normal dead traction, $\mathbf{S}^l\mathbf{n}_0^l=-S_{11}^l\mathbf{e}_1$, while the boundary $\partial\mathcal{B}_0^r$ is left   traction-free, $\mathbf{S}^r\mathbf{n}_0^r=\mathbf{0}$. Equilibrium imposes the reaction force $R_1$ to be the negative of the resultant of the applied tractions 
    \begin{equation}\label{R1bvp1eq}
        R_1=S_{11}^l h_0. 
    \end{equation}
    
    \item{- For \tt BVP2:}  the boundary $\partial\mathcal{B}_0^l$ is left traction-free, $\mathbf{S}^l\mathbf{n}_0^l=\mathbf{0}$, while the boundary $\partial\mathcal{B}_0^r$ is loaded through a normal dead traction, $\mathbf{S}^r\mathbf{n}_0^r=S_{11}^r\mathbf{e}_1$. Equilibrium imposes the reaction force $R_1$ to be the negative of the resultant of the applied tractions 
    \begin{equation}
    \label{R1bvp2eq}
        R_1=-S_{11}^r h_0. 
    \end{equation}
\end{enumerate}
The end point of the flat punch  (corresponding to either a sharp corner or the initial point of a  rounded corner) is located at the back-transformed point $\mathbf{y}_0=\left(\ell_{in}^0,h_0/2\right)$, belonging to $\mathcal{B}_0^a$, at  a distance $\ell_{in}^0\in[0,\ell_0]$. 

The assumptions  
\begin{equation}
\ell_0/h_0>2 ~~\mbox{ and }~~
    h_0<   \ell_{in}^0 <\ell_0-h_0,
\end{equation}
allow to neglect 
the perturbation introduced by the end (sharp or smooth) of the punch 
on the two lateral 
boundaries $\partial\mathcal{B}_0^l$ and $\partial\mathcal{B}_0^r$, where the deformed state is approximated as uniform and therefore independent of $x_2^0$. 
Under this approximation, eqn (\ref{stock79}) can be used to obtain the horizontal force $R_1$ in the two following cases. 
\begin{enumerate}[]
    \item{- For \tt BVP1}: the part of the boundary $\partial\mathcal{B}_0^a$ outside the constraint and $\partial\mathcal{B}_0^r$ are traction-free; moreover, $\partial\mathcal{B}_0^r$ is unloaded,  $\lambda_1^r=\lambda_2^r=1$ and $\Phi^r=0$. Thus, the reaction force $R_1$,  eqn (\ref{reactionR1smoothENERGY}), becomes 
    \begin{equation}\label{R1bvp1conf}
R_1=\dfrac{\Phi^l\left(\lambda_1^l,\lambda_2^l=\overline{\lambda}_2\right)}{\lambda_1^l}h_0,
    \end{equation}
    an equation showing that the  force is always positive.
    The two unknowns $R_1$ and $\lambda_1^l$ (the latter enforced to be coincident to $1\left/\overline{\lambda}_2\right.$ in the incompressible case) can be evaluated by solving eqns (\ref{R1bvp1eq})   and (\ref{R1bvp1conf}), together with the constitutive equation (\ref{conststor}) [or  eqn (\ref{constINCOMP}) in the incompressible case]. The result comes in a closed form for the incompressible case as 
        \begin{equation}\label{R1bvp1incom}
R_1=\displaystyle\dfrac{\displaystyle  \left(\overline{\lambda}_2^{\,2}-1\right)^2}
    {\displaystyle\overline{\lambda}_2}
    \dfrac{\mu\, h_0}{2}.
   \end{equation}
   
Although not expressible in a     
closed form, in the compressible case,
for a small strain $\overline{\varepsilon}_2<0$  defining $\overline{\lambda}_2=1+\overline{\varepsilon}_2$,  the following series expansions,  truncated at the fourth-order can be obtained
    \begin{equation}
    \label{R1bvp1com}
    \begin{array}{lll}
    R_1=\left[1-\dfrac{(3-\alpha) (1-2 \nu) }{3 (1-\nu)}\overline{\varepsilon}_2
    +\dfrac{(\alpha  (3 \alpha -20)+35) \nu ^2+\alpha  (22-3 \alpha ) \nu +(\alpha -6) \alpha -34 \nu +8}{12 (1-\nu)^2}\overline{\varepsilon}_2^{\,2} \right]\dfrac{ \mu h_0   }{1-\nu }\overline{\varepsilon}_2^{\,2}+\mbox{o}\left(\overline{\varepsilon}_2^{\,4}\right),
    \\[4mm]
    \begin{array}{rl}\lambda_1^l=& 1+\left[-1+\dfrac{\overline{\varepsilon}_2}{2 \nu }-\dfrac{ 3+\alpha  \left(2 \nu ^2+\nu -1\right)-\nu  (4+\nu)}{6 (1-\nu)^2 \nu }\overline{\varepsilon}_2^{\,2}\right.\\
    [4mm]
    &\left.+\dfrac{17\alpha ^2 \left(2 \nu ^3+\nu ^2+\nu -1\right)-\alpha  (\nu  (4 \nu  (\nu +2)-23)+9)+4 \nu  (\nu  (\nu +6)-11)}{24  \nu (1-\nu )^3}\overline{\varepsilon}_2^{\,3} \right]\dfrac{\nu  \overline{\varepsilon}_2 }{1-\nu}+\mbox{o}\left(\overline{\varepsilon}_2^{\,4}\right);
    \end{array}
\end{array}
\end{equation}
    \item{- For \tt BVP2:} 
    the reaction force $R_1$ is given by eqn (\ref{reactionR1smoothENERGY}) as
    \begin{equation}\label{R1bvp2conf}
R_1=\dfrac{\Phi^l\left(\lambda_1^l,\lambda_2^l=\overline{\lambda}_2\right)-\Phi^r\left(\lambda_1^r,\lambda_2^r\right)}{\lambda_1^r}h_0.
    \end{equation}
    
    The four unknowns $R_1$, $\lambda_2^r$, $\lambda_1^l$ and $\lambda_1^r$ (the last two enforced to be coincident to 1$\left/\overline{\lambda}_2\right.$ and to 1/$\lambda_2^r$ in the incompressible case)  can be evaluated by solving eqns (\ref{R1bvp2eq})   and (\ref{R1bvp2conf}), together with the constitutive equation (\ref{conststor}) 
[or eqn (\ref{constINCOMP}) in the incompressible case]. 
Similarly to {\tt BVP1}, the result is provided  in a closed form for the incompressible case as
\begin{equation}
\label{R1bvp2incom}
R_1=\displaystyle\dfrac{\displaystyle  1-10 \overline{\lambda}_2^{\,4}+\overline{\lambda}_2^{\,8}+\left(1+\overline{\lambda}_2^{\,4}\right)\sqrt{1+14 \overline{\lambda}_2^{\,4}+\overline{\lambda}_2^{\,8}} }
    {\displaystyle\overline{\lambda}_2^{\,3} \sqrt{1+\overline{\lambda}_2^{\,4}+\sqrt{1+14 \overline{\lambda}_2^{\,4}+\overline{\lambda}_2^{\,8}}}}
    \dfrac{\mu\,h_0}{ 3 \sqrt{6} },\qquad
    \lambda_2^r=\dfrac{\displaystyle\sqrt{1+\overline{\lambda}_2^{\,4}+\sqrt{1+14 \overline{\lambda}_2^{\,4}+\overline{\lambda}_2^{\,8}}}}{\displaystyle\sqrt{6} \overline{\lambda}_2},
\end{equation}
   and as the following expansions in $\overline{\varepsilon}_2$ truncated at the fourth-order for the compressible case
   \begin{equation}
   \label{R1bvp2com}
   \begin{array}{lll}
        R_1=\left[1+\dfrac{\alpha  (1-2 \nu )-3(1- \nu)}{3 (1-\nu)}\overline{\varepsilon}_2 +\dfrac{\alpha ^2 (3 (\nu -1) \nu +1)-6 \alpha  (\nu -1) (2 \nu -1)+14 (\nu -1)^2}{12 (1-\nu)^2}\overline{\varepsilon}_2^{\,2} \right]\dfrac{\mu  h_0}{1-\nu }\overline{\varepsilon}_2^{\,2} +\mbox{o}\left(\overline{\varepsilon}_2^{\,4}\right),\\[4mm]
    \lambda_2^r=1+\left[1+\dfrac{ \alpha  (1-2 \nu)-3(1-\nu)}{3 (1-\nu)}\overline{\varepsilon}_2+\dfrac{(3 (\alpha -2) \alpha +5) \nu ^2-3 (\alpha -3) \alpha  \nu +(\alpha -3) \alpha -13 \nu +8}{12 (1-\nu)^2}\overline{\varepsilon}_2^{\,2} \right]\dfrac{\nu   }{2 (1-\nu)}\overline{\varepsilon}_2^{\,2}+\mbox{o}\left(\overline{\varepsilon}_2^{\,4}\right),
    \\[4mm]
        \lambda_1^l=1-\left[1-\dfrac{1}{2 (1-\nu)}\overline{\varepsilon}_2+\dfrac{(2-\nu) }{6 (1-\nu)^2}\overline{\varepsilon}_2^{\,2}-\dfrac{(2-\nu) (3-2 \nu ) }{24 (1-\nu )^3}\overline{\varepsilon}_2^{\,3}\right]\dfrac{\nu }{1-\nu }\overline{\varepsilon}_2 +\mbox{o}\left(\overline{\varepsilon}_2^{\,4}\right),
    \\[4mm]
    \lambda_1^r=1-\left[1+\dfrac{ \alpha  (1-2 \nu)-3(1- \nu)}{3 (1-\nu)}\overline{\varepsilon}_2+\dfrac{(3 (\alpha -2) \alpha +5) \nu ^2-3 (\alpha -3) \alpha  \nu +(\alpha -3) \alpha -10 \nu +5}{12 (1-\nu)^2}\overline{\varepsilon}_2^{\,2} \right]\dfrac{\overline{\varepsilon}_2^{\,2}}{2}  +\mbox{o}\left(\overline{\varepsilon}_2^{\,4}\right).
\end{array}
\end{equation}
\end{enumerate}

It is noted that the reaction force  $R_1$ and the stretches solving   {\tt BVP1} and {\tt BVP2} are both characterized by the same structure of their expansions,
\begin{equation}
\left\{
\begin{array}{cc}
       R_1\\
       \lambda_1^r -1\\
       \lambda_2^r-1
       \end{array}\right\}
       =\cdots  \overline{\varepsilon}_2^{\,2} +\cdots  \,\overline{\varepsilon}_2^{\,3}+\cdots \, \overline{\varepsilon}_2^{\,4},\qquad
       \lambda_1^l-1=\cdots\,  \overline{\varepsilon}_2+\cdots \, \overline{\varepsilon}_2^{\,2}+\cdots \, \overline{\varepsilon}_2^{\,3}+\cdots  \,\overline{\varepsilon}_2^{\,4}.
\end{equation}

The value of the reaction force component $R_1$, obtained as the numerical solution of the system of nonlinear equations 
eqns (\ref{R1bvp1eq}), (\ref{R1bvp1conf}), and (\ref{conststor}) [or  eqn (\ref{constINCOMP}) in the incompressible case] 
for {\tt BVP1} (continuous lines) and eqns (\ref{R1bvp2eq})    (\ref{R1bvp2conf}), and  (\ref{conststor}) 
[or eqn (\ref{constINCOMP}) in the incompressible case] 
for {\tt BVP2} (dashed lines), is reported  as a function of the imposed stretch $\overline{\lambda}_2$ 
in Fig. \ref{R1_solnonlin} on the left 
(for different values of the ground-state Poisson's ratio $\nu$) and as a function of the  ground-state Poisson's ratio $\nu$ (for different values of the imposed nominal stretch  $\overline{\lambda}_2$)  on the right. Note that both neo-Hookean and Pence and Gou models are reported in the figure on the left, while the former model corresponds to the limit of $\nu=0.5$ on the right. 

\begin{figure}[!htb]
\centering
\includegraphics[width=1\textwidth]{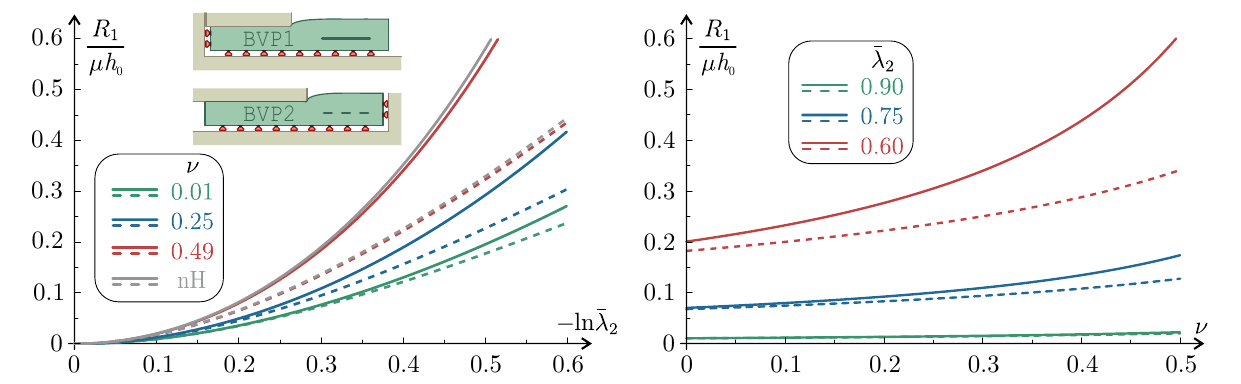}
\caption{Normalized reaction force  $R_1/(\mu h_0)$  as a function of the imposed nominal stretch $\overline{\lambda}_2$ (left, for the neo-Hookean material \lq nH' and for the Pence-Gou material) and of  the  ground-state Poisson's ratio $\nu$ (right) (for the Pence-Gou material). The continuous and dashed lines respectively correspond to the solution of {\tt BVP1} and {\tt BVP2}, eqns (\ref{R1bvp1incom}) and (\ref{R1bvp2incom}) for the neo-Hookean material and the numerical solution of  
eqns (\ref{R1bvp1eq}), (\ref{R1bvp1conf}), and (\ref{conststor}), and eqns (\ref{R1bvp2eq})    (\ref{R1bvp2conf}), and  (\ref{conststor}) for the Pence-Gou material. 
}
\label{R1_solnonlin}
\end{figure}

 \paragraph{Reliability of the uniform state assumption at the boundaries $\partial\mathcal{B}_0^l$ and $\partial\mathcal{B}_0^r$.}
The curves describing the reaction force $R_1$ in Fig. \ref{R1_solnonlin} are obtained by considering the uniformity of the stretches, and therefore of the strain energy, along each of the two lateral boundaries $\partial\mathcal{B}_0^l$ and $\partial\mathcal{B}_0^r$. The reliability of this assumption is assessed by comparing $R_1$ with the corresponding value $R_1^{\text{FE}}$ numerically evaluated through  finite element simulations, where both the neo-Hookean and the  \lq first-term' Stor{\aa}kers models are available. 

The map of relative difference $\left(R_1^{\text{FE}}-R_1\right)/R_1^{\text{FE}}$ is reported for a neo-Hookean material  with $\ell_0/h_0=20$, under an imposed nominal stretch $\overline{\lambda}_2=0.7$ in Fig. \ref{errorMAP} (lower part, on the left) for {\tt BVP1}, for different   ratios $\ell_{in}/\ell_0\in[0,1]$ and $r/h_0\in[0.1,0.5]$, 
being $r$ the radius  of  the rounded corner of the constraint (Fig. \ref{errorMAP}, upper part, right).
\begin{figure}[!h]
\centering
\includegraphics[width=1\textwidth]{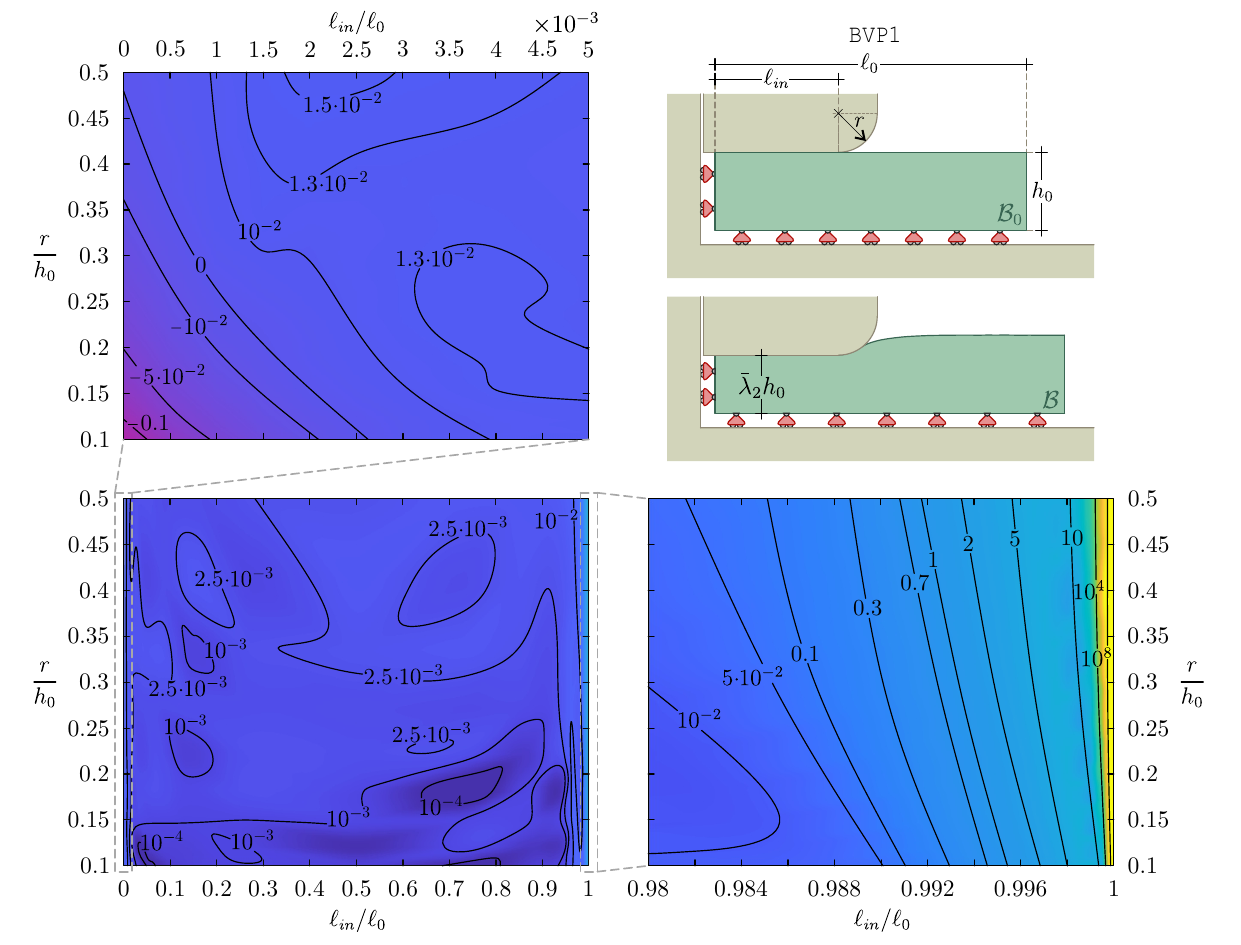}
\caption{(Lower part, on the left) Map of relative difference $\left(R_1-R_1^{\text{FE}}\right)/R_1^{\text{FE}}$, for different ratios $\ell_{in}/\ell_0$ and $r/h_0$ for a neo-Hookean elastic material with $\ell_0/h_0=20$ at an imposed nominal stretch $\overline{\lambda}_2=0.7$ for {\tt BVP1} (Upper part, on the  right). Magnified map for  small (upper part, on the left) and  large (lower part, on the  right) values of $\ell_{in}/\ell_0$, respectively  showing the possibility of negative values and very large positive values for the relative difference.}
\label{errorMAP}
\end{figure}
The map shows that the relative difference    $\left(R_1^{\text{FE}}-R_1\right)/R_1^{\text{FE}}$ is confined to positive and very low values (less than 3\scalp 10$^{-2}$) for $\ell_{in}/\ell_0\in[0.005,0.98]$ (Fig. \ref{errorMAP}, bottom left), a minimum negative value (approximately -0.12) is attained for $\{\ell_{in}/\ell_0, r/h_0\}=\{0,0.1\}$ (Fig. \ref{errorMAP}, above left), while a steep gradient arises and large (infinite in the limit case) values occur for $\ell_{in}/\ell_0\approx1$ since $R_1^{\text{FE}}$ vanishes when $\ell_{in}/\ell_0=1$ (Fig. \ref{errorMAP}, bottom right). Therefore, the map confirms  the reliability of evaluating the reaction force $R_1$ through eqn (\ref{reactionR1smoothENERGY}), based on the approximation of homogeneity for the strain at each lateral sides of the elastic solid, except when one of the two sides  $\partial\mathcal{B}^l$ or $\partial\mathcal{B}^r$ is located very close to the punch corner. Note that the results from FE are not included in Fig. \ref{R1_solnonlin}, because they are simply superimposed to the curves.

\subsection{Eulerian buckling induced by transverse compression}\label{bucklingbyconf}

The rectangular elastic body $\mathcal{B}_0$, eqn (\ref{rectangulardomain}), is  partially subject to a symmetric transverse compression at its edges, corresponding to a nominal transverse stretch $\overline{\lambda}_{2}<1$, imposed  by two parallel pairs of mirrored punches, spaced at a fixed distance $d$ from each other, Fig. \ref{criticalstretchevolution} (right). 
Horizontal reaction forces are  acting at the corners of each of the punches, $\left\{R_1^{al},-R_1^{ar},R_1^{bl},-R_1^{br}\right\}\mathbf{e}_1$. 
Assuming a symmetric response with respect to the vertical direction  at $x_1^0=\ell_0/2$, a symmetry in the forces follows
\begin{equation}
R_1^{al}=R_1^{ar},\qquad
R_1^{bl}=R_1^{br},
\end{equation}
introducing a compressive state in the central part of the elastic body. 
The nominal transverse stretch $\overline{\lambda}_{2}$ can be  decreased starting from   the undeformed state ($\overline{\lambda}_{2}=1$), until a  critical value $\overline{\lambda}_2^{\,cr}$ is reached, for which the buckling of the elastic body occurs. 
This occurrence is always possible before a surface instability when the elastic body is sufficiently slender. 

The buckling condition is investigated through the reduced  model of the extensible elastica with varying domain, as sketched in the inset of Fig. \ref{criticalstretchevolution} (left) and described in Sect.   \ref{sezstrutture}.
Under the mentioned symmetry condition, the internal force component along $\mathbf{e}_2$ vanishes ($N_2(x_1^0)=0$) and, by further restricting the treatment  to a quadratic strain energy density, eqn (\ref{ElEnQuadratic}), for the rod, the equilibrium equations (\ref{eqtravefuori}) reduce to 
\begin{equation}\label{extelasticaode}
   \left\{
   \begin{array}{lll}
B\theta''\left(x_1^0\right)+R_1\left[1+\eta\left(x_1^0\right)\right]\sin\theta\left(x_1^0\right)=0,\\[3mm]
K \eta\left(x_1^0\right)=-R_1\cos \theta\left(x_1^0\right),
\end{array}
\right.
\qquad x_1^0\in\left(\ell_{in}^0,\ell_0-\ell_{in}^0\right),
\end{equation}
where $\ell_{in}^0$ defines the undeformed coordinate $x_1^0$ corresponding to the sliding sleeve exit, in the deformed configuration. The equilibrium equations (\ref{extelasticaode}) are complemented by  the following boundary conditions and isoperimetric constraints
\begin{equation}\label{bcextelasticbuck}
   \theta\left(\ell_{in}^0\right)=\theta\left(\ell_0-\ell_{in}^0\right)=0,\qquad 
   \int_{\ell_{in}^0}^{\ell_0-\ell_{in}^0} \left[1+\eta\left(x_1^0\right)\right]\cos\theta\left(x_1^0\right)\mbox{d}x_1^0=d,\qquad 
\int_{\ell_{in}^0}^{\ell_0-\ell_{in}^0} \left[1+\eta\left(x_1^0\right)\right]\sin\theta\left(x_1^0\right) \mbox{d}x_1^0=0. 
\end{equation}
In the equilibrium equations (\ref{extelasticaode}), $R_1$ is the resultant reaction force aligned parallel to $\mathbf{e}_1$ and  exerted by each pair of sliding sleeves on the elastic solid,  $R_1=R_1^{al}+R_1^{bl}$.

\begin{figure}[!h]
\centering
\includegraphics[width=.85\textwidth]{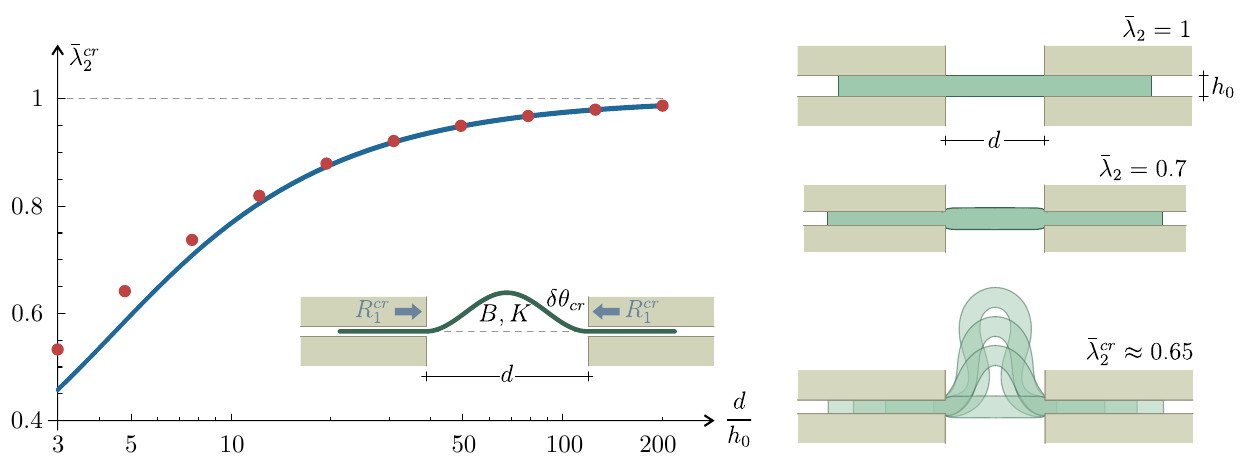}\caption{Eulerian buckling from transverse compression exerted by two pairs of symmetric (frictionless, flat, and rigid) punches. Left: Critical value of the nominal transverse stretch 
$\overline{\lambda}_2^{\,cr}$ for a Pence and Gou elastic material with $\nu=0.25$ as a function of the slenderness  parameter $d/h_0$. 
Semi-analytical prediction 
(blue curve) 
from the extensible elastica of varying-length 
(sketched in the inset)
vs FE results (red dots).
Right: Deformed shape evolution at decreasing  transverse stretch $\overline{\lambda}_2$ 
for an elastic rectangular solid with $d/h_0$=5. 
Unstable ejection occurs at the critical stretch   $\overline{\lambda}_2^{\,cr}
\approx 0.65
$. 
Symmetry with respect to the central vertical axis has been enforced. 
}
\label{criticalstretchevolution}
\end{figure}

The buckling condition can be investigated by analyzing small  perturbations $\delta\theta\left(x_1^0\right)$ and $\delta\eta\left(x_1^0\right)$ in the rotation and axial deformation fields around the trivial straight configuration defined by
\begin{equation}
    \theta_{eq}\left(x_1^0\right)=0,\qquad
    \eta_{eq}\left(x_1^0\right)=-\dfrac{R_1}{K},
    \qquad x_1^0\in\left(\ell_{in}^0,\ell_0-\ell_{in}^0\right),
\end{equation}
for which a linearization of the equilibrium equations (\ref{extelasticaode}) leads to
\begin{equation}
\left\{
\begin{array}{lll}
B\delta\theta''\left(x_1^0\right)+R_1\left[1-\dfrac{R_1}{K}\right]\delta\theta\left(x_1^0\right)=0,\\
\delta\eta\left(x_1^0\right)=0,
\end{array}
\right.
\qquad x_1^0\in\left(\ell_{in}^0,\ell_0-\ell_{in}^0\right),
\end{equation}
while a linearization of the  isoperimetric constraints to
\begin{equation}
     \left[1-\dfrac{R_1}{K}\right]\left(\ell_0-2\ell_{in}^0\right)=d,\qquad 
\int_{\ell_{in}^0}^{\ell_0-\ell_{in}^0} \delta\theta\left(x_1^0\right) \mbox{d}x_1^0=0. 
\end{equation}



A non-trivial equilibrium configuration can be found for the critical reaction force $R_1^{cr}$ and the corresponding critical rotation field $\delta\theta_{cr}\left(x_1^0\right)$, which can be evaluated as
\begin{equation}
\label{R1crext}
    R_1^{cr}=\dfrac{4\pi^2 B}{d^2+\dfrac{4\pi^2 B}{K}},\qquad
    \delta\theta_{cr}\left(x_1^0\right)=\overline{\theta}\cos\left[\dfrac{2\pi\left(x_1^0-\ell_{in}^0\right)}{\ell_0-2\ell_{in}^0}\right],
    \qquad x_1^0\in\left(\ell_{in}^0,\ell_0-\ell_{in}^0\right),
\end{equation}
where $\overline{\theta}$ represents a small amplitude, which remains arbitrary within the limits of a  linear perturbation analysis.

By assuming $B=E h_0^3/12$ and $K=E h_0$, with $E=2 \mu /(1-\nu)$ being the ground-state Young modulus under plane strain for the Pence and Gou model (\ref{NHC}), the critical reaction force $R_1^{cr}$ (\ref{R1crext})
reduces to
\begin{equation}\label{R1crnonlin}
    R_1^{cr}=\dfrac{\pi^2  h_0^2}{3 \,d^2+\pi^2 h_0^2} \dfrac{2 \mu h_0}{1-\nu}.
\end{equation}

The critical force $R_1^{cr}$ 
is reached for the straight configuration of the extensible elastica, $\theta(s)=0$, equivalent to  a \lq trivial' configuration of the
elastic body of rectangular shape, satisfying vertical  symmetry at $x_2^0$. The four tangential reactions at the punch corners have all the same values, 
\begin{equation}
R_1^{al}=R_1^{ar}=R_1^{bl}=R_1^{br},
\end{equation}
so that the elastic rectangular domain can be reduced to its quarter, loaded as for  {\tt BVP2}. 

The critical force $R_1^{cr}$ 
can be evaluated using eqn. (\ref{R1crnonlin}), but has to be expressed in terms of a critical 
transverse stretch $\overline{\lambda}_2^{\,cr}$, through the $J$--integral, which can in turn be approximated as the solution of nonlinear equation (\ref{R1bvp2conf}).
The corresponding  \lq semi-analytical' critical stretch $\overline{\lambda}_2^{\,cr}$ is reported as a continuous curve in Fig. \ref{criticalstretchevolution} (left)  for a Pence and Gou material with $\nu=0.25$, as a function of the slenderness parameter $d/h_0$.

As a complement to the above, the second-order expansion of the reaction force $R_1$ for {\tt BVP2}, eqn (\ref{R1bvp2com})$_1$,
leads to an approximation for the critical stretch $\overline{\lambda}_2^{\,cr}$  referred to a \lq first-term' Stor{\aa}kers material, obtained for large values of $d/h_0$ as
\begin{equation}
\begin{array}{lll}
    \overline{\lambda}_2^{\,cr}=& 1-\pi\sqrt{\dfrac{2}{3}}\,\dfrac{h_0}{d}
    +
   \pi^2  \dfrac{3 (1 - \nu) - \alpha (1 - 2 \nu)}{9 (1 - \nu)}\dfrac{h_0^2}{d^2}
   +
   \pi^3\dfrac{ \alpha ^2 [11 (1-\nu ) \nu -2]+12 \alpha  (1-\nu) (1-2 \nu)+15 (1-\nu)^2}{54 \sqrt{6} (1-\nu)^2}\dfrac{h_0^3}{d^3}\\[4mm]
   &\ds-\pi^4\dfrac{ [3 (1-\nu )-\alpha  (1-2 \nu )]\left\{\alpha ^2 [1+5\nu  (1-\nu ) ]-6 \alpha  (1-\nu ) (1-2 \nu )+72 (1-\nu )^2\right\}}{486  (1-\nu )^3}\dfrac{h_0^4}{d^4}
   +o(h_0^4/d^4),
   \end{array}
\end{equation}
which reduces to a mere  geometric relation when the approximation is truncated at first-order.

\paragraph{Critical stretch $\overline{\lambda}_2^{\,cr}$ for buckling  from FE  simulations.}
The  FE model  described in the previous Section is here adopted by considering that the punches end with a rounded corner, rounded with $r=d/50$.
Symmetry is imposed with respect to the vertical direction at  $x_1^0=\ell_0/2$. A symmetric imperfection is introduced in the initial undeformed geometry 
to trigger the bifurcation in the numerical analysis. In particular, instead of rectangular shape, the central portion of 
the elastic domain, $x_1^0\in\left[(\ell_0-d)/2,(\ell_0+d)/2\right]$,  has been implemented as a parallelogram with internal angles very close to $\pi/2$, namely, equal to $\pi/2\pm \pi/10^4$. 
Results are reported 
in Fig. \ref{criticalstretchevolution} (left)
for the Pence and Gou model,  characterized by $\nu=0.25$ (for which surface instability is estimated to occur for  $\overline{\lambda}_2^{\,si}\approx 0.473$), for a constant ratio $\ell_0/d=3$ and for different slenderness $d/h_0$. The critical transverse stretches  $\overline{\lambda}_2^{\,cr}$,  numerically evaluated through a Riks analysis, are reported as dots. The numerical results are in excellent agreement with the \lq semi-analytical' predictions obtained from the extensible elastica model. 
The evolution of the deformed shape is reported in Fig. \ref{criticalstretchevolution} (right) for 
a rectangular elastic domain of initial slenderness $d/h_0=5$, at three decreasing levels of transverse stretch, $\overline{\lambda}_2=\{1,0.7,0.65\}$. 
The smallest of the reported stretch corresponds to the critical value $\overline{\lambda}_2^{\,cr}$, for which  the system buckles following an unstable branch and therefore suffers a spontaneous and uncontrolled ejection from the compressing constraints. This is similar to the response of  other structural systems constrained by sliding sleeves  investigated in  \cite{BosiInjection} and \cite{elasticasling}.

\subsection{Dynamic longitudinal ejection of incompressible solids through transverse compression}\label{dynamicejesect}

Equilibrium has been so far  enforced, as a consequence of the constraint applied on $\partial\mathcal{B}^l$ in {\tt BVP1} or on $\partial\mathcal{B}^r$ in {\tt BVP2}. 
When such a constraint is removed after imposing $\overline{\lambda}_2<1$, the punch reaction $R_1$ is unbalanced. As a consequence, the elastic solid is pushed by the reaction component $R_1(t)$ away from the constraint, thus producing its complete ejection. The Newton’s second law can be expressed by \cite{toupin}
\begin{equation}\label{Newtonseconda}
    R_1(t)= \rho_0 \ell_0 h_0 \ddot{x}_1^c(t),
\end{equation}
where $\rho_0$ is the mass density of the elastic solid in the undeformed state, $x_1^c(t)$ is the  horizontal coordinate of the center of mass, and a superimposed dot represents the derivative with respect to time $t$.

While the extension of the $J$--integral measure to dynamics and the analysis of the influence of inertia 
are left to future investigation,  
 the ejection process of the elastic body is simply analyzed, assuming that the constraint on $\partial\mathcal{B}^l$ in {\tt BVP1} is instantaneously  removed, after transverse loading, $\overline{\lambda}_2<1$. 
 A neo-Hookean material is assumed and a simplified approach is developed, based on the following two assumptions.
\begin{enumerate}
\item  The reaction force $R_1(t)$ maintains the constant value $R_1$ evaluated during a quasi-static transverse compression, eqn (\ref{R1bvp1conf}), in which $\lambda_1^l=1\left/\overline{\lambda}_2\right.$.

\item The coordinate of the center of mass $x_1^c(t)$ of the deformed elastic body is approximated with the center of mass of an equivalent domain realized as the discontinuous union of two rectangular solids, one with the thickness of the space between the constraints 
(so that $\lambda_1=1/\lambda_2=1\left/\overline{\lambda}_2\right.$)
and the other with the height of the unloaded elastic solid
(so that $\lambda_1=\lambda_2=1$).

Under the above assumptions, the position of the center of mass is approximated as
\begin{equation}
\label{centromassaapproxdyn}
x_1^c\left(t\right)=\frac{\ell_0}{2}-\dfrac{\overline{\lambda}_2 \ell_{in}\left(t\right) \left[
2\ell_0-\left(1+\overline{\lambda}_2 \right)\ell_{in}\left(t\right)
\right]}{2\ell_0},
\end{equation}
\end{enumerate}
and 
 the Newton’s second law (\ref{Newtonseconda}) can be rewritten as a nonlinear second-order differential equation,  
\begin{equation}\label{nonlinearodedynamics}
\rho_0\left[\left(1-\overline{\lambda}_2\right)\ell_{in}\left(t\right)+ \ell_0 \right]\ddot{\ell}_{in}\left(t\right)
+\rho_0  \left(1-\overline{\lambda}_2\right)\dot{\ell}_{in}^2\left(t\right)+\Phi^l\left(\overline{\lambda}_2\right)
=0,
\end{equation}
where $\ell_{in}\left(t\right)$ is the portion of elastic solid inside the constraint at time $t$.

Under the initial conditions
\begin{equation}
 \ell_{in}\left(0\right)=\overline{\ell}_{in},\qquad
  \dot{\ell}_{in}\left(0\right)=0,
\end{equation}
the evolution in time of $\ell_{in}\left(t\right)$ is obtained as the solution of the nonlinear ordinary differential equation (\ref{nonlinearodedynamics})  as
\begin{equation}
\ell_{in}\left(t\right)=\sqrt{\left(
\frac{
\ell_0}{
1-\overline{\lambda}_2}
+ \,\overline{\ell}_{in}\right)^2-
\frac{
\Phi^l\left(\overline{\lambda}_2\right)\, t^2}{\rho_0(1-\overline{\lambda}_2)}}-
\frac{\ell_0 }{1-\overline{\lambda}_2}.
\label{linejegeneral}
\end{equation}

The time $t_{eje}$ for which the complete ejection is predicted, $\ell_{in}\left(t_{eje}\right)=0$, follows as
\begin{equation}
t_{eje}=\sqrt{\frac{\rho_0 \overline{\ell}_{in} \left[2 \ell_0+\left(1-\overline{\lambda}_2\right)\overline{\ell}_{in}\right]}{\Phi^l\left(\overline{\lambda}_2\right)}}.
\label{tejegeneral}
\end{equation}

For a 
neo-Hookean material (\ref{mrmat}), the evolution in time of $\ell_{in}(t)$, eqn (\ref{linejegeneral}), and the corresponding  ejection time $t_{eje}$, eqn (\ref{tejegeneral}),  simplify as 
\begin{equation}
\ell_{in}\left(t\right)=
\sqrt{ 
\left(
\frac{\ell_0}{1-\overline{\lambda}_2}+ \overline{\ell}_{in}
\right)^2-\frac{\mu \, t^2
(1+\overline{\lambda}_2)^2 (1-\overline{\lambda}_2)   }{2\rho_0 \overline{\lambda}_2^{\,2}} 
} - 
\frac{\ell_0}
{1-\overline{\lambda}_2},
\qquad
t_{eje}=\frac{ \overline{\lambda}_2 \overline{\ell}_{in}}{1-\overline{\lambda}_2 }
\sqrt{\frac{2 \rho_0 }{\mu } 
\left(
\dfrac{2 \ell_0}{\overline{\ell}_{in}}+1-\overline{\lambda}_2 
\right)}.
\label{linetejeNK}
\end{equation}

\begin{figure}[!h]
\centering
\includegraphics[width=1\textwidth]{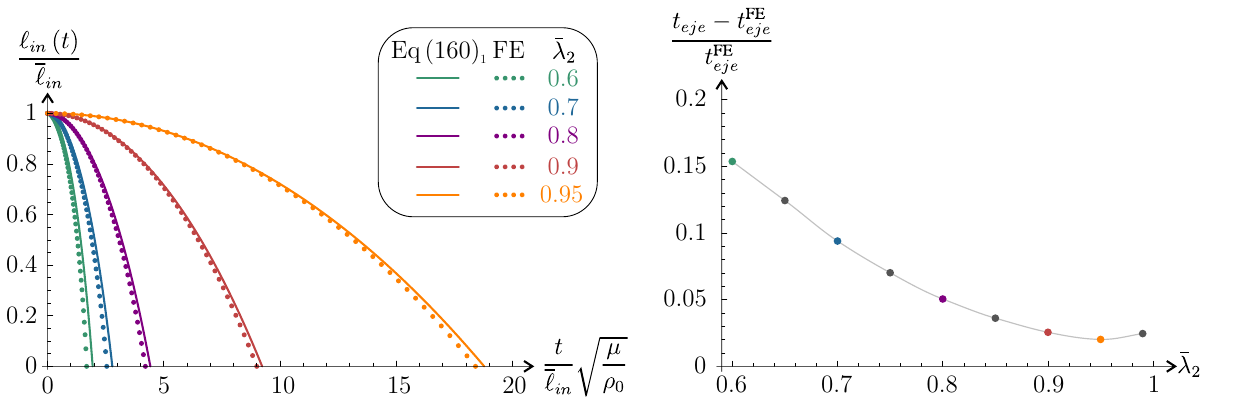}
\caption{Left: Dimensionless representation of the 
instantaneous length 
$\ell_{in}(t)$ 
of material inside the constraint as a function of time, for different nominal transverse  stretch $\overline{\lambda}_2$,  applied to a neo-Hookean material with  $\overline{\ell}_{in}=0.9 \ell_0$. Results from the estimation  (\ref{linetejeNK})$_1$ (continuous lines) are compared with FE simulations (dots). Right: Relative error in the time for complete ejection, where $t_{eje}$ is given by the estimate   (\ref{linetejeNK})$_2$, while $t_{eje}^{\text{FE}}$ by the FE simulations. }
\label{dynamiceje}
\end{figure}

\paragraph{Dynamic ejection from FE simulations.}
The previously described  FE model  is exploited below to analyze the dynamic ejection problem for a neo-Hookean material, subject to a transverse stretch larger than that corresponding to surface instability,  $\overline{\lambda}_2 < \lambda_2^{si}\approx 0.544$. 
The simulations are carried out in two steps, the  first is the static analysis already presented for {\tt BVP1}, while in the second step 
the elastic solid is instantaneously released and the motion analyzed.  
The first step is performed  to start from  $\overline{\ell}_{in}=0.9 \ell_0$, for an undeformed geometry of the material defined by $\ell_0/h_0=20$.  The corner of the punch is modeled as rounded with $r=h_0/10$.

The evolution in time of $\ell_{in}(t)$ is reported in Fig. \ref{dynamiceje} (left), 
with a continuous line for the approximate solution  (\ref{linetejeNK})$_1$ and 
with dots for the FE simulations. Different initial stretches
$\overline{\lambda}_2$
are investigated. 
The agreement is excellent, as can also be observed from the curve on the right, reporting the relative difference in the ejection time 
$(t_{eje}^{\text{FE}}-t_{eje})/t_{eje}^{\text{FE}}$ as a function of $\overline{\lambda}_2$.

Four snapshots of representative configurations at different instants of time (including the  configuration at the release time $t=0$) are reported in Fig. \ref{dynamicejeene}. 
The contour plots reported inside the deformed elastic solid depict the kinetic energy density 
$\tau$ 
(in the upper half-part of the solid) 
and 
elastic strain energy density $\Phi$ (in the lower half-part).
At the release time ($t=0$), the kinetic energy density $\tau$  is null over the whole solid (blue region), while the strain energy density $\Phi$ is almost piecewise uniform, because the  elastic body is slender and the corners of the rigid constraints introduce only a small perturbation. 
As shown by the three snapshots taken after the release time, the spatial piecewise uniformity is essentially maintained for both kinetic and strain energy densities throughout the ejection process, during which a continuous transfer of elastic to kinetic energy occurs.

\begin{figure}[!h]
\centering
\includegraphics[width=1\textwidth]{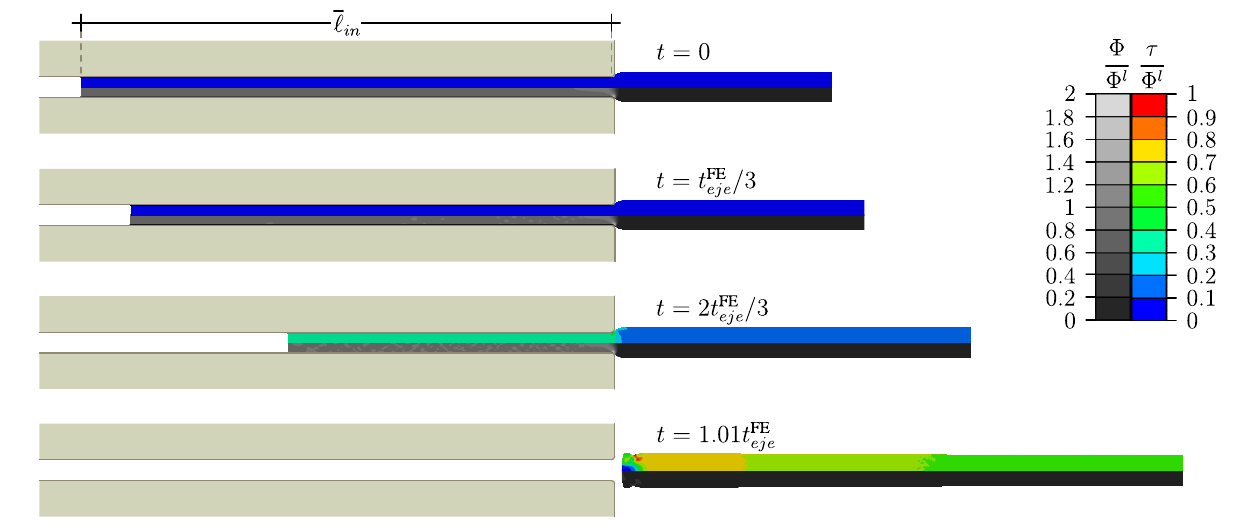}
\caption{
Four snapshots (obtained with a FE analysis) of representative configurations of a neo-Hookean solid (of undeformed aspect ratio $\ell_0/h_0=20$),  during its ejection from a rigid frictionless constraint as the effect of a nominal transverse  stretch $\overline{\lambda}_2=0.7$. 
Different instants of time $t$ 
are considered, including the release time $t=0$, where the kinetic energy is null.
Inside the solid, contour plots are  reported of the 
kinetic energy density $\tau$
(on the upper half-part of the solid) and of 
the elastic strain energy density $\Phi$ (on the lower half-part), both divided by the value of the latter at the left edge  of the elastic domain, $\Phi^l$.
}
\label{dynamicejeene}
\end{figure}

\section{Conclusions}

Concepts developed within the framework of defect mechanics, involving use of energy-momentum tensor and leading to path independent $J$--integral to be equal to the energy release rate for defect movement, have been extended to the mechanics of frictionless contact 
between an elastic solid and a rigid, flat punch. Within a quasi-static setting, it has been shown that a sharp or smooth corner, present at the end of a frictionless punch indenting a planar surface of an elastic solid,  produces a (concentrated, for sharp corner) Newtonian force aligned parallel to the surface and coincident with the configurational force in that direction. This force component, so far passed unnoticed, may influence failure of material at contact or may induce Eulerian buckling or movement in a transversely compressed elastic slab. The investigated deformation mechanisms can be used for soft actuation or to explain migration of soft matter from stiffer to more compliant constrained environment.

\paragraph{CRediT authorship contribution statement.}  F. Dal Corso: Conceptualization; Project administration; Formal analysis; Investigation; Methodology;  Supervision;  Visualization; Writing – original draft; Writing – review
and editing. 
M. Amato:  Formal analysis; Investigation; Methodology; Software; Data curation; Validation; Visualization; Writing – original draft; Writing – review
and editing.
D. Bigoni: Conceptualization; Project administration; Formal analysis; Investigation; Methodology; Supervision; Writing – original draft, Writing – review and editing;
Funding acquisition.

\paragraph{Acknowledgements.} The authors gratefully acknowledge financial support from the European Research Council (ERC)
under the European Union’s Horizon 2020 research and innovation programme, Grant agreement No. 
ERC-ADG-2021-101052956-BEYOND. 
 The methodologies developed in the present work fall within the aims of the GNFM (Gruppo Nazionale per la Fisica Matematica) of the INDAM (Istituto Nazionale di Alta Matematica).

\appendix
\renewcommand{\theequation}{{\thesection.}\arabic{equation}}
\renewcommand\thefigure{\thesection.\arabic{figure}}   

\section{Reaction force at a frictionless contact from the  principle of virtual works}\label{AppendixA}
\setcounter{equation}{0}
\setcounter{figure}{0}

With reference to Fig. \ref{target}, a body subject to (referential) dead volume force $\mathbf{b}_0$,  under prescribed displacement on $\partial \mathcal{B}^u_0$, dead loading $\boldsymbol{\sigma}_0$ on $\partial \mathcal{B}^\sigma_0$, and in frictionless contact with the smooth and rigid constraint on $\partial \mathcal{B}_0^{\scriptsize \mbox{tou}}$, 
is assumed to be in equilibrium with the displacement field $\mathbf{u}^*$,  deformation gradient $\mathbf{F}^*$, first Piola-Kirchhoff stress  $\mathbf{S}^*$. The total potential energy $\mathcal{V}$ of the system at equilibrium  is given by
\begin{equation}
    \mathcal{V}(\mathbf{F}^*, \mathbf{u}^*, 
    \boldsymbol{\sigma}_0 , \mathbf{b}_0) = \int_{\mathcal{B}_0} \Phi(\mathbf{F}^*) - \int_{\partial \mathcal{B}^\sigma_0} 
\boldsymbol{\sigma}_0  \cdot \mathbf{u}^* - \int_{\mathcal{B}_0} \mathbf{b_0} \cdot  \mathbf{u}^*.
\end{equation}
 
A rigid-body displacement vector $\delta \boldsymbol{\xi}$ is applied 
to the rigid constraint, so that as a consequence,  all the fields in the elastic solid are perturbed and, in particular, the displacement and its gradient 
as follows
\begin{equation}
\mathbf{u}(\delta \boldsymbol{\xi}) = \mathbf{u}^* + \delta\mathbf{u}(\delta \boldsymbol{\xi}), ~~~~
\mathbf{F}(\delta \boldsymbol{\xi})= \mathbf{F}^* + \delta\mathbf{F}(\delta \boldsymbol{\xi}) .
\end{equation}
Accordingly, 
the total potential energy of the system at equilibrium after the rigid-body perturbation of the frictionless constraint
becomes
\begin{equation}
    \mathcal{V}(\mathbf{F}(\delta \boldsymbol{\xi}), \mathbf{u}(\delta \boldsymbol{\xi}), \boldsymbol{\sigma}_0, \mathbf{b}_0) = \int_{\mathcal{B}_0} \Phi(\mathbf{F}^*+ \delta\mathbf{F}(\delta \boldsymbol{\xi})) - \int_{\partial \mathcal{B}_0^\sigma} 
\boldsymbol{\sigma}_0  \cdot (\mathbf{u}^*+\delta\mathbf{u}(\delta \boldsymbol{\xi})) - \int_{\mathcal{B}_0} \mathbf{b}_0 \cdot  (\mathbf{u}^*+\delta\mathbf{u}(\delta \boldsymbol{\xi})). 
\end{equation}
The change in the total potential energy $\delta \mathcal{V}$ between the configurations is given by
\begin{equation}
    \delta \mathcal{V}(\mathbf{F}^*, \mathbf{u}^*, \boldsymbol{\sigma}_0 , \mathbf{b}_0, \delta\boldsymbol{\xi})=
    \mathcal{V}(\mathbf{F}(\delta \boldsymbol{\xi}), \mathbf{u}(\delta \boldsymbol{\xi}), \boldsymbol{\sigma}_0, \mathbf{b}_0)
    -
    \mathcal{V}(\mathbf{F}^*, \mathbf{u}^*, 
    \boldsymbol{\sigma}_0 , \mathbf{b}_0).
\end{equation}
Under the assumption that the perturbation $\delta \boldsymbol{\xi}$ is small, implying that $\delta \mathbf{u}$ and $\delta \mathbf{F}$ are of the same order, 
the change in the total potential energy  $\delta \mathcal{V}$ 
can be approximated at first-order as
\begin{equation}
\label{sganapa}
    \delta\mathcal{V}(\mathbf{F}^*, \mathbf{u}^*, \boldsymbol{\sigma}_0 , \mathbf{b}_0, \delta\boldsymbol{\xi}) = \int_{\mathcal{B}_0} 
    \mathbf{S}^* \cdot \boldsymbol{\delta}{\mathbf{F}}(\delta \boldsymbol{\xi}) - \int_{\partial \mathcal{B}^\sigma_0} 
\boldsymbol{\sigma}_0  \cdot \delta \mathbf{u}(\delta \boldsymbol{\xi})- \int_{\mathcal{B}_0} \mathbf{b}_0 \cdot  \delta\mathbf{u}(\delta \boldsymbol{\xi}), 
\end{equation}
where, from  eqn (\ref{const}), 
\begin{equation}
    \mathbf{S}^*=\left.\dfrac{\partial \Phi(\mathbf{F})}{\partial \mathbf{F}}\right|_{\mathbf{F}^*}.
\end{equation}

The  virtual work principle, expressed with reference to the unperturbed static fields and to the perturbed kinematic fields,  yields
\begin{equation}
    \int_{\partial \mathcal{B}_0^{\scriptsize \mbox{tou}}} \delta\mathbf{u} (\delta \boldsymbol{\xi})\cdot \mathbf{S}^*\mathbf{n}_0 = 
\int_{\mathcal{B}_0} 
    \mathbf{S}^* \cdot \delta\mathbf{F}(\delta \boldsymbol{\xi}) - \int_{\partial \mathcal{B}^\sigma_0} 
\boldsymbol{\sigma}_0  \cdot \delta\mathbf{u}(\delta \boldsymbol{\xi}) - \int_{\mathcal{B}_0} \mathbf{b}_0 \cdot  \delta\mathbf{u}(\delta \boldsymbol{\xi}). 
\end{equation}
Note that the detaching from the contact can only occur from the grazing zone, where $\mathbf{S}^*\mathbf{n}_0=\mathbf{0}$, eq. (\ref{grazing}), so that 
\begin{equation}
    \int_{\partial \mathcal{B}_0^{\scriptsize \mbox{tou}}} \delta\mathbf{u}(\delta \boldsymbol{\xi}) \cdot \mathbf{S}^*\mathbf{n}_0 = 
    \int_{\partial \mathcal{B}_0^{C}} \delta\mathbf{u}(\delta \boldsymbol{\xi}) \cdot \mathbf{S}^*\mathbf{n}_0 ,
\end{equation}
proving that detaching does not affect the integral. 

A comparison with equation (\ref{sganapa}) leads to 
\begin{equation}
\label{sganapa4}
    \delta\mathcal{V}(\mathbf{F}^*, \mathbf{u}^*, \boldsymbol{\sigma}_0 , \mathbf{b}_0, \delta \boldsymbol{\xi}) =
\int_{\partial \mathcal{B}_0^{C}} \delta\mathbf{u}(\delta \boldsymbol{\xi}) \cdot \mathbf{S}^*\mathbf{n}_0 .
\end{equation}

It is  noted that in general the perturbed displacement $\delta\mathbf{u}(\delta \boldsymbol{\xi})$ does not coincide with  $\delta\boldsymbol{\xi}$ along $\partial \mathcal{B}^{C}_0$, because the elastic 
body may slip along the boundary in contact. However,  the contact condition
\begin{equation}
    \Sigma(\mathbf{x}^*- \delta\boldsymbol{\xi}) = 0,
\end{equation}
 for small perturbations provides
\begin{equation}
 \mathbf{q}(\boldsymbol{x}^*) \cdot  \left(\delta\mathbf{u}(\delta \boldsymbol{\xi}) -\delta\boldsymbol{\xi} \right)=0,
\end{equation}
where  $\mathbf{q}(\boldsymbol{x}^*)$ is the normal vector to $\Sigma$ at $\boldsymbol{x}^*$
\begin{equation}
    \mathbf{q}(\boldsymbol{x}^*) = \left. \frac{\partial \Sigma (\mathbf{x})}{\partial \boldsymbol{x}} \right|_{\mathbf{x}^*}, 
\end{equation}
which is  parallel to $\mathbf{S}^*\mathbf{n_0}$ because of the frictionless contact condition, eq. (\ref{pugnettone}). Therefore, equation (\ref{sganapa4}) can be rewritten as
\begin{equation}
\label{sganapa8}
    \delta\mathcal{V}(\mathbf{F}^*, \mathbf{u}^*, \boldsymbol{\sigma}_0 , \mathbf{b}_0, \delta\boldsymbol{\xi})  =
\delta \boldsymbol{\xi}\cdot \int_{\partial \mathcal{B}_0^{C}} \mathbf{S}^*\mathbf{n}_0 ,
\end{equation}
because vector $\delta\boldsymbol{\xi}$ is constant.

Equation (\ref{sganapa8}) shows that the total potential energy variation due to a small perturbation $\delta \boldsymbol{\xi}$  in the position of the frictionless constraint is the negative of the scalar product between ${\delta\boldsymbol{\xi}}$ and the resultant of the force $\mathbf{R}^c$ that the constraint applies to the  body,
\begin{equation}
\label{sganapa10}
    \mathbf{R}^c=
\int_{\partial \mathcal{B}_0^{C}} \mathbf{S}^*\mathbf{n}_0 ,
\end{equation}
namely, 
\begin{equation}
\label{sganapa12}
\delta\mathcal{V} =
    \mathbf{R}^c \cdot \delta\boldsymbol{\xi}.
\end{equation}

Finally, noticing that 
\begin{equation}
  \delta \mathcal{V}=  \frac{\partial \mathcal{V}}{\partial \boldsymbol{\xi}} \cdot \delta \boldsymbol{\xi}, 
\end{equation}
due to the arbitrariness of $\delta \boldsymbol{\xi}$, the reaction force transmitted by frictionless constraint to the solid is obtained 
\begin{equation}
\label{suanonna}
    \mathbf{R}^c=\frac{\partial \mathcal{V}}{\partial \boldsymbol{\xi}}.
\end{equation}

\section*{References}
\printbibliography[heading=none]

@incollection{gurtin1999nature,
  title={The nature of configurational forces},
  author={Gurtin, ME},
  booktitle={Fundamental Contributions to the Continuum Theory of Evolving Phase Interfaces in Solids: A Collection of Reprints of 14 Seminal Papers},
  pages={281--314},
  year={1999},
  publisher={Springer}
}

@book{kienzler,
  title={Mechanics
in Material Space},
  author={Kienzler, R and  Herrmann, G},
  year={2000},
  publisher={Springer}
}

@article{mauginino,
  title={Material forces: concepts and applications},
  author={Maugin, GA},
  journal={ASME Applied Mechanics Re-
views},
  volume={48},
  number={5},
  pages={213--245},
  year={1995}
}

@article{locascio,
  title={Cell movement is guided
by the rigidity of the substrate},
  author={Lo, C M and Wang, H B and Dembo, M and Wang, Y L},
  journal={Biophysical Journal},
  volume={79},
  number={},
  pages={144--152},
  year={2000}
}

@article{benvenuti,
  title={Mechanics of tensegrity cell units
incorporating asymmetry and insights
into mollitaxis},
  author={Benvenuti, E and Reho, G A and Palumbo, S and Fraldi, M},
  journal={Journal of the Royal Society Interface},
  volume={20},
  number={20230082},
  pages={},
  year={2023}
}

@incollection{toupin,
title = {The classical Field theories},
editor = {Fl\"{u}gge  S},
series = {Encyclopedia of Physics},
publisher = {Springer-Verlag},
volume = {III/1},
pages = {},
year = {1960},
issn = {},
author = {Truesdell, C and Toupin, RA}
}

@article{cellulona,
  title={Directed cell migration towards softer
environments},
  author={Isomursu, A and  Park, K-Y, and Hou, J and Cheng, B and Mathieu, M and  Shamsan, GA and Fuller, B and Kasim, J and Mahmoodi, MM and Lu, TJ and Genin, G.M and Xu, F and Lin, M and Distefano, MD and Ivaska, J. and Odde, DJ},
  journal={Nature Materials},
  volume={21},
  number={877},
  pages={1081--1090},
  year={2022},
  publisher={}
}

@article{flanders,
  title={Differentiation under the integral sign},
  author={Flanders, H},
  journal={The American Mathematical Monthly},
  volume={80},
  number={},
  pages={615--627},
  year={1973},
  publisher={}
}

@article{Cherepanov,
  title={The propagation of cracks in a continuous medium},
  author={Cherepanov, GP},
  journal={Journal of Applied Mathematics and Mechanics},
  volume={31},
  number={3},
  pages={503--512},
  year={1967},
  publisher={}
}

@article{bigoniRLI,
  title={The stress concentration near a rigid line inclusion in a prestressed, elastic material.
Part II. Implications on shear band nucleation, growth and energy release rate},
  author={Bigoni, D. and Dal Corso, F and Gei, M},
  journal={Journal of the Mechanics and Physics of Solids},
  volume={56},
  pages={839--857},
  year={2008}
}

@article{goudarzi,
  title={Dispersion of rigid line inclusions as stiffeners and shear band instability triggers},
  author={Goudarzi, M and Dal Corso, F and Bigoni, D and Simone, A},
  journal={International Journal of Solids and Structures},
  volume={210--211},
  pages={255--272},
  year={2021}
}

@article{eshelby1951force,
  title={The force on an elastic singularity},
  author={Eshelby, JD},
  journal={Philosophical Transactions of the Royal Society of London. Series A, Mathematical and Physical Sciences},
  volume={244},
  number={877},
  pages={87--112},
  year={1951},
  publisher={The Royal Society London}
}

@incollection{eshelby1956lattice,
title = {The Continuum Theory of Lattice Defects},
editor = {Seitz, F and Turnbull, D},
booktitle = {Progress in Solid State Physics},
publisher = {Academic Press},
volume = {3},
pages = {79-144},
year = {1956},
issn = {0081-1947},
author = {Eshelby, J.D.}
}

@article{bigonitorsional,
  title={Torsional locomotion },
  author={Bigoni, D and Dal Corso, F and Misseroni, D and Bosi, F},
  journal={Proceedings of the Royal Society A},
  volume={470},
  pages={20140599},
  year={2014}
}

@Article{storaakers1986material,
  author    = {Stor{\aa}kers, B},
  journal   = {Journal of the Mechanics and Physics of Solids},
  title     = {On material representation and constitutive branching in finite compressible elasticity},
  year      = {1986},
  number    = {2},
  pages     = {125--145},
  volume    = {34},
  publisher = {Elsevier},
}

@Book{bigoni2012nonlinear,
  author    = {Bigoni, Davide},
  publisher = {Cambridge University Press},
  title     = {Nonlinear solid mechanics: bifurcation theory and material instability},
  year      = {2012},
}

@Book{barbercontactbook,
  author    = {Barber, J },
  publisher = {Springer},
  title     = {Contact Mechanics},
  year      = {2018},
}

@Article{pence2015compressible,
  author    = {Pence, TJ and Gou, K},
  journal   = {Mathematics and Mechanics of Solids},
  title     = {On compressible versions of the incompressible neo-Hookean material},
  year      = {2015},
  number    = {2},
  pages     = {157--182},
  volume    = {20},
  publisher = {Sage Publications Sage UK: London, England},
}

@article{ballarini,
  title={A Newtonian interpretation of configurational forces on dislocations and cracks},
  author={Ballarini, R and Royer-Carfagni, G},
  journal={Journal of the Mechanics and Physics of Solids},
  volume={95},
  pages={602--620},
  year={2016}
}

@article{koutso2023,
  title={Stabilization against gravity and self-tuning of an elastic variable-length rod through an oscillating sliding sleeve.},
  author={Koutsogiannakis, P and Misseroni, D and Bigoni, D and Dal Corso, F},
  journal={Journal of the Mechanics and Physics of Solids},
  volume={181},
  pages={105452},
  year={2023}
}

@article{eshelby1975,
  title={The elastic energy-momentum tensor},
  author={Eshelby, JD},
  journal={Journal of Elasticity},
  volume={5},
  number={3-4},
  pages={321--335},
  year={1975},
  publisher={Noordhoff}
}

@article{rice1968path,
  title={A path independent integral and the approximate analysis of strain concentration by notches and cracks},
  author={Rice, JR},
journal={Journal of Applied Mechanics},
  volume={35},
  number={2},
  pages={379--386},
  year={1968}
}

@incollection{rice1968,
title = {Mathematical Analysis in the Mechanics of Fracture},
editor = {Liebowitz, H},
series = {Fracture: An Advanced Treatise},
publisher = {Academic Press},
volume = {2},
pages = {191-311},
year = {1968},
issn = {},
doi = {},
url = {},
author = {Rice, JR}
}

@article{Xie2009,
title = {Applications of conservation integral to indentation with a rigid punch},
journal = {Engineering Fracture Mechanics},
volume = {76},
number = {7},
pages = {949--957},
year = {2009},
author = {Xie, YJ and Lee, KY and Hu, XZ and  Cai, YM},
}

@article{Xie2003,
  title={Crack initiation at contact surface},
  author={Xie, YJ and Hills, DA},
  journal={Theoretical and Applied Fracture Mechanics},
  volume={40},
  number={3},
  pages={279--283},
  year={2003}
}

@article{Ma2006,
  title={Vector J-integral analysis of crack initiation at the edge of complete sliding contact},
  author={Ma, L and Korsunsky, AM},
  journal={Proceedings of the Royal Society A: Mathematical, Physical and Engineering Sciences},
  volume={462},
  pages={1805--1820},
  year={2006}
}

@article{Ma2008,
title = {Surface dislocation nucleation from frictional sliding contacts},
journal = {International Journal of Solids and Structures},
volume = {45},
number = {22--23},
pages = {5936--5945},
year = {2008},
author = {Ma, L and Korsunsky, AM}
}

@article{Giannakopoulos1998,
  title={Aspects of equivalence between contact mechanics and fracture mechanics: theoretical connections and a life-prediction methodology for fretting-fatigue},
  author={Giannakopoulos, AE and Lindley, TC and Suresh, S},
  journal={Acta Materialia},
  volume={46},
  number={9},
  pages={2955--2968},
  year={1998}
}

@book{Johnsonbook,
  title={Contact mechanics},
  author={Johnson, KL},
  year={1985},
  publisher={Cambridge University Press }
}

@book{Saddbook,
  title={Elasticity: Theory, Applications, and Numerics},
  author={Sadd, MH},
  year={2020},
  publisher={Academic Press }
}

@article{ARMANINI201982,
title = {Configurational forces and nonlinear structural dynamics},
journal = {Journal of the Mechanics and Physics of Solids},
volume = {130},
pages = {82-100},
year = {2019},
issn = {0022-5096},
author = {Armanini, C  and Dal Corso, F  and Misseroni, D and Bigoni, D},
}

@article{bigoni2015eshelby,
  title={Eshelby-like forces acting on elastic structures: theoretical and experimental proof},
  author={Bigoni, D and Dal Corso, F and Bosi, F and Misseroni, D},
  journal={Mechanics of Materials},
  volume={80},
  pages={368--374},
  year={2015},
  publisher={Elsevier}
}

@article{o2015some,
  title={Some perspectives on Eshelby-like forces in the elastica arm scale},
  author={O'Reilly, OM},
  journal={Proceedings of the Royal Society A: Mathematical, Physical and Engineering Sciences},
  volume={471},
  number={2174},
  pages={20140785},
  year={2015},
  publisher={The Royal Society Publishing}
}

@article{dalcorsosnake,
  title={Serpentine locomotion through elastic energy release},
  author={Dal Corso, F and Misseroni, D and Pugno, NM and Movchan, AB and Movchan, NV, and Bigoni, D},
  journal={Journal of the Royal Society Interface},
  volume={14},
  pages={20170055},
  year={2017},
  publisher={The Royal Society Publishing}
}

@article{ciavarella1998,
  title={The influence of rounded edges on indentation by a flat punch},
  author={Ciavarella, M and Hills, DA and Monno, G},
  journal={Proceedings of the Institution of Mechanical Engineers, Part C: Journal of Mechanical Engineering Science},
  volume={212},
  number={4},
  pages={319--328},
  year={1998},
  publisher={Sage}
}

@article{ciavarella2002,
  title={On stress concentration on nearly flat contacts},
  author={Ciavarella, M and Macina, G and Demelio, GP},
  journal={The Journal of Strain Analysis for Engineering Design},
  volume={37},
  number={6},
  pages={493--501},
  year={2002},
  publisher={Sage}
}

@article{chadwick,
  title={Applications of an energy-momentum tensor in non-linear elastostatics},
  author={Chadwick, P},
  journal={Journal of Elasticity},
  volume={5},
  pages={249--258},
  year={1975},
  publisher={Springer}
}

@article{kim2023,
  title={Study on the edge rounding with consulting contact mechanics},
  author={Kim, HK},
  journal={Tribology International},
  volume={183},
  pages={108426},
  year={2023},
  publisher={Elsevier}
}

@article{MAGNUSSON20018441,
title = {Behaviour of the extensible elastica solution},
journal = {International Journal of Solids and Structures},
volume = {38},
number = {46},
pages = {8441--8457},
year = {2001},
author = {Magnusson, A and Ristinmaa, M and Ljung, C},
}

@article{BosiInjection,
  title={Development of configurational forces during the injection of an elastic rod},
  author={Bosi, F and Misseroni, D and Dal Corso, F  and Bigoni, D},
  journal={Extreme Mechanics Letters},
  volume={471},
  pages={83--88},
  year={2015}
}

@article{Liakou2018,
  title={Constrained buckling of variable length elastica: Solution by geometrical segmentation},
  author={Liakou, A and Detournay, E},
  journal={International Journal of Non-Linear Mechanics},
  volume={99},
  pages={204--217},
  year={2018}
}

@article{Wang2022,
  title={Eshelbian force on a steadily moving liquid blister},
  author={Wang, ZQ and Detournay, E},
  journal={International Journal of Engineering Science},
  volume={170},
  pages={103591},
  year={2022}
}

@article{elasticasling,
  title={The elastica sling},
  author={Cazzolli, A and Dal Corso, F},
  journal={European Journal of Mechanics, A/Solids},
  volume={105},
  number={3},
  pages={105273},
  year={2024},
  publisher={Elsevier BV}
}

@article{o2007material,
  title={A material momentum balance law for rods},
  author={O’Reilly, OM},
  journal={Journal of Elasticity},
  volume={86},
  pages={155--172},
  year={2007},
  publisher={Springer}
}

@book{o2017book,
  title={Modeling nonlinear problems in the mechanics of strings and rods},
  author={O'Reilly, OM},
  year={2017},
  publisher={Springer}
}

@article{hanna2018partial,
  title={Partial constraint singularities in elastic rods},
  author={Hanna, JA and Singh, H and Virga, EG},
  journal={Journal of Elasticity},
  volume={133},
  number={1},
  pages={105--118},
  year={2018},
  publisher={Springer}
}

\end{document}